\newcommand{\dpr}{{\prime \prime}}

\newcommand{\ret}{Re_\tau}
\newcommand{\reb}{Re_{bulk}}
\newcommand{\utau}{u_\tau}
\newcommand{\ubulk}{U_{bulk}}

\newcommand{\upp}{{u_x^{\dpr}}}
\newcommand{\upr}{{u_x^\prime}}
\newcommand{\vpp}{{u_r^{\dpr}}}
\newcommand{\vpr}{{u_r^\prime}}
\newcommand{\wpp}{{u_\theta^{\dpr}}}

\newcommand{\av}[1]{\left\langle {#1} \right\rangle}

\newcommand{\rstar}{{r^\star}}

\documentclass{jfm}
\usepackage[svgnames]{xcolor}
\usepackage{graphicx}
\usepackage{subcaption}
\usepackage{hyperref,url}
\usepackage{tikz,natbib}
\usepackage{nicefrac}
\usepackage{float,morefloats}
\usepackage[mathscr]{euscript}
\usepackage[export]{adjustbox}
\usepackage{scalerel,stackengine}
\usepackage[normalem]{ulem}
\usepackage{mathrsfs}
\usepackage{mathtools}
\usepackage{bbding}

\restylefloat{table}
\stackMath
\newcommand\reallywidehat[1]{%
\savestack{\tmpbox}{\stretchto{%
  \scaleto{%
    \scalerel*[\widthof{\ensuremath{#1}}]{\kern-.6pt\bigwedge\kern-.6pt}%
    {\rule[-\textheight/2]{1ex}{\textheight}}
  }{\textheight}%
}{0.5ex}}%
\stackon[1pt]{#1}{\tmpbox}%
}

\shorttitle{Reynolds number effect on drag reduction}
\shortauthor{D. J. Coxe, Y. T. Peet, R. J. Adrian}

\title{Reynolds number effect on drag reduction in pipe flows by a transverse wall oscillation}

\author{Daniel Coxe\aff{1
  \corresp{\email{dcoxe@asu.edu}}},
  Yulia Peet\aff{1}
 \and Ronald Adrian\aff{1}}

\affiliation{\aff{1}Arizona State University, School for Engineering of Matter Transport and Energy,
Tempe, AZ 85281, USA}

\begin{document}

\maketitle

\begin{abstract}
\end{abstract}

\begin{keywords}
Turbulent Pipe Flow, Transverse wall oscillation, drag reduction, energy spectra, large-scale motions
\end{keywords}
\abstract {Direct numerical simulations of turbulent pipe flow with transverse wall oscillation (WWO) and with no transverse wall oscillation (NWO) are carried out at friction Reynolds numbers   $Re_\tau = 170, 360, \text{ and } 720$. The period and amplitude of the oscillation  are selected to achieve high drag reduction in this Reynolds number range, and the effect of increasing Reynolds number on the amount of drag reduction achievable is analyzed. Of a particular interest in this study is the identification of the scales of motion most affected by drag reduction at different Reynolds numbers. To answer this question, both one-dimensional and two-dimensional spectra of different statistical quantities are analyzed with and without transverse wall oscillation. The effect of wall oscillation is found to suppress the intermediate- and large-scale motions in the buffer layer of the flow, while large-scale and very-large-scale motions in the log layer and the wake region are enhanced. While suppression of the near-wall turbulence promotes drag reduction, enhancement of the large-scale motions in the log and the wake region is found to oppose drag reduction. Since higher Reynolds number flows support development of a growing range of large-scale structures, it is suggested that their prevalence in the energy spectra combined with their negative effect on drag reduction account for a reduced effectiveness of wall oscillation as a drag reduction mechanism with increasing Reynolds numbers.}

\section{Introduction}\label{sec:intro}
Reduction of skin friction drag in turbulent flows is a highly-sought outcome of passive and active flow control techniques, especially given that skin friction drag contributes approximately 50\%, 90\% and 100\% of the total drag on airplanes, submarines and pipelines, respectively ~\citep{gad1994interactive, abbassi2017skin}. Most of the studies on drag reduction, both numerical and experimental, were performed in a relatively low Reynolds number regime, typically around $Re_{\tau}\sim 200-400$ in simulations, and $Re_{\tau} < 2000$ in experiments~\citep{gatti2016reynolds, ricco2021review, marusic2021energy}, where $Re_{\tau}$ denotes the friction Reynolds number. In contrast, typical Reynolds numbers encountered in applications attain the values of $Re_{\tau}=4000$ on a wind turbine blade or in a long-distance pipeline, $6000$ at the mid-span of a Boeing 787 wing, and $10^4-10^5$ along the length of a 787 fuselage during cruise~\citep{leschziner2020friction, marusic2021energy}.  Consequently, understanding the effects of Reynolds number on drag reducing techniques becomes the matter of the utmost importance.

A decrease of effectiveness with increasing Reynolds number was reported for drag reduction (DR) techniques that employ transversely-oscillated walls~\citep{touber2012near,yao2019reynolds}, streamwise-traveling waves~\citep{gatti2013performance,hurst2014effect,gatti2016reynolds,marusic2021energy},  opposition control~\citep{choi1994active,chang2002viscous,iwamoto2002reynolds,deng2016origin}, superhydrophobic surfaces~\citep{rastegari2019drag}, anisotropic
permeable substrates~\citep{gomez2019turbulent}, and micro-bubble injections~\citep{ferrante2005reynolds} as the methods of flow control. While the trend seems to persist across a variety of established DR methods, the physical reasons behind this loss of performance remain elusive.

Earlier studies have attributed the decrease of performance of DR techniques with Reynolds number to geometrical effects of ``shrinking'' of the near-wall layer affected by control (which is constant in wall units with $Re_{\tau}$) with respect to the overall domain height (which increases in wall units  with $Re_{\tau}$)~\citep{ferrante2005reynolds,gatti2016reynolds}. Consistent with this reasoning, the arguments have also been proposed explaining drag reduction via an upward shift of the mean velocity profile in the logarithmic region~\citep{gad2000flow}. This shift, in wall units, was postulated to be independent of the Reynolds number. Therefore, scaled with bulk mean velocity in wall units of uncontrolled flow, this constant shift would lead to a lower percentage of drag reduction as Reynolds number increases~\citep{gatti2016reynolds,rastegari2019drag,gomez2019turbulent}. While these arguments offer important insights regarding the observable outcomes of drag reduction and their trends, further elucidation is needed to explain why the applied mechanisms of flow control affect primarily the near-wall layer of the flow and what modifications are required to increase their effectiveness at higher Reynolds numbers.

In recent years, attention has turned to investigating the contributions of different scales of motion both to skin friction~\citep{deck2014large,agostini2019connection,duan2021contributions}, and to skin friction reduction~\citep{agostini2014spanwise,cormier2016interaction,deng2016origin,zhang2020characteristics,agostini2021statistical,chan2022large}. It has been shown that large-scale motions~\citep{guala2006large,balakumar2007large,hutchins2007evidence} grow stronger in high-Reynolds number flows~\citep{marusic2010high,smits2011high} and, consequently, their contribution to skin friction drag increases~\citep{yao2019reynolds,marusic2021energy}.  In their recent investigation, \cite{yao2019reynolds} hypothesized that a decrease of drag reduction effectiveness with an increase in Reynolds number in a channel flow controlled by a transverse wall oscillation with a non-dimensional oscillation period of $T^+=100$ may be attributed to a weakened effectiveness of control in suppressing the near-wall large-scale turbulence, whose contribution to skin friction drag progressively increases. \cite{marusic2021energy} have come to a similar conclusion for a turbulent boundary layer controlled by a streamwise-inhomogeneous transverse wall oscillation and investigated the effectiveness of surface motions conducted at much larger oscillation periods of $T^+\approx 600-1000$ geared towards a targeted manipulation of the large-scale structures.  

The current manuscript investigates the effect of Reynolds number on drag reduction in a turbulent pipe flow using transverse wall oscillations. 
This method of flow control has received considerable attention in the literature~\citep{leschziner2020friction,ricco2021review}, aided by the following advantages: 1) It is relatively easy to set up in experiments and simulations; 2) It is void of additional complicated physics  as occurs, for example, with the injection of micro-bubbles~\citep{kodama2000experimental,ferrante2005reynolds}, polymers~\citep{warholic1999influence,kim2007effects}, or with the utilization of compliant~\citep{gad2002compliant,esteghamatian2022spatiotemporal}, superhydophobic~\citep{lee2011underwater,rastegari2019drag}, or porous surfaces~\citep{gomez2019turbulent,du2022experimental}; 3) It provides relatively high values of drag reduction (on the order of 30\%--50\%~\citep{quadrio2004critical,hurst2014effect}). Despite a large number of studies devoted to the effect of wall oscillations, majority of these studies have been performed in a setting of a plane wall geometry, such as in a channel or a canonical boundary layer~\citep{leschziner2020friction,ricco2021review}. In fact, for pipe flows, the majority of investigations were limited to $Re_{\tau}\lessapprox 170$~\citep{quadrio2000numerical,choi2002drag,duggleby2007dynamical,auteri2010experimental,liu2022turbulence}, with the exception of an experimental study by~\cite{choi1998drag}, where two cases with $Re_{\tau}=652$ and $Re_{\tau}=962$ were reported; however, they varied the non-dimensional amplitude of wall oscillations with the Reynolds number, which makes it hard to separate the effect of Reynolds number from that of an increased amplitude of wall oscillations  in this setting.  Turbulent pipe flow, with an obvious application to pipeline transport, is a canonical configuration, which has some similarities, but also significant differences with   channel and boundary layer flows~\citep{monty2009comparison,smits2011high}. It is therefore important to assess whether previous conclusions drawn predominantly from flat wall configurations (channel flows and boundary layers) on the effect of Reynolds number on drag reduction with transversely oscillated walls, hold in pipes. The current study aims to fill this gap. We are particularly interested in characterizing the dominant modifications that transverse wall oscillations inflict on different scales of motion in a pipe flow, including their energy content, shear stress spectra, net turbulent force and a turbulent contribution to the bulk mean velocity at different Reynolds numbers, with the ultimate goal of explaining the reason for a reduction in DR effectiveness with Reynolds number in turbulent pipe flows.

The paper is organized as follows. Section~\ref{sec:setup} presents the problem setup and the details of the numerical methodology. Section~\ref{sec:results} summarizes results, including validation and a detailed spectral analysis of turbulent quantities in a pipe flow with and without wall oscillation at three Reynolds numbers, $Re_{\tau}=170, 360$ and $720$. Section~\ref{sec:conclusions} presents conclusions.

\section{Problem Setup}\label{sec:setup}
\subsection{Geometry and domain configuration}

\begin{figure}
    \centering
    \begin{subfigure}{0.48\linewidth}
        \includegraphics[width=0.95\linewidth]{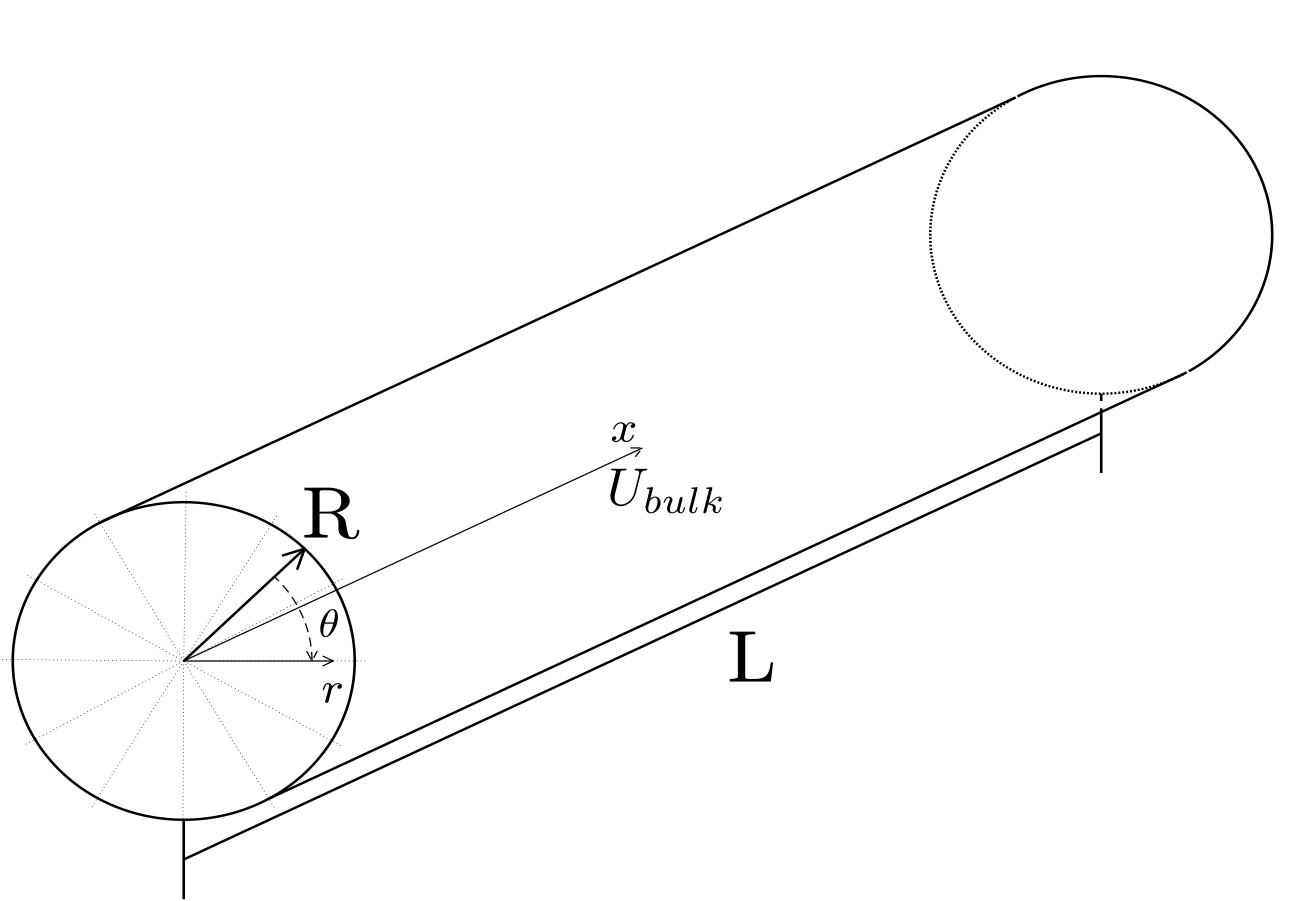}
        \caption{}
        \label{fig:domStd}
    \end{subfigure}
    \begin{subfigure}{0.48\linewidth}
        \includegraphics[width=0.95\linewidth]{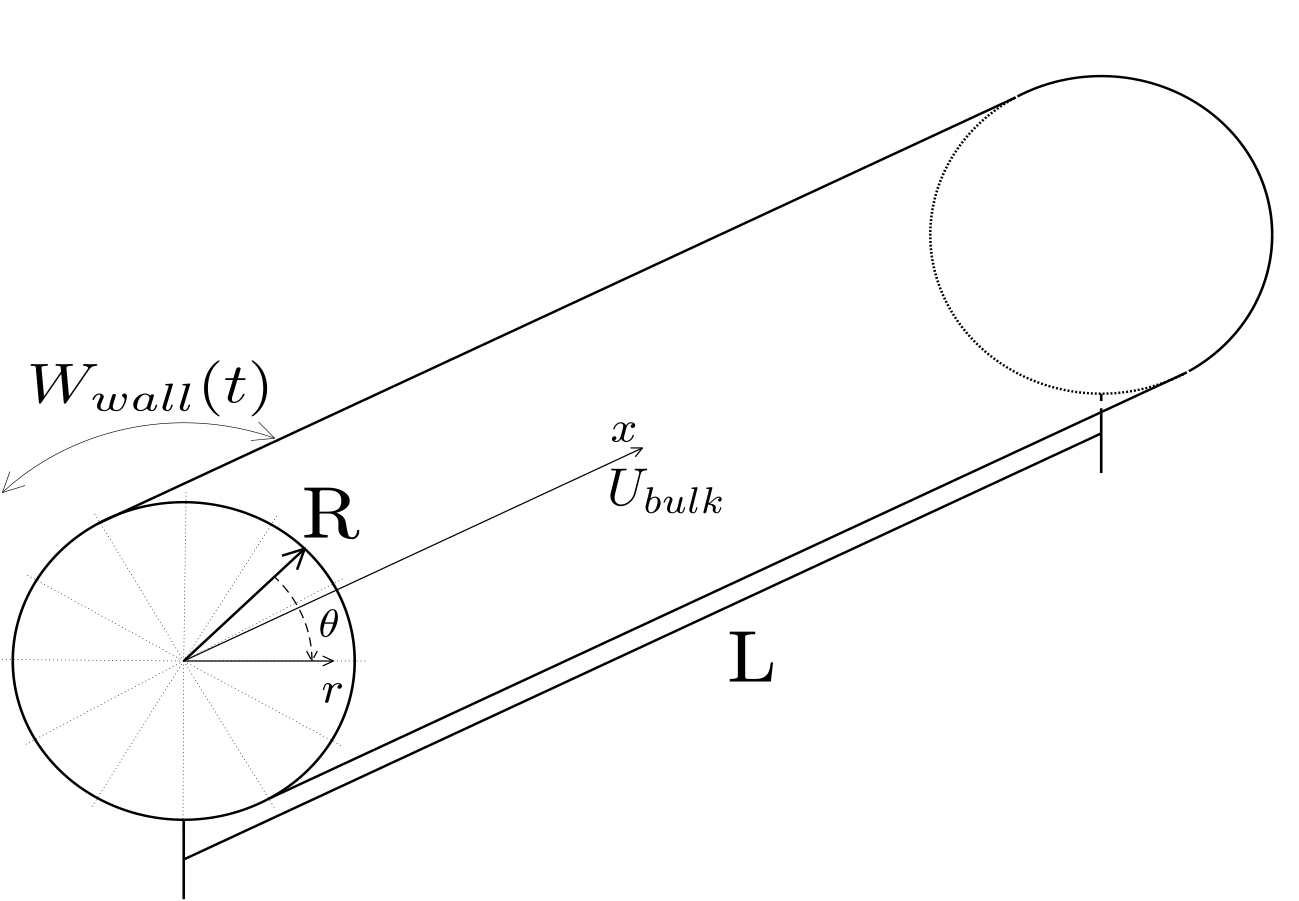}
        \caption{}
        \label{fig:domOsc}
    \end{subfigure}
    \caption{Pipe flow configuration for (a) NWO case, and (b) WWO case. Cylindrical coordinate system $(x,r,\theta)$ represents streamwise, radial and azimuthal directions, respectively. $R$ is the pipe radius, and $L$ is the pipe length. }
    \label{fig:domShape}
\end{figure}

In this study, a pipe flow with an azimuthally oscillated wall (WWO case) is considered as a prototypical configuration to achieve drag reduction, and compared to a standard pipe flow without a wall oscillation (NWO case), each having the same mean pressure drop over the length, $L=24R$, where $R$, is the radius of the pipe,  Figure \ref{fig:domShape}. The cylindrical coordinate system $(x,r,\theta)$ represents streamwise, radial and azimuthal directions, respectively, with the unit vectors $\left(\Vec{\mathbf{e}}_x,\,\Vec{\mathbf{e}}_r,\,\Vec{\mathbf{e}}_{\theta}\right)$ in each direction, and the corresponding velocity vector is $\mathbf{u}=(u_x,u_r,u_{\theta})$. The wall oscillation is achieved in the WWO case by specifying an azimuthal pipe wall velocity as
\begin{equation}
W_{wall}(t)=W_0 \sin(\frac{2\pi}{T_0}t),
\end{equation}
where $W_0$ is the amplitude, and $T_0$ is the period of the wall oscillation.

We define the following notations for the globally averaged quantities. Angle brackets without the indices will represent the quantities averaged over streamwise and azimuthal directions, and in time:
\begin{equation}
\av{f}(r)=\frac{1}{2\pi L T}\int_{0}^T\int_0^L
\int_0^{2\pi} f(x,r,\theta,t) \,dt\,dx \,d\theta,
\end{equation}
with $T$ being the averaging time.
For averaging in time only, angle brackets with the subscript $t$ will be used:
\begin{equation}
\av{f}_t(x,r,\theta)=\frac{1}{T}\int_{0}^T f(x,r,\theta,t) \,dt.
\end{equation}
The functional dependencies in parentheses for the averaged quantities will be omitted for brevity when it is clear from the context.

We define the friction Reynolds number, $Re_\tau = u_\tau R/\nu$,  where $u_\tau = \sqrt{\av{\tau_w}/\rho}$ is the friction velocity, $\av{\tau_w}$ is the mean wall shear stress, $\rho$ is the density, and $\nu$ is the kinematic viscosity. The bulk Reynolds number is $Re_{bulk} =2\, U_{bulk} R/\nu$, with  
\begin{equation}\label{eqn:Ubulk}
    U_{bulk} = \frac{2}{R^2} \int_0^R \av{u_x}(r) \,r \,dr
\end{equation} 
being the bulk mean velocity. 

In the current setup, the mean wall shear stress, $\av{\tau_w}$, and hence $Re_{\tau}$, are fixed between the NWO and WWO cases, while the volumetric flow rate, hence bulk mean velocity, is allowed to vary. Consequently, the drag reduction  which arises as a result of wall oscillation (WWO case) is manifested by an increase in $U_{bulk}$ and a concomitant increase in $Re_{bulk}$. As is well known~\citep{panton1984incompressible}, the mean wall shear stress and the mean pressure gradient are related by a simple force balance that leads to 
\begin{equation}\label{eqn:dpdx}
\av{\frac{\partial\, p}{\partial x}}=-2 \frac{\av{\tau_w}}{R}.
\end{equation} Hence, the NWO and WWO cases also have identical mean pressure gradients.

\subsection{Test cases and flow parameters}
This study considers three Reynolds numbers, $Re_\tau = 170, 360, \text{ and } 720$, leading to six total cases considered (two cases - NWO and WWO - per Reynolds number). Viscous wall units are defined by introducing the friction velocity $u_\tau$, the viscous wall length scale, $l_{\tau} = \nu/ \utau$ and viscous wall time scale $t_\tau = \nu/u_\tau^2$, and non-dimensionalized variables are denoted by superscript `+': $W^+=W/u_{\tau}$, $L^+=L/l_{\tau}$, and $T^+=T/t_{\tau}$. Whence,
\begin{equation}
    W^+_{wall}(t) = W_0^+ \sin \left( \frac{2\pi}{T_0^+} t \right).
\end{equation}
We set the values of $W_0^+=10$ and $T_0^+=100$ as the non-dimensional amplitude and  period of wall oscillations, fixed across all three Reynolds numbers. These values are chosen from a set of parametric studies~\citep{jung1992suppression,baron1995turbulent, choi1998drag,  quadrio2004critical}  that demonstrated high values of drag reduction with these parameters within the range of Reynolds numbers $\ret=200-500$.

Setting viscosity and density in all the cases as $\nu=10^{-6} \,m^2/s$ and $\rho=1000 \,kg/m^3$ (considering water as a carrier fluid)  and specifying a nominal pipe radius as $R=0.1\,m$ allows us to calculate dimensional and non-dimensional wall scaling parameters for the cases as listed in Table~\ref{tab:viscunits}. Note that with these definitions, the  pipe radius in viscous wall units is $R^+ = R/l_{\tau}=Re_\tau$, and $L^+ = 24 Re_\tau$. Since $\av{\tau_w}$, hence $u_{\tau}$ is kept constant between the NWO and WWO cases, all the values listed in Table~\ref{tab:viscunits} are the same between the NWO and WWO cases for the same $Re_{\tau}$. 

\begin{table}
    \centering
    \begin{tabular}{c c c c c c c}
         $\ret$ & $\utau (m/s)$ & $ -\av{\frac{\partial p}{\partial x}} (Pa) $ & $l_\tau (m) $ & $t_\tau (s)$  &  $R^+$ & $L^+$ \\ \hline
         170 & $0.0034$ & $0.02312$ & $2.9\times 10^{-4}$ & $0.0865$  &  170 & 4080\\
         360 & $0.0072$ & $0.10368$ & $1.4\times 10^{-4}$ & $0.0193$  & 360 & 8640\\
         720 &$ 0.0144$ & $0.41472$ & $6.9\times 10^{-5}$ & $0.0048$  & 720 & 17280\\ 
    \end{tabular}
    \caption{List of wall scaling parameters for a pipe with the radius $R=0.1m$, with water flowing through it of viscosity $\nu = 10^{-6}\, m^2/s$ and density $\rho = 1000\,kg/m^3$.}
    \label{tab:viscunits}
\end{table}
\label{sec:method}
A turbulent flow representation in drag-reduced flow with wall oscillation (WWO) and non-drag reduced flow with no wall oscillation (NWO) configurations  across the three Reynolds numbers is obtained via Direct Numerical Simulations (DNS). The governing equations are the incompressible Navier-Stokes' equations
\begin{equation}
	\nabla \cdot \mathbf{u} = 0,
	\label{eqn:continuity}
\end{equation}
\begin{equation}
	\frac{\partial \mathbf{u}}{\partial t} + (\mathbf{u}\cdot\nabla)\mathbf{u} = -\frac{1}{\rho}\nabla p\,' + \nu \nabla^2\mathbf{u} -\frac{1}{\rho}\,\av{\frac{\partial p}{\partial x}} \Vec{\mathbf{e}}_x,
	\label{eqn:momentum}       
\end{equation}
where $\mathbf{u}$ is the velocity, $\nu$ is the kinematic viscosity, $\rho$ is the density, $p\,'$ is the fluctuating pressure defined as $p'=p-\av{\partial p/\partial x}  x$, where $\av{\partial p/\partial x}$ is the constant mean pressure gradient supplied externally to balance the wall shear stress (See equation (\ref{eqn:dpdx})). Numerically, external pressure gradient is treated as a spatially-invariant body force.  

Equations (\ref{eqn:continuity}), (\ref{eqn:momentum}) are numerically solved using an open-source spectral element solver Nek5000~\citep{fischer_book, fischer2015nek5000}. The spectral element method (SEM), similar to a finite element methodology, decomposes a computational domain into a collection of elements,  but it utilizes high-order basis functions within each element, specifically, high-order Lagrange interpolating polynomials associated with the Gauss-Legendre-Lobatto quadrature points. The method employed in the current study leverages the chosen polynomial approximation in a tensor-product form that allows for a fast convergence in multiple dimensions. Thus, Nek5000 solves the governing equations (\ref{eqn:continuity}), (\ref{eqn:momentum}) in a Cartesian coordinate system on hexahedral grids. For the temporal integration, a third-order backward-differentiation formula is employed for the viscous terms, and an explicit third-order extrapolation for the convective terms. To achieve a divergence-free solution, incompressible Navier-Stokes equations are solved with the operator splitting technique. The resulting Helmholtz and Poisson equations are solved with the preconditioned conjugate gradient (PCG) method, and the generalized mean residual  (GMRES) method, respectively~\citep{fischer1997overlapping,tufo2001fast}. For an approximation of a cylindrical geometry, an unstructured hexahedral grid is employed with an exact curved edge treatment of a cylindrical surface~\citep{fischer2015nek5000}. The results are subsequently transferred onto a cylindrical grid using high-order polynomial interpolation~\citep{fischer2015nek5000,merrill2016spectrally} to perform spectral analysis of the data. Spectral-element methods provide minimal dissipation and dispersion errors and are advantageous for DNS of turbulent flows~\citep{kreiss1972comparison, Wang2012}. Previous application of Nek5000 to DNS simulations of turbulent pipe flows can be found in~\cite{duggleby2007effect,duggleby2007dynamical,el2013direct,merrill2016spectrally}.

\subsection{Numerical grids and boundary conditions}

Numerical grid parameters employed in the current study for the three different
Reynolds numbers are listed in Table \ref{table:gridres}. All simulations were executed using ninth-order polynomials as basis functions. Periodic boundary conditions are applied in a streamwise direction on both the velocity and the fluctuating pressure $p'$. 
No-slip boundary conditions are set up at the pipe wall, with $\mathbf{u}_{wall}(t)=(u_x,u_r,u_\theta)=0$ in the NWO case, and $\mathbf{u}_{wall}(t)=(u_x,u_r,u_\theta)=(0,0,W_{wall}(t))$ in the WWO case.

To initialize the simulations, the lowest Reynolds number NWO case, $Re_\tau = 170$, was started by superimposing wave-like perturbations onto a mean velocity profile $(U_{mean}(r),0,0)$~\citep{hillewaert2017ws2} as
\begin{align}
    u_x(x,r,\theta,0) = U_{mean}(r) + 0.01 (\beta)\sin{\left(\alpha x\right)}\sin{\left(\beta \,\theta\right)}, \nonumber \\ 
    u_r(x,r,\theta,0) = 0.01 \sin{\left(\alpha x\right)}\sin{\left(\beta\, \theta\right)}, \\
    u_\theta(x,r,\theta,0) = - 0.01 (\beta)\sin{\left(\alpha x\right)}\sin{\left(\beta\, \theta\right)} \nonumber, 
\end{align}
with $\alpha = 14\,\pi R/L$, $\beta = 4$, and $U_{mean}(r)$ approximated using $\left(1/7 \right)^{th}$ power law for turbulent pipe flows~\citep{nikuradse1966laws,schlichting2016boundary}.
The simulations were run until turbulence was fully developed, which took approximately 15$T_{flow}$, with $T_{flow} = L/U_{bulk}$ being a flow through time.  The simulation was then run for additional $10000t_\tau$ to collect statistics. Each subsequent higher Reynolds number NWO case was initialized from a fully developed lower Reynolds number NWO case, mapped onto a corresponding finer grid. The WWO cases for each Reynolds number were initialized from fully-developed NWO cases corresponding to the same Reynolds number. In each case, we allowed for the simulations to reach a statistically-steady state (which was monitored through a time series of bulk mean velocity) and subsequently collected statistics for additional $10000t_\tau$.

\begin{table}[H]
    \centering
	\begin{tabular}{ c c c c c c c}
		$\text{Re}_\tau$ & $N_{el}$ & $N_{gp} $ & $\Delta x^+$ min/max & $\Delta r^+$ min/max & $\Delta (r\theta)^+$ min/max   & $\Delta t^+$ \\ \hline
		170 & 36848 & 27M & 3.5/14.05 & .1/1.6 & .67/2.75   & 0.0125\\ 
		360  & 238336 & 173M & 2.9/11.90 & .15/2.5 & .81/3.3  & 0.0125\\ 
		720 & 860160 & 627M & 3.3/13.6 & .22/4.18 & 1.3/5.4   & 0.0125\\ 
	\end{tabular}
	\caption{Numerical grid parameters for the presented DNS studies. $N_{el}$ denotes the number of elements, and $N_{gp}$ -- the total number of grid points within each grid. NWO and WWO cases utilize identical grids for each $Re_{\tau}$.}
	\label{table:gridres}
\end{table}


\subsection{Post-processing and notation}
\subsubsection{Phase averaging}
For turbulent flows with a periodically varying component, the turbulent fluctuations ($u_i^{\prime \prime}$) are defined to be the deviations from the long time mean ($ \av{u_i}$) plus the periodically varying component of the mean ($u_i^\phi$) \citep{hussain1970mechanics} as:
\begin{equation}
    u_i = \av{u_i} + u_i^\phi + u_i^{\prime \prime}.
    \label{eqn:triple}
\end{equation}

The component $u_i^\phi$ is equal to the phase mean minus the long time average:

\begin{equation}
    u_i^\phi = \av{u_i \mid \phi } - \av{u_i} , \phi = \tau + n T \:\forall (n, \tau \in \left[ 0 , T\right)), 
\end{equation}
with $T$ being the period of oscillation, $n$ is an integer number, and the notation $\av{u_i \mid \phi }$ denoting a conditional average given the phase.

When we report turbulent fluctuations, they shall be reported as $u_i^\dpr$ for both the WWO and NWO cases. Note, that for the NWO case, the turbulent fluctuation $u_i''$ thus defined is  equal to a standard turbulent fluctuation as obtained from Reynolds decomposition, $u_i'=u_i-\av{u_i}$, but for the WWO case, these quantities are different. 
\subsubsection{Fourier decomposition}
To capture length scales of motion, we use Fourier decomposition to decompose the flow field as:
\begin{equation}\label{eq:fourier1}
    u_i(x,r,\theta,t) = \int_{-\infty}^\infty \int_{-\infty}^\infty \hat{u}_i (k_x,r,k_\theta,t) \exp\left(i \mathbf{k} \cdot \mathbf{x} \right) d k_x d k_\theta,
\end{equation}
where $\hat{u}_i(k_x,r,k_\theta,t)$ is the Fourier coefficient, $\mathbf{k} = k_x\Vec{\mathbf{e}}_x + k_\theta \Vec{ \mathbf{e}}_\theta$ is the vector-valued wavenumber, with ($k_x$,$k_{\theta}$) denoting its streamwise and azimuthal components, and $\mathbf{x} = x\Vec{\mathbf{e}}_x+\theta\Vec{\mathbf{e}}_{\theta}$ is a shorthand notation for the projection of the position vector onto the axial-azimuthal plane. Throughout the paper, we will use the bold symbols to denote  vectors, and the bold symbols with arrows to denote unit vectors.

The Fourier coefficient $\hat{u}_i (k_x,r,k_\theta,t)$ is given by:
\begin{equation}\label{eq:fourier2}
    \hat{u}_i (k_x,r,k_\theta,t) = \frac{1}{(2\pi)^2}\int_{-\infty}^{\infty}\int_{-\infty}^{\infty} u_i (x,r,\theta,t) \exp{\left( -i ( \mathbf{k}\cdot\mathbf{x}) \right)} dx d\theta.
\end{equation}

We define a two dimensional (co-)spectrum of velocity as a time-averaged product of the spectrum of the turbulent fluctuations of velocity:
\begin{equation}
    \Phi_{u_i u_j} (k_{x},r,k_{\theta}) = \av{ \overline{\widehat{u''}}_i (k_{x},r,k_{\theta},t) {\widehat{u''}}_j (k_x,r,k_\theta,t)}_t,
\end{equation}
where $\overline{(\cdot)}$ denotes a complex conjugate. The one-dimensional (co-)spectrum is a subset of the two dimensional (co-)spectrum taken by integrating over the wavenumbers along the opposite direction:
\begin{gather}
    \Phi_{u_i u_j} (k_x,r) = \int_{k_\theta} \Phi_{u_i u_j} (k_x,r,k_\theta) dk_\theta, \label{eqn:1ds1} \\
    \Phi_{u_i u_j} (r,k_\theta ) = \int_{k_x} \Phi_{u_i u_j} (k_x,r,k_\theta) dk_x. 
\end{gather}

Furthermore, the Parseval's theorem can be used to express the second-order moments of turbulent statistics via the integration of the co-spectra in the wavenumber space:
\begin{equation}
    \av{u_i^\dpr u_j^\dpr} = \int_{k_x} \int_{k_\theta}  \Phi_{u_i u_j} (k_x,r,k_\theta) \,dk_xdk_{\theta} .
    \label{eqn:parseval}
\end{equation}

To compute the corresponding Fourier transforms, we leverage periodicity of the flow in streamwize and azimuthal directions, which allows us to replace Fourier integral in Equation (\ref{eq:fourier1}) with Fourier series as
\begin{gather}
     u_i(x,r,\theta,t) = \sum_{n=-\infty}^\infty \sum_{m=-\infty}^{\infty} \hat{u}_i (k_{x_n},r,k_{\theta_m},t) \exp\left(i \mathbf{k} \cdot \mathbf{x} \right), \label{eqn:fourier_infsum}\\
     \hat{u}_i (k_{x_n},r,k_{\theta_m},t) = \frac{1}{2\pi L} \int_{0}^{L} \int_{0}^{2\pi} u_i (x,r,\theta,t) \exp{\left( -i (\mathbf{k}_{mn}\cdot\mathbf{x}) \right)} dx d\theta,
\end{gather}
with 
\begin{equation}\label{eq:wavenum}
   k_{x_n} = \frac{2\pi n}{L},\:\:\:k_{\theta_m}=m, 
\end{equation}
and $L$ being the domain length. Equations (\ref{eqn:1ds1})--(\ref{eqn:parseval}) are replaced correspondingly with their Fourier series counterparts.

In a numerical computation with a finite number of samples, the infinite series in (\ref{eqn:fourier_infsum}) is approximated by a finite sum over wavenumbers as
\begin{gather}
     u_i(x,r,\theta,t) = \sum_{n=-N_x/2}^{N_{x}/2} \sum_{m=-N_{\theta}/2}^{N_{\theta}/2} \hat{u}_i (k_{x_n},r,k_{\theta_m},t) \exp\left(i \mathbf{k}_{mn} \cdot \mathbf{x} \right), \label{eqn:fourier}\\
     \hat{u}_i (k_{x_n},r,k_{\theta_m},t) = \frac{1}{2\pi L} \int_{0}^{L} \int_{0}^{2\pi} u_i (x,r,\theta,t) \exp{\left( -i (\mathbf{k}_{mn}\cdot\mathbf{x}) \right)} dx d\theta.
\end{gather}
Approximations for the co-spectra and the second-order moments are updated accordingly. 

\begin{table}
    \centering
    \begin{tabular}{c c c c c}
         $Re_{\tau}$ & $N_x$& $N_\theta$ & $\lambda_{x\,\text{min}}^{+}$ &  $ \lambda_{s\,\text{min,wall}}^+$ \\ \hline
         170 & 384 & 80& 21.2 &  26.8\\
         360 & 768 & 160 & 22.6 & 28.4\\
         720 & 2048 & 320 & 15.6 & 24.4
    \end{tabular}
    \caption{Computational parameters employed for a Fourier analysis of the DNS data. $N_x, N_{\theta}$ -- number of terms carried in the Fourier series in streamwise and azimuthal directions, respectively; $\lambda_{x\,\text{min}}^{+}$ - minimum streamwise wavelength; $\lambda_{s\,\text{min,wall}}^{+}$ -- minimum azimuthal wavelength at the wall.}
    \label{tab:fourierGrid}
\end{table}
Fourier coefficients, spectra and co-spectra defined above can be equivalently represented in terms of the wavelengths, $\lambda_{x} =2\pi/k_{x},  \lambda_{\theta} = 2\pi/k_{\theta}$ (subscripts $n$, $m$ in a discrete representation of wavenumbers and wavelenghts will be omitted for brevity throughout the remainder of the manuscript). Note that from the definition of wavenumbers in Equation (\ref{eq:wavenum}), it is seen that the wavelength $\lambda_{x}$ has a dimension of length, but $\lambda_{\theta}$ is adimensional. Therefore, we can define a dimensional azimuthal wavelength 
$\lambda_{s}(r) = r \lambda_{\theta}$, where $r$ is the local radial location of the variable to be considered. The number of terms, $N_x$ and $N_{\theta}$, carried in a Fourier analysis of the DNS data for each Reynolds number (equation (\ref{eqn:fourier})) is specified in Table~\ref{tab:fourierGrid}, together with the smallest wavelengths that are computed as a result of the Fourier analysis. It can be seen that the smallest computed wavelengths are above the DNS grid resolution, to avoid any spurious oscillations potentially caused by interpolation from the DNS grid and under-resolution.

\subsubsection{Correlation functions}

Correlation functions in space are calculated using Wiener-Khinchin theorem for a periodically extended signal and are presented in the following form:
\begin{equation}
    R_{u_i u_j}(\Delta x, \Delta \theta, r_0, r) = \frac{1}{2\pi L T} \int_0^T \int_0^L \int_0^{2\pi} u_i(x,r_0,\theta,t) u_j(x+\Delta x, r, \theta+\Delta \theta,t) \,dt\, dx\, d\theta.
\end{equation}
The correlation coefficient is defined as:
\begin{equation}
    \rho_{u_i u_j}(\Delta x, \Delta \theta, r_0, r) = \frac{R_{u_i u_j} (\Delta x, \Delta \theta, r_0, r) - \av{u_i}(r_0)\av{u_j}(r_0) }{ \sqrt{\av{{u_i^\dpr}^2} (r_0)\av{{u_j^\dpr}^2}(r_0) } }.
\end{equation}

For compactness, the correlation coefficient at zero separation along a homogeneous direction shall be written as:

\begin{gather}
    \rho_{u_i u_j} (\Delta x = 0, \Delta \theta, r_0,r) = \rho_{u_i u_j} (\Delta \theta, r_0,r), \\
    \rho_{u_i u_j} (\Delta x, \Delta \theta = 0, r_0,r) = \rho_{u_i u_j} (\Delta x, r_0,r).
\end{gather}

\section{Results}\label{sec:results}
\subsection{Validation}
Although validation of Nek5000 in application to DNS of turbulent pipe flows was already documented elsewhere~\citep{el2013direct,merrill2016spectrally}, Appendices \ref{sec:validation_nwo} and \ref{sec:validation_wwo} present its validation using the present computational setup for the NWO and WWO cases, respectively, versus available published data.

\subsection{Effect of Reynolds number on drag reduction}

Table \ref{tble:Ub_Cf} presents the global quantities for the computed drag-reduced (WWO) cases, including the achieved bulk Reynolds number, $Re_{bulk}$, percent increase in bulk mean velocity, $U_{bulk}$, and percent reduction in skin friction coefficient, $C_f$, with respect to the corresponding standard (NWO) cases. From Table \ref{tble:Ub_Cf}, it is evident that the mechanism of
drag reduction with transverse wall oscillations becomes less effective as Reynolds number increases.

\begin{table}[H]
    \centering
    \begin{tabular}{  c c c c c } 
     $Re_\tau$ & $(U_{bulk}^+$,$\bar{U}_{c})_{NWO}$ & $(U_{bulk}^+$,$\bar{U}_{c})_{WWO}$ & \%  $\Delta U_{bulk}$ &  \% $\Delta C_f$ \\ \hline
     170 & (14.4,1.32) & (17.1,1.31)  & 18.54 & 28.8 \\ 
     360 & (16.2,1.27) & (18.8,1.25) & 16.25 & 26.0 \\ 
     720 & (18.0,1.26) & (20.5,1.23)  & 13.9 & 22.9 \\ 
    \end{tabular}
    \caption{Bulk quantities and their percent change for the WWO cases as compared to the NWO cases. $\bar{U}_c=U_c/U_{bulk}$ denotes the centerline velocity scaled with $U_{bulk}$.  The percent increase in bulk mean velocity is defined as: $ \left(U_{bulk_{WWO}}-U_{bulk_{NWO}}\right)/U_{bulk_{NWO}}\cdot 100\%$.  The skin friction is calculated as: $C_f = 2 \av{\tau_w}/\left( \rho \,U_{bulk}^2 \right)$, and the percent reduction in $C_f$ as: $- \left(C_{f_{WWO}}-C_{f_{NWO}}\right)/C_{f_{NWO}}\cdot 100\%$. }
    \label{tble:Ub_Cf} 
\end{table}

\begin{figure}
    \centering
    \includegraphics[width=0.5\linewidth]{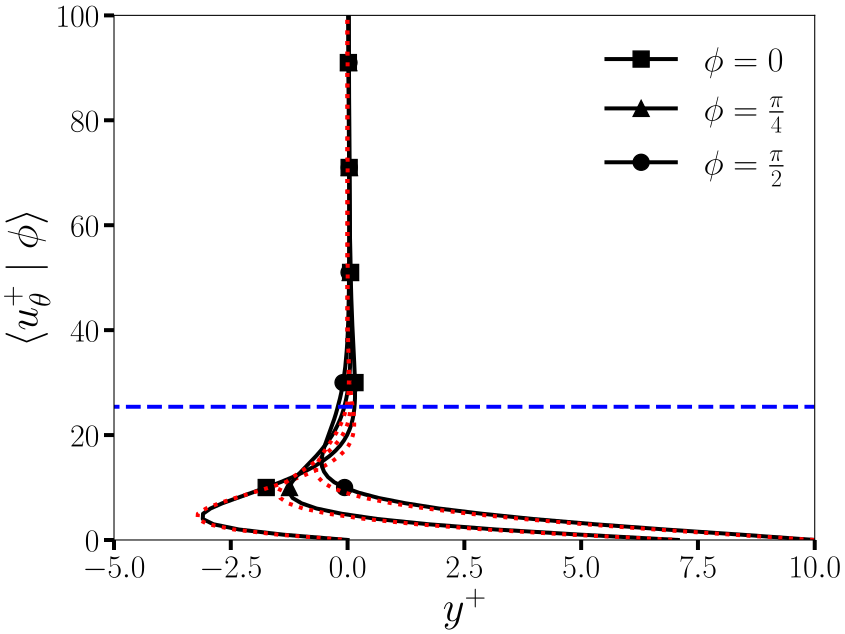}
    \caption{Comparison between phase mean azimuthal velocity profile (black solid lines) for $Re_\tau =720$ and a laminar solution (red dashed lines) for a pipe with its wall oscillating about its axis in cylindrical coordinates~\citep{song2020viscous,coxe2022stokes} at phases $0$, $\pi/4$, $\pi/2$, and $\pi$. Blue horizontal dashed line indicates the Stokes' layer thickness, $\delta_{Sl}^+\approx 25$.}
    \label{fig:stokesazmComp}
\end{figure}

It is generally accepted that drag reduction with azimuthally oscillated walls occurs due to an interaction of turbulence with the so-called Stokes' layer, which refers to a layer of  non-zero phase mean azimuthal velocity developed over an oscillating wall. Figure \ref{fig:stokesazmComp} demonstrates
 the impact of the transverse wall oscillations on the phase mean turbulent azimuthal velocity profile for the highest simulated $Re_{\tau}=720$, in comparison with the corresponding laminar solution for a pipe with its wall oscillating about its axis in cylindrical coordinates~\citep{song2020viscous,coxe2022stokes}.  The blue horizontal line indicates the Stokes' layer thickness, defined as the location where the azimuthal velocity of the laminar solution drops below 1\% of the maximum wall velocity. For the current frequency of wall oscillation, it can be shown that the Stokes' layer thickness over a cylindrical pipe wall is equal to the one obtained in a classical Stokes' second problem solution, $\delta_{Sl}=4.6\sqrt{T_0\,\nu/\pi}$~\citep{panton1984incompressible,coxe2022stokes}. In wall units, this quantity equals to $\delta_{Sl}^+=4.6\sqrt{T_0^+/\pi}\approx 25$ for the current wall oscillation parameters, independent of the Reynolds number.

 In describing the remainder of results, we will frequently be referring to different regions commonly identified within the turbulent boundary layer~\citep{schlichting2016boundary}, in addition to the Stokes' layer, as summarized in Table~\ref{tab:locations}, where we use the notation $y^+=y/l_{\tau}$, with  $y=R-r$. 
 We will be using the asterisk to denote the quantities scaled with the outer dimension, such that $r^{\star}=r/R$ and $y^{\star}=1-r/R$.

\begin{table}[H]
    \centering
    \begin{tabular}{c c}
         Region & Range  \\ \hline 
         Viscous layer & $y^+ \le 5$ \\
         Buffer layer & $5 < y^+ \leq 30$ \\
         Stokes' layer & $ y^+ \le 25 $ \\
         Log layer & $30 < y^+ \leq 0.2 \ret $ \\
         Wake region & $y^+ > 0.2 \ret$ \\
    \end{tabular}
    \caption{Regions within the turbulent boundary layer, including the Stokes' layer, referred to throughout this work.  The inner layer is the composite of the viscous, buffer and log layers while the outer layer is the composite of the log layer and the wake region of the flow~\citep{panton1984incompressible}.}
    \label{tab:locations}
\end{table}

\subsection{Effect of Reynolds number on single point statistics}
\begin{figure}
    \centering
    \begin{subfigure}{0.48\linewidth}
    \includegraphics[width=0.98\linewidth]{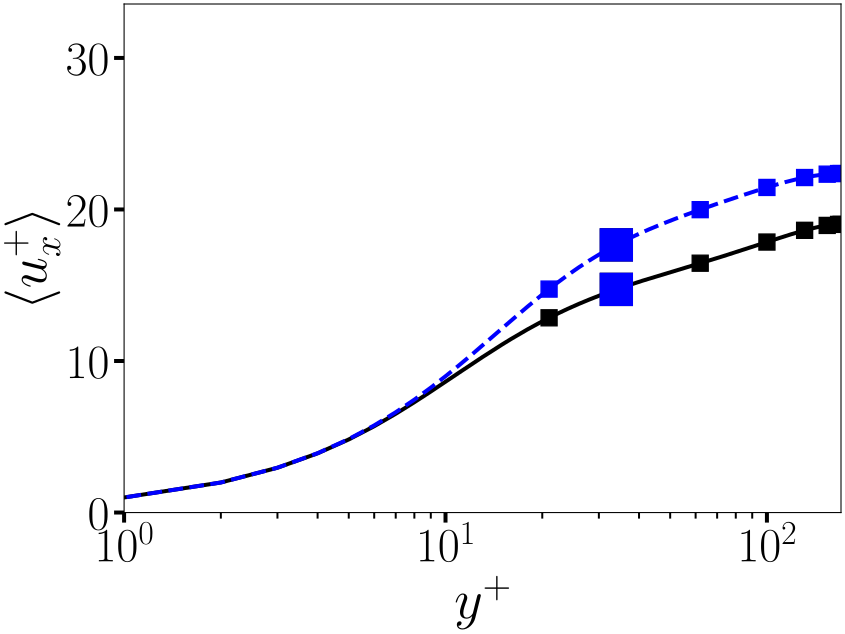}
    \caption{$\ret=170$}
    \label{fig:shift170}
    \end{subfigure}%
    \begin{subfigure}{0.48\linewidth}
    \includegraphics[width=0.98\linewidth]{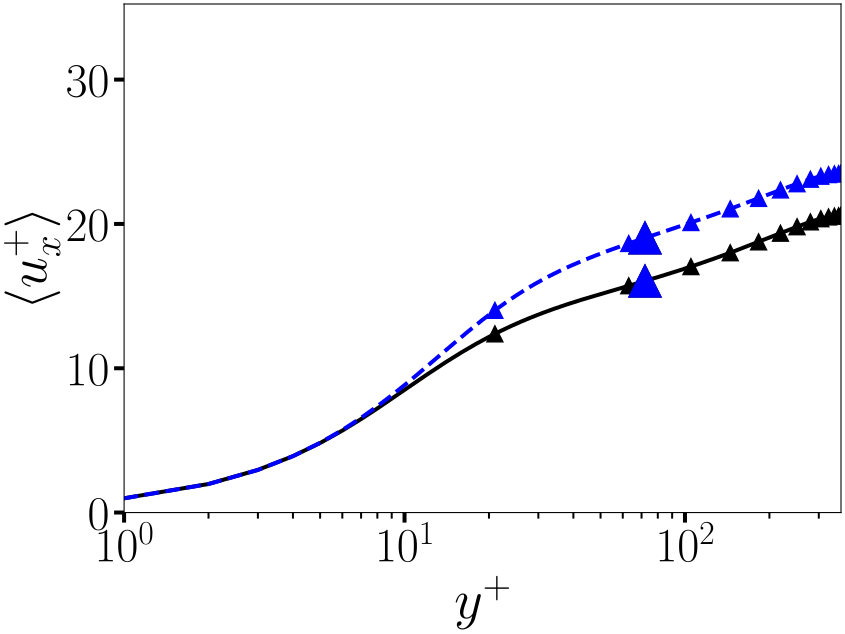}
    \caption{$\ret=360$}
    \label{fig:shift360}
    \end{subfigure}
     \begin{subfigure}{0.48\linewidth}
    \includegraphics[width=0.98\linewidth]{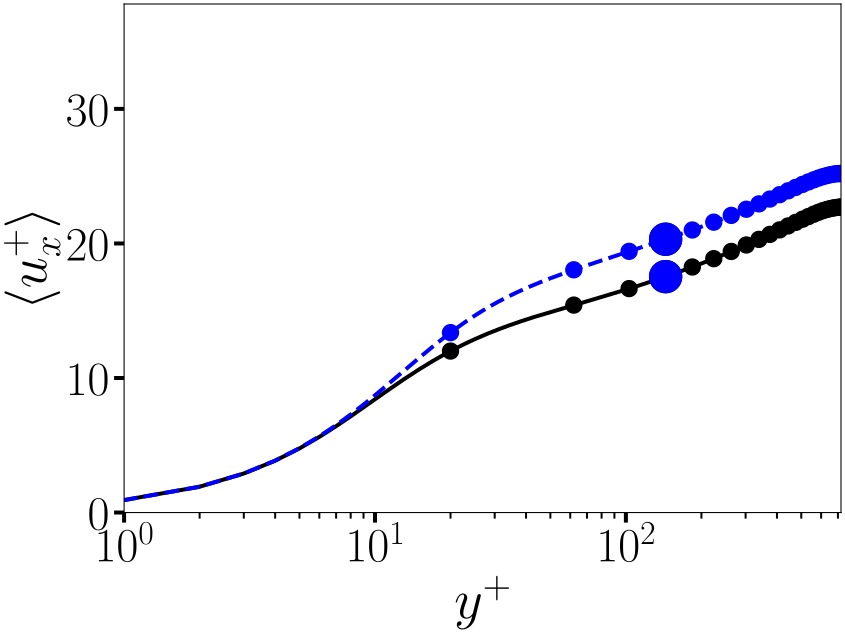}
    \caption{$\ret=720$}
    \label{fig:shift720}
    \end{subfigure}
    \begin{subfigure}{0.48\linewidth}
    \includegraphics[width=0.98\linewidth]{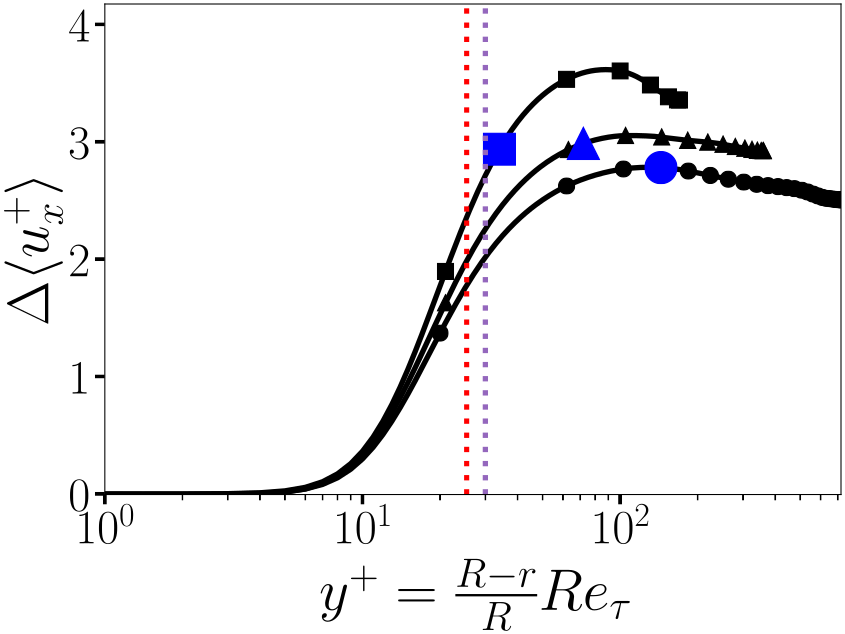}
    \caption{Change in $\av{u_x^+}$}
    \label{fig:ubar}
    \end{subfigure}%
    \caption{Mean streamwise velocity profiles at (a) $\ret=170$,  (b) $\ret=360$, and  (c) $\ret=720$. Black solid line, NWO; blue dashed line, WWO. (d) Change in mean  velocity between NWO and WWO. The red dotted line indicates the top of the Stokes' layer, the purple dotted line is the top of the buffer layer, and the location of the enlarged blue markers indicates the top of the log layer with respect to each Reynolds number. }
    \label{fig:meanVelComp}
\end{figure}

\begin{figure}
    \centering
    \begin{subfigure}{0.33\linewidth}
    \includegraphics[width=0.98\linewidth]{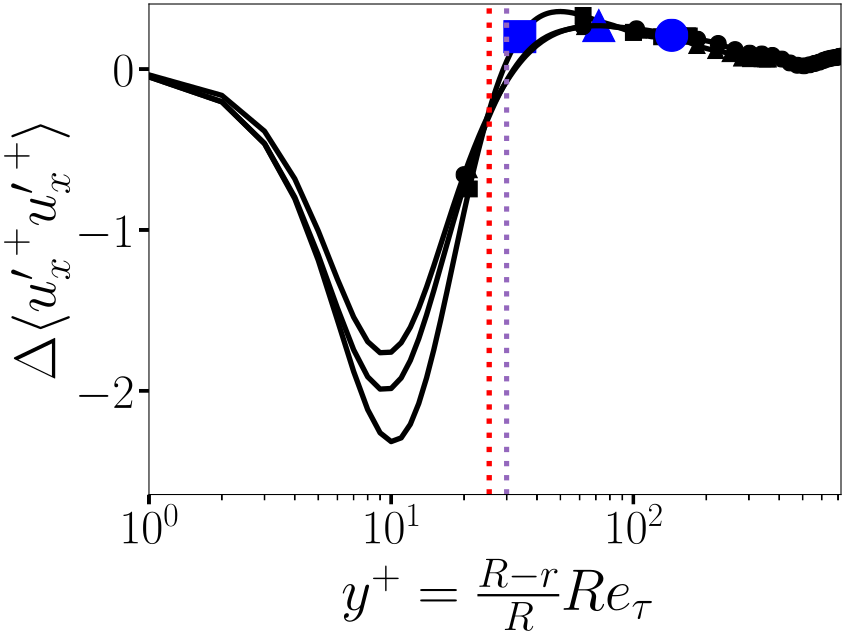}
    \caption{Streamwise velocity}
    \label{fig:uubar}
    \end{subfigure}
    \begin{subfigure}{0.33\linewidth}
    \includegraphics[width=0.98\linewidth]{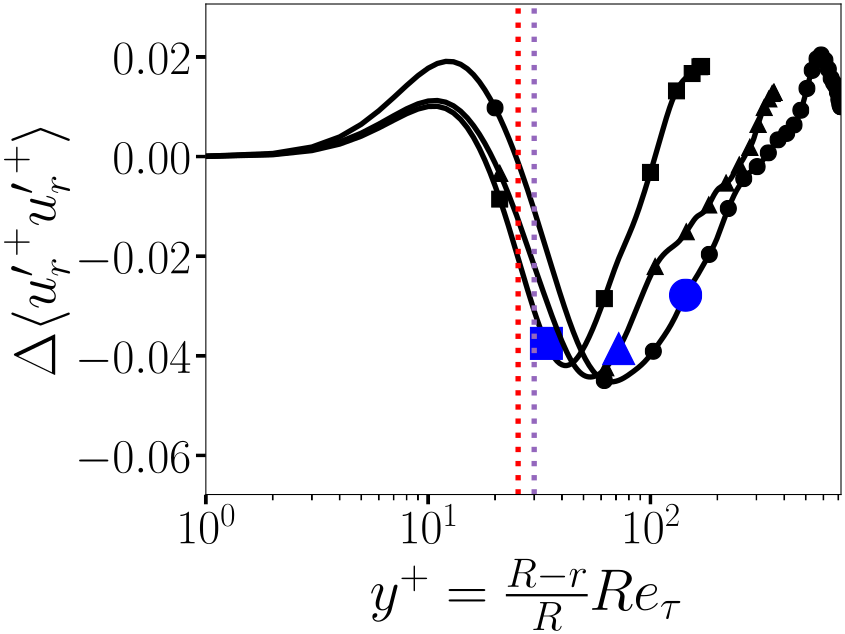}
    \caption{Radial velocity}
    \label{fig:vvbar}
    \end{subfigure}%
    \begin{subfigure}{0.33\linewidth}
    \includegraphics[width=0.98\linewidth]{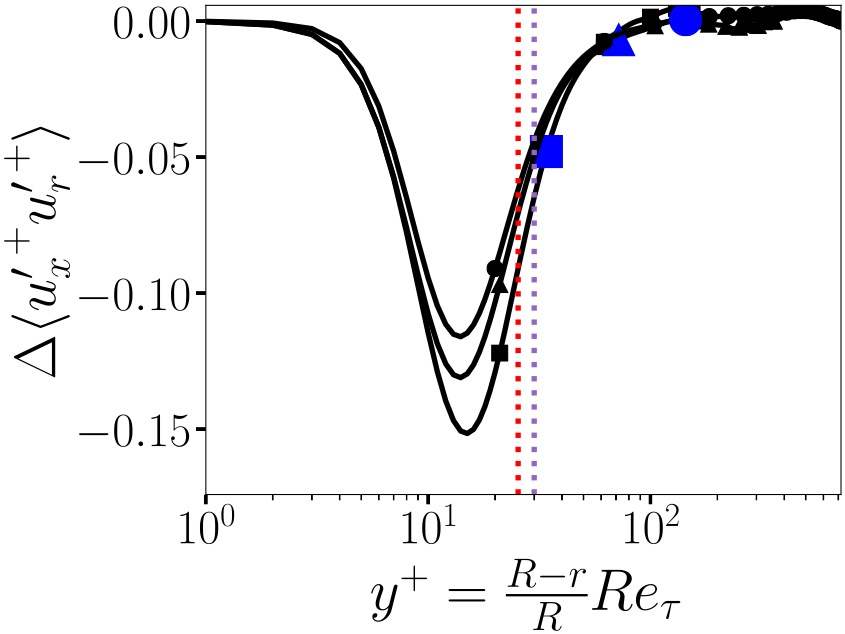}
    \caption{Shear stresses}
    \label{fig:uvbar}
    \end{subfigure}
    
    \caption{Comparison of the change in the second-order statistics as influenced by the Reynolds number.  The red dotted line indicates the top of the Stokes' layer, the purple dotted line is the top of the buffer layer, and the location of the enlarged blue markers indicates the top of the log layer with respect to each Reynolds number.  The location of changes to the statistical profiles shows a reasonable collapse with viscous units indicating that the wall oscillations impact turbulence scales within the log layer of the flow.  While the Stokes' layer is confined to a region bounded by the buffer layer of the flow, effects on statistics are also observed above the log layer.}
    \label{fig:meanTurbComp}
\end{figure}

A well-known consequence of drag reduction is an upward shift of the mean velocity profile in the log region of the flow~\citep{hurst2014effect,gatti2016reynolds}. Figure \ref{fig:meanVelComp} documents this shift for the three Reynolds numbers considered in this study. The upward shift is approximately $\Delta \av{u_x^+}=3.4, 2.5$ and 2.3 for $\ret=170, 360$ and 720, respectively. Consistent with the observations in~\cite{hurst2014effect}, the shift decreases with $\ret$, but the amount of decrease is diminishing. \cite{gatti2016reynolds} argue that the shift becomes independent from the Reynolds number once it reaches a high enough value, and the observed trend supports this argument.
Figure \ref{fig:ubar} documents the change in the mean velocity as a result of transverse wall oscillation for the three Reynolds numbers.  The change is defined as $\Delta f = f(WWO) - f(NWO)$, i.e. quantity evaluated with transverse wall oscillations minus quantity evaluated with no wall oscillation.  This convention will be maintained throughout the remainder of the work.  For all three Reynolds numbers the maximum change in mean streamwise velocity occurs around $y^+\approx 100$.  This location happens to be in the top half of the log layer for the highest Reynolds number, above the log layer for $Re_\tau = 360$ and approaching the centerline of the pipe for $\ret = 170$.   Consistent with the reduction in the value of log-layer shift, as the Reynolds number increases, the peak change in the mean streamwise velocity reduces.
Figure \ref{fig:meanTurbComp} subsequently documents the change in the second-order statistics as a result of transverse wall oscillation. 
Wall oscillations suppress the streamwise turbulent kinetic energy within the buffer layer for all three Reynolds numbers.  Above the buffer layer, streamwise turbulent kinetic energy is slightly increased.  Radial turbulent kinetic energy changes are two orders of magnitude smaller than the changes in streamwise turbulent kinetic energy. Its trends in the buffer and the log-layer are the reverse of those of the streamwise turbulence kinetic energy. The change in the Reynolds shear stress is one order of magnitude smaller than the change in streamwise kinetic energy. The result of wall oscillations is to suppress the Reynolds shear stress through the top of the log layer for all three Reynolds numbers.

\subsection{Effect of Reynolds number on energy spectra}
\begin{figure}
    \centering
    \begin{subfigure}{0.32\linewidth}
        \includegraphics[width=0.99\linewidth]{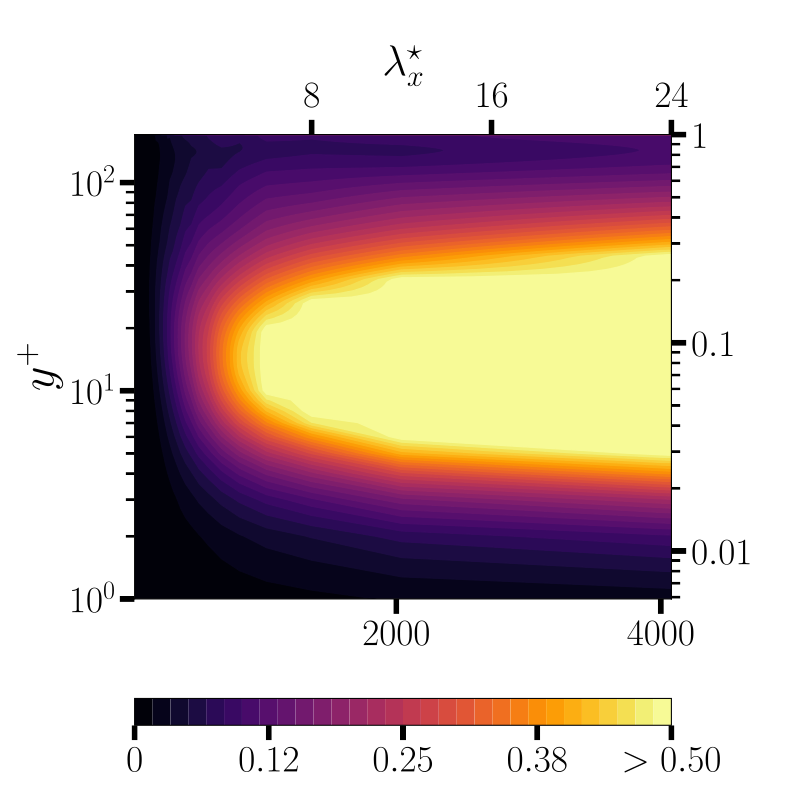}
        \caption{$\ret=170$, NWO}
        \label{fig:UuStwYp170nwo}
    \end{subfigure}
    \begin{subfigure}{0.32\linewidth}
        \includegraphics[width=0.99\linewidth]{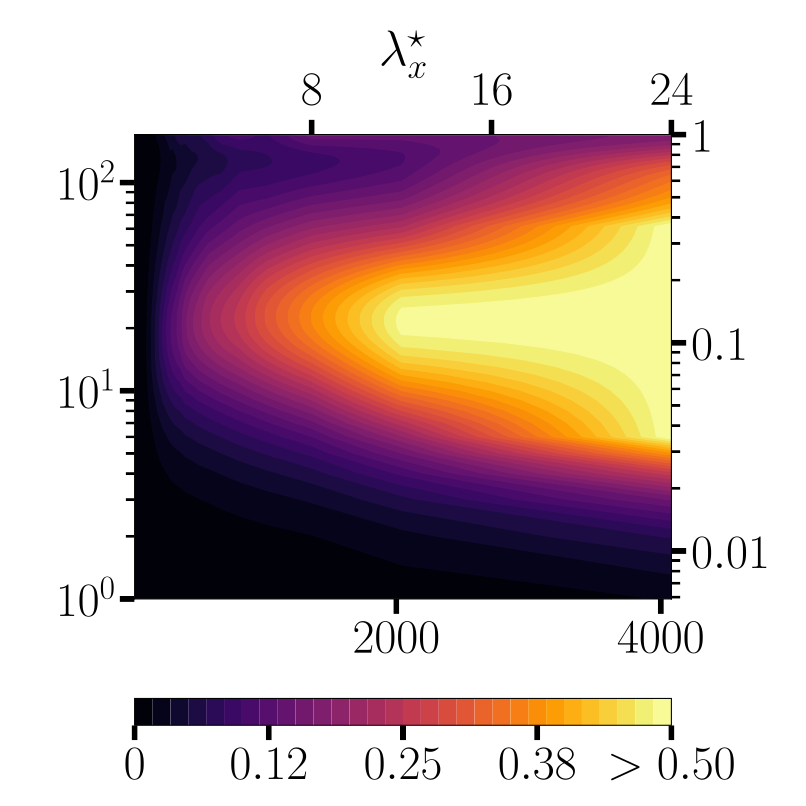}
        \caption{$\ret=170$, WWO}
        \label{fig:UuStwYp170wwo}
    \end{subfigure}
    \begin{subfigure}{0.32\linewidth}
        \includegraphics[width=0.99\linewidth]{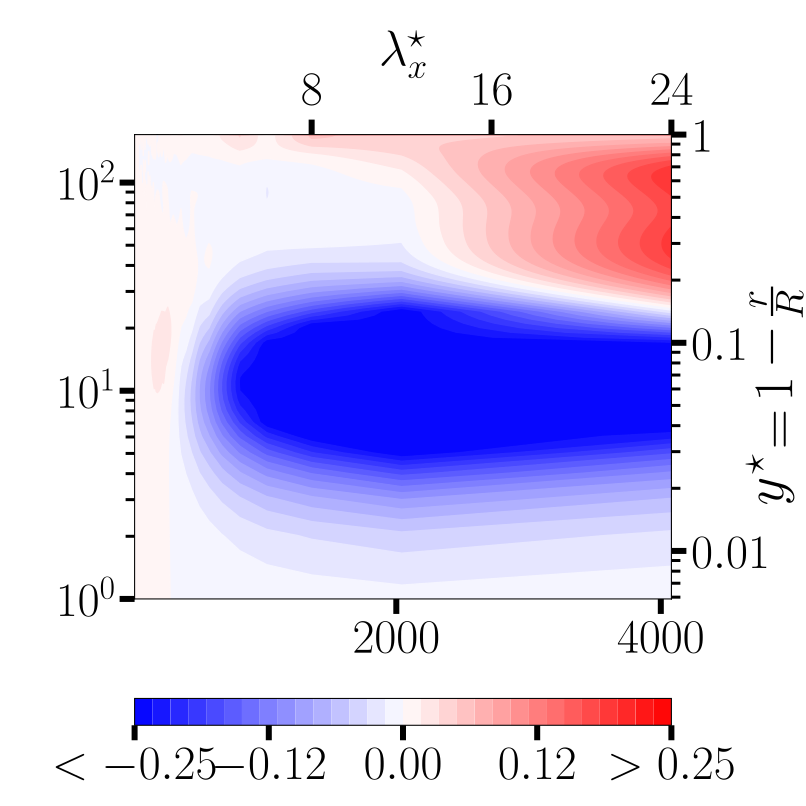}
        \caption{$\ret=170$, WWO-NWO}
        \label{fig:DuuStwRe170}
    \end{subfigure}
    
    \begin{subfigure}{0.32\linewidth}
        \includegraphics[width=0.99\linewidth]{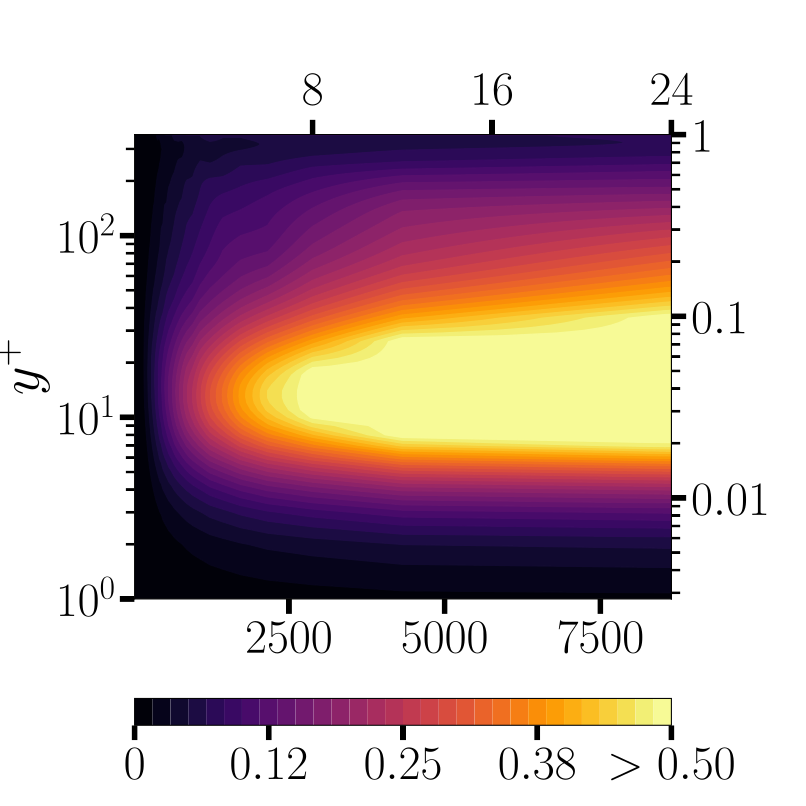}
        \caption{$\ret=360$, NWO}
        \label{fig:UuStwYp360nwo}
    \end{subfigure}%
    \begin{subfigure}{0.32\linewidth}
        \includegraphics[width=0.99\linewidth]{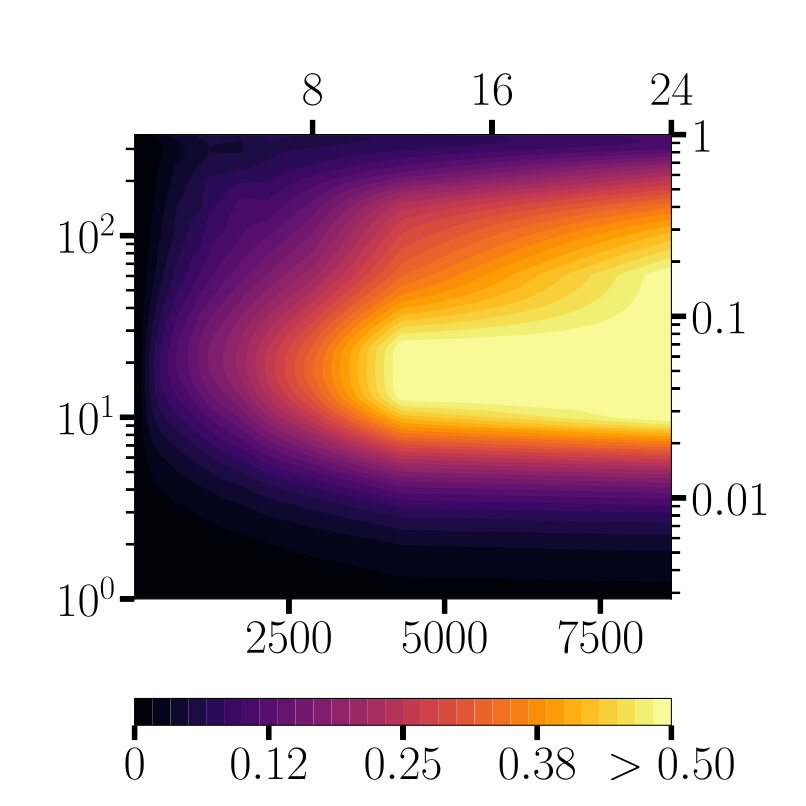}
        \caption{$\ret=360$, WWO}
        \label{fig:UuStwYp360wwo}
    \end{subfigure}
    \begin{subfigure}{0.32\linewidth}
        \includegraphics[width=0.99\linewidth]{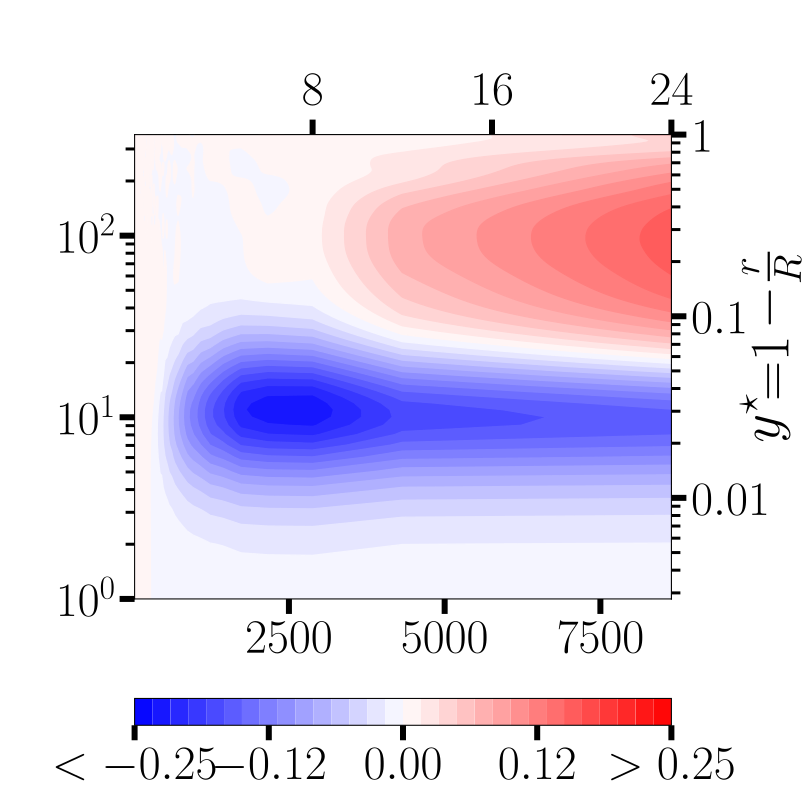}
        \caption{$\ret=360$, WWO-NWO}
        \label{fig:DuuStwRe360}
    \end{subfigure}%
    
    \begin{subfigure}{0.32\linewidth}    
        \includegraphics[width=0.99\linewidth]{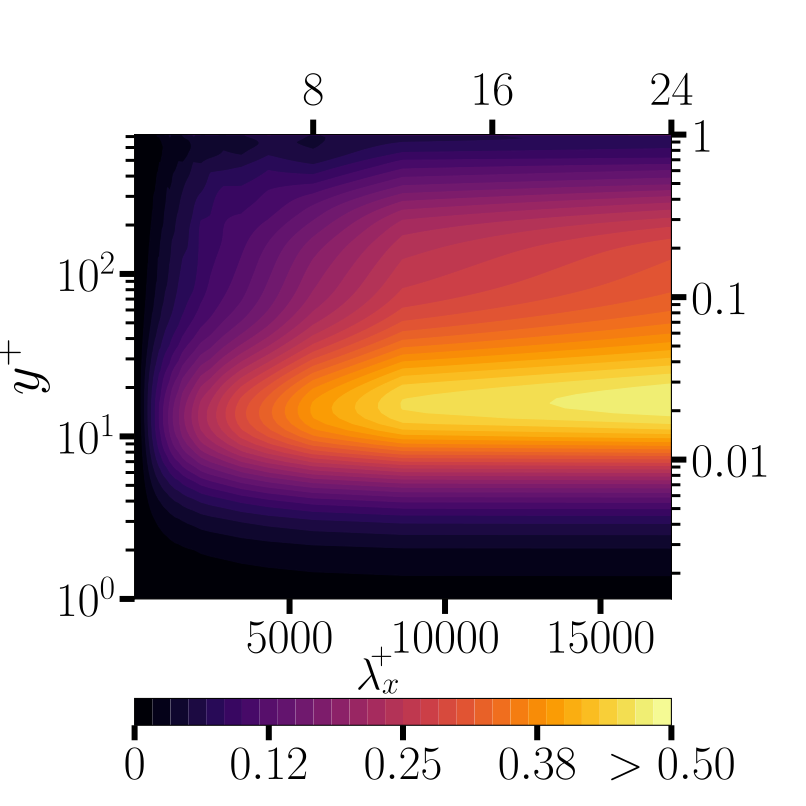}
        \caption{$\ret=720$, NWO}
        \label{fig:UuStwYp720nwo}
    \end{subfigure}%
    \begin{subfigure}{0.32\linewidth}    
        \includegraphics[width=0.99\linewidth]{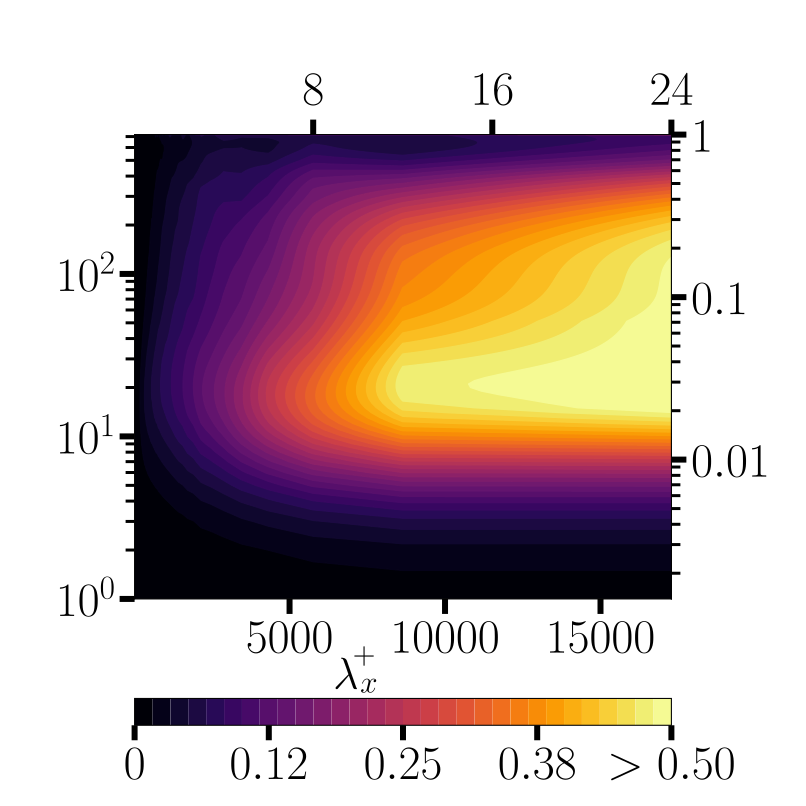}
        \caption{$\ret=720$, WWO}
        \label{fig:UuStwYp720wwo}
    \end{subfigure}  
    \begin{subfigure}{0.32\linewidth}
        \includegraphics[width=0.99\linewidth]{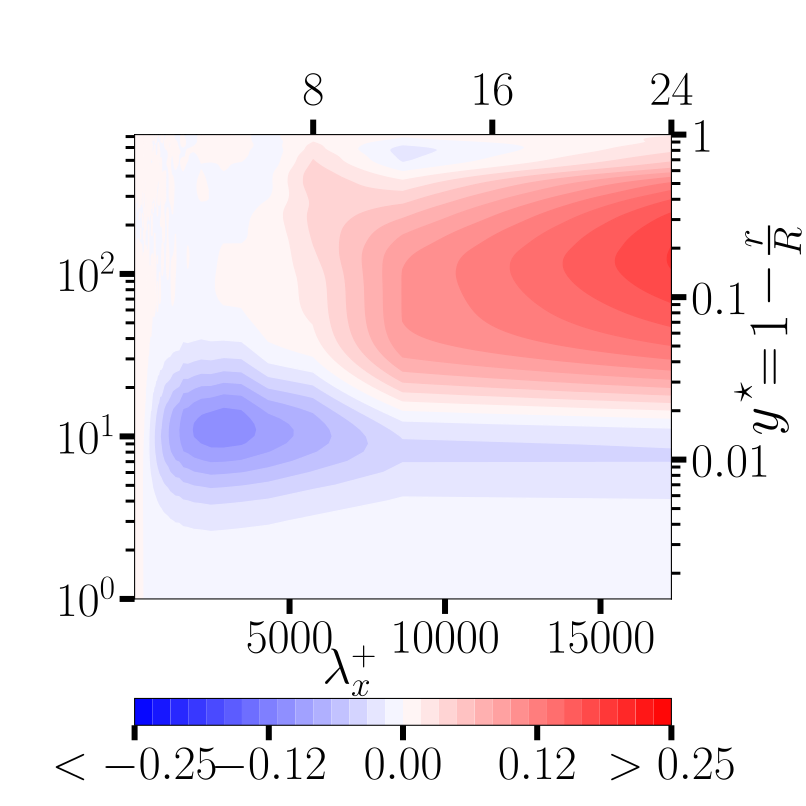}
        \caption{$\ret=720$, WWO-NWO}
        \label{fig:DuuStwRe720}
    \end{subfigure}%

    \caption{Streamwise kinetic energy as a function of wall normal location and streamwise wavelength,  $\Phi_{u_x u_x}(\lambda_x^+,y^+)/u_\tau^2$: (a,d,g) NWO spectra; (b,e,h) WWO spectra; (c,f,i) change in spectra, $\Delta \Phi_{u_x u_x}(\lambda_x^+,y^+)/u_\tau^2$.  From top to bottom: (a,b,c) $\ret = 170$, (d,e,f) $\ret = 360$, and (g,h,i) $\ret = 720$. }
    \label{fig:DeltaUUstw}
\end{figure}

Figure~\ref{fig:DeltaUUstw} documents the streamwise spectra of streamwise kinetic energy in the NWO and WWO cases, as well as its change, as a result of wall oscillations.  Wall oscillations  enhance the energy in large streamwise wavelengths in and above the log layer of the flow. 
This is consistent with drag reduction mechanisms which suppress near-wall turbulence \citep{kim2008dynamics,ricco2021review}.  It is worth noting that at the low Reynolds number ($Re_\tau = 170$) and moderately low Reynolds number ($Re_\tau = 360$), these large outer-layer structures are restricted by the vertical height of the domain.  
However, across all Reynolds numbers wall oscillation suppresses the energy of the streamwise structures in the buffer layer having wavelengths less than $\lambda_x^+ \approx 10,000$.  These large-scale structures in the buffer layer are thought to correspond to the concatenation of hairpin packets~\citep{adrian2007hairpin,lee2019space}, whose legs produce a signature of the well-known streamwise streaks in the boundary layer~\citep{jimenez2022streaks}.  The trend of large scales of motion being suppressed versus enhanced exhibits a clearly defined boundary located between the buffer and the log layer, at $y^+ \approx 30$. Of interest is a subtle amplification of very short streamwise scales in the buffer layer ($\lambda_x^+<500$) by wall oscillations.

\begin{figure}
    \centering
    \begin{subfigure}{0.33\linewidth}
\includegraphics[width=0.99\linewidth]{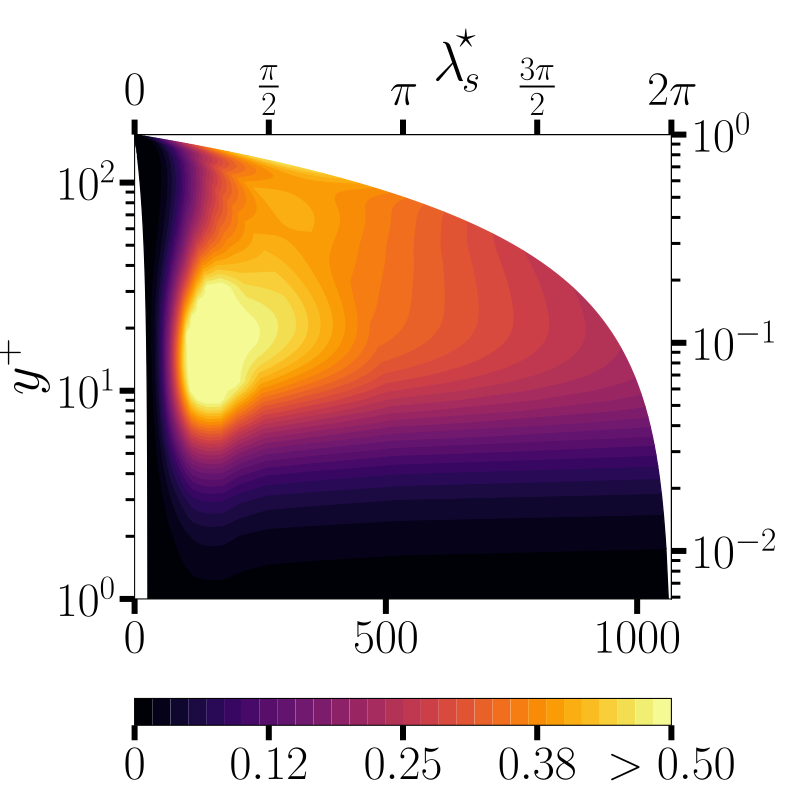}
        \caption{$\ret=170$, NWO}
        \label{fig:UuAzmYp170nwo}
    \end{subfigure}%
    \begin{subfigure}{0.33\linewidth}
\includegraphics[width=0.99\linewidth]{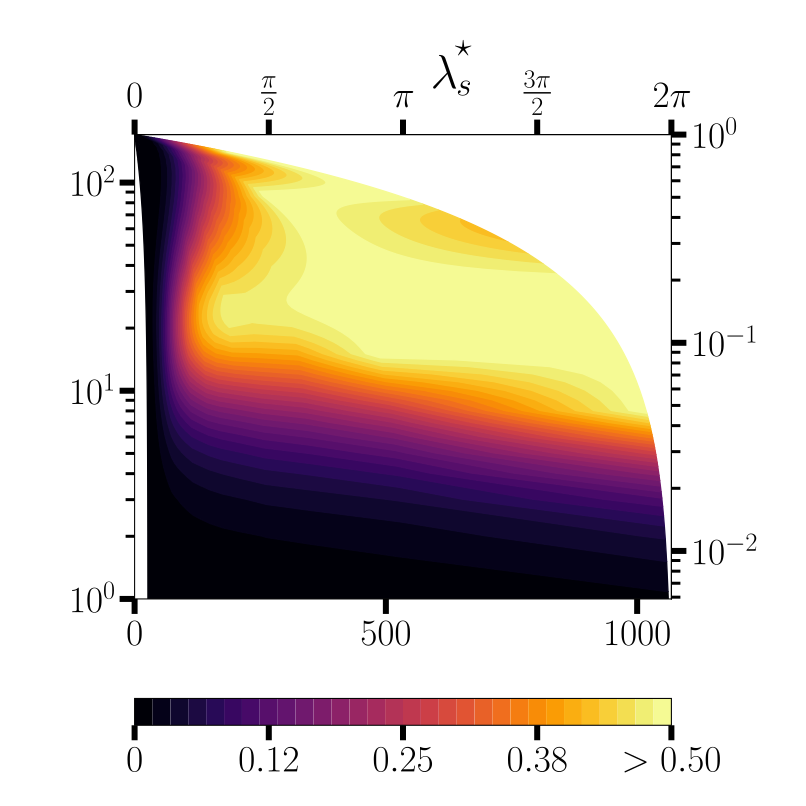}
        \caption{$\ret=170$, WWO}
        \label{fig:UuAzmYp170wwo}
    \end{subfigure}
    \begin{subfigure}{0.33\linewidth}
\includegraphics[width=0.99\linewidth]{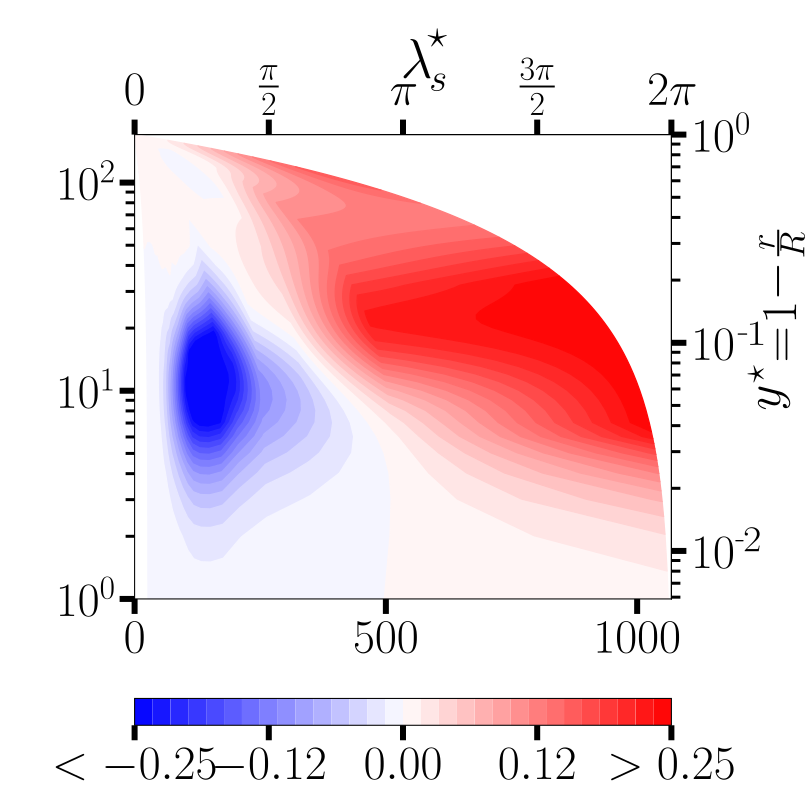}
        \caption{$\ret=170$, WWO-NWO}
        \label{fig:DuuAzmRe170}
    \end{subfigure}
    
    \begin{subfigure}{0.33\linewidth}
\includegraphics[width=0.99\linewidth]{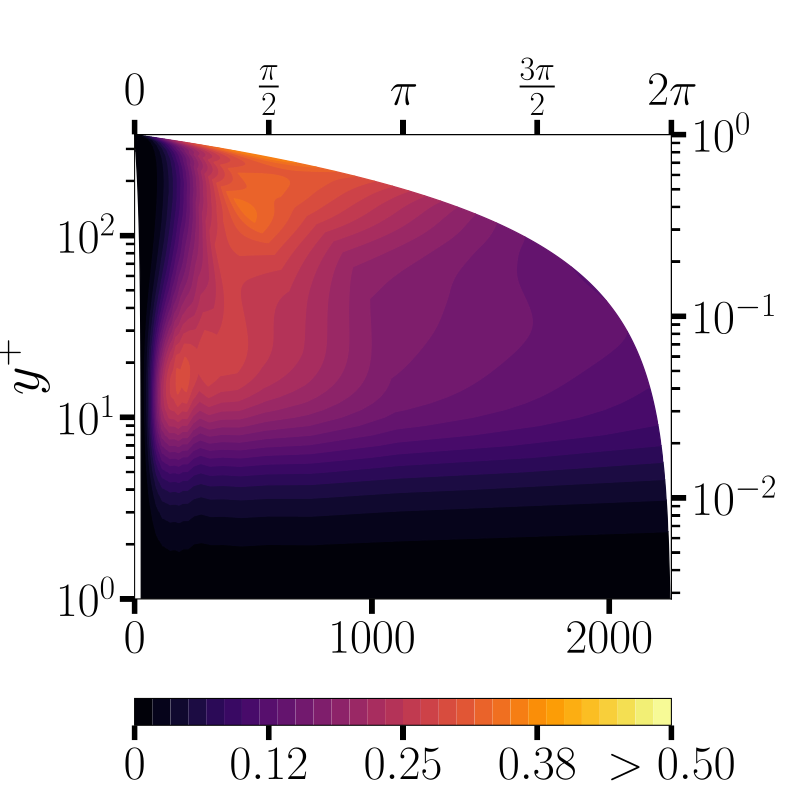}
        \caption{$\ret=360$, NWO}
        \label{fig:UuAzmYp360nwo}
    \end{subfigure}%
    \begin{subfigure}{0.33\linewidth}
\includegraphics[width=0.99\linewidth]{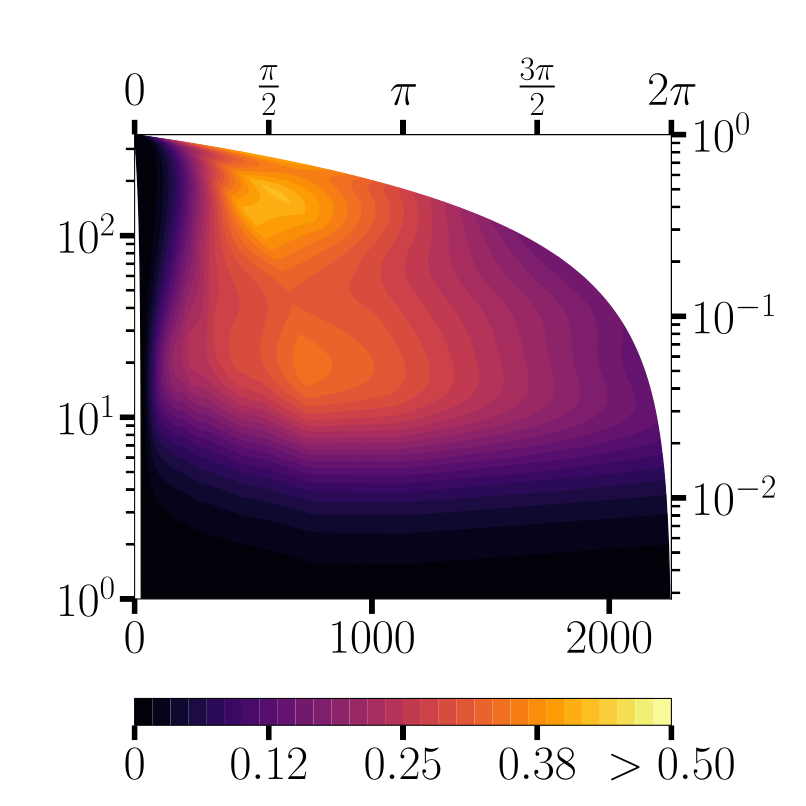}
        \caption{$\ret=360$, WWO}
        \label{fig:UuAzmYp360wwo}
    \end{subfigure}
     \begin{subfigure}{0.33\linewidth}
\includegraphics[width=0.99\linewidth]{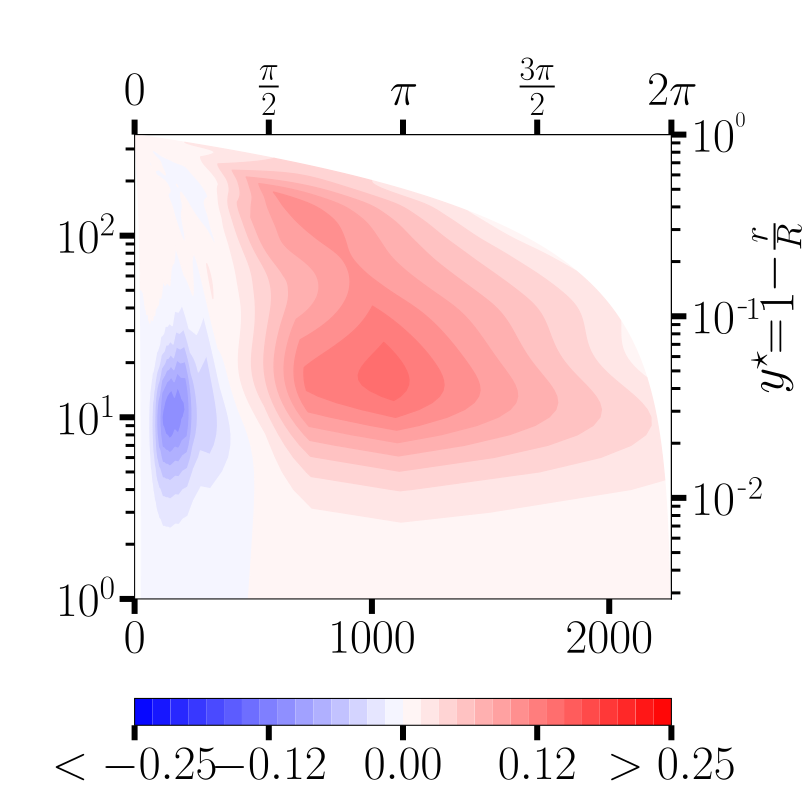}
        \caption{$\ret=360$, WWO-NWO}
        \label{fig:DuuAzmRe360}
    \end{subfigure}

    \begin{subfigure}{0.33\linewidth}
\includegraphics[width=0.99\linewidth]{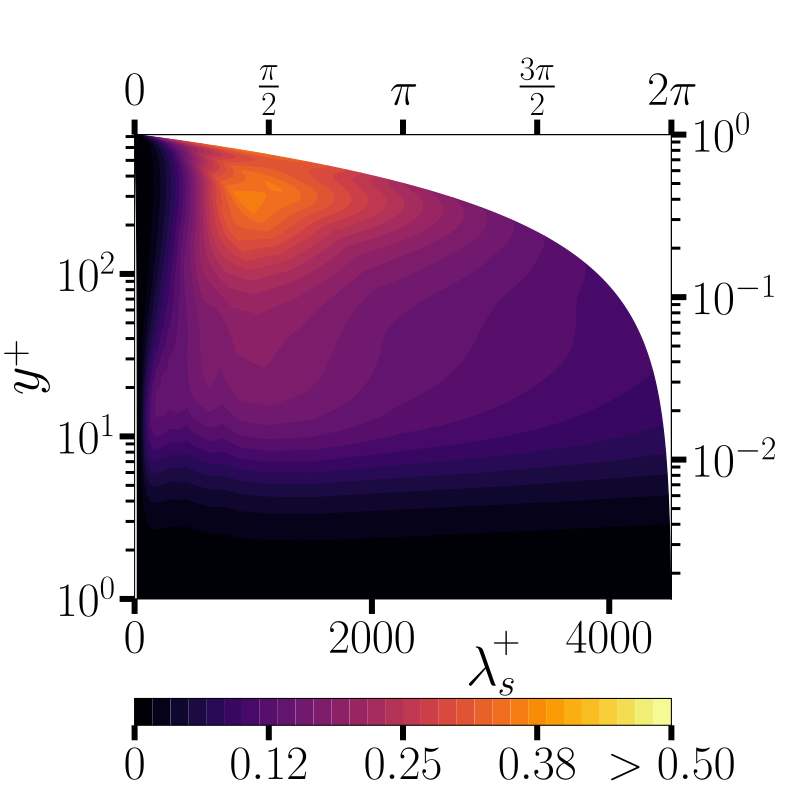}
        \caption{$\ret=720$, NWO}
        \label{fig:UuAzmYp720nwo}
    \end{subfigure}%
    \begin{subfigure}{0.33\linewidth}
\includegraphics[width=0.99\linewidth]{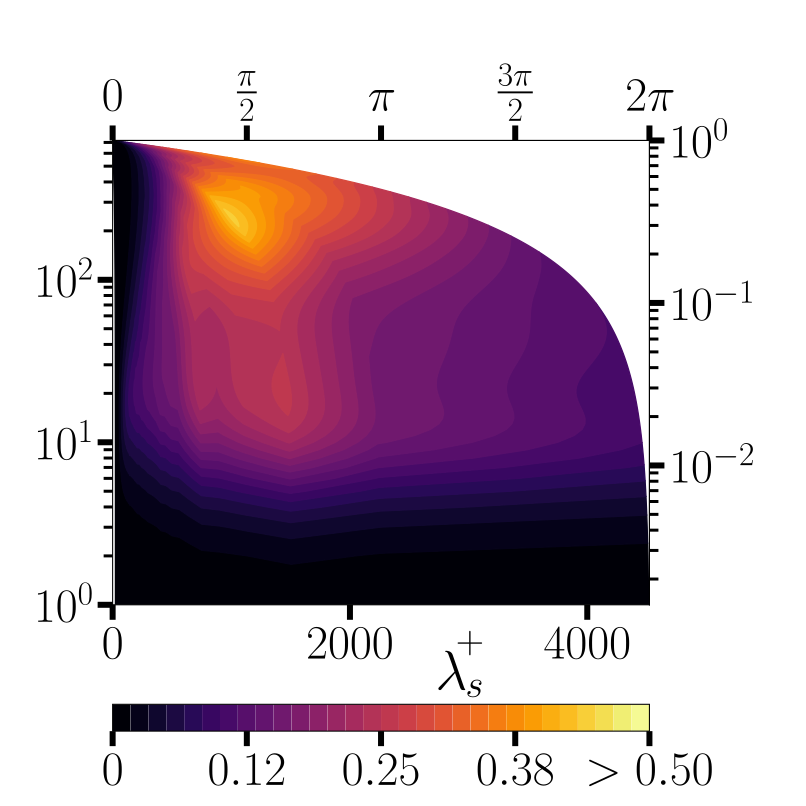}
        \caption{$\ret=720$, WWO}
        \label{fig:UuAzmYp720wwo}
    \end{subfigure}
     \begin{subfigure}{0.33\linewidth}
\includegraphics[width=0.99\linewidth]{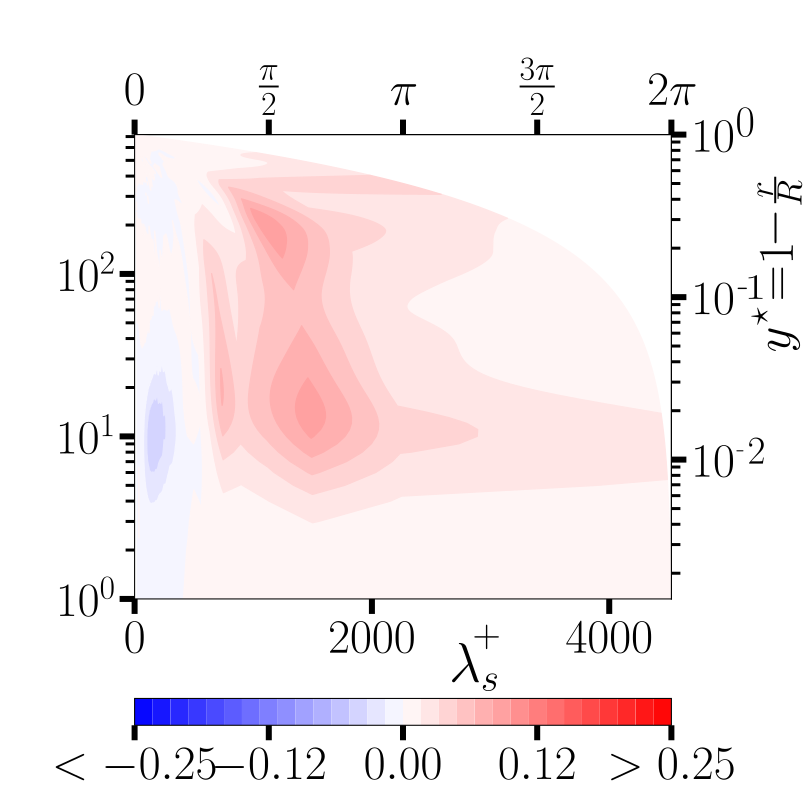}
        \caption{$\ret=720$, WWO-NWO}
        \label{fig:DuuAzmRe720}
    \end{subfigure}
    \caption{
    Streamwise kinetic energy as a function of wall normal location and arclength,  $\Phi_{u_x u_x}(\lambda_s^+,y^+)/u_\tau^2$: (a,d,g) NWO spectra; (b,e,h) WWO spectra; (c,f,i) change in spectra, $\Delta \Phi_{u_x u_x}(\lambda_s^+,y^+)/u_\tau^2$.  From top to bottom: (a,b,c) $\ret = 170$; (d,e,f) $\ret = 360$; (g,h,i) $\ret = 720$.}
    \label{fig:DeltaUUAzm}
\end{figure}

\begin{figure}
    \centering
    \begin{subfigure}{0.33\linewidth}
        \includegraphics[width=0.95\linewidth]{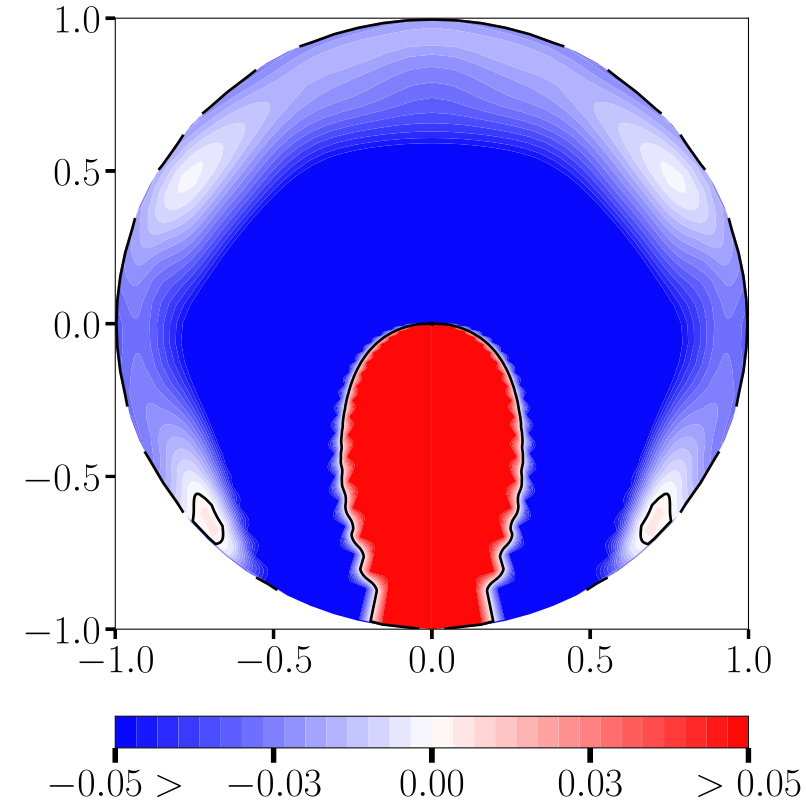}
        \caption{$\ret=170$, NWO}
        \label{fig:pUU170STD}
    \end{subfigure}%
    \begin{subfigure}{0.33\linewidth}
        \includegraphics[width=0.95\linewidth]{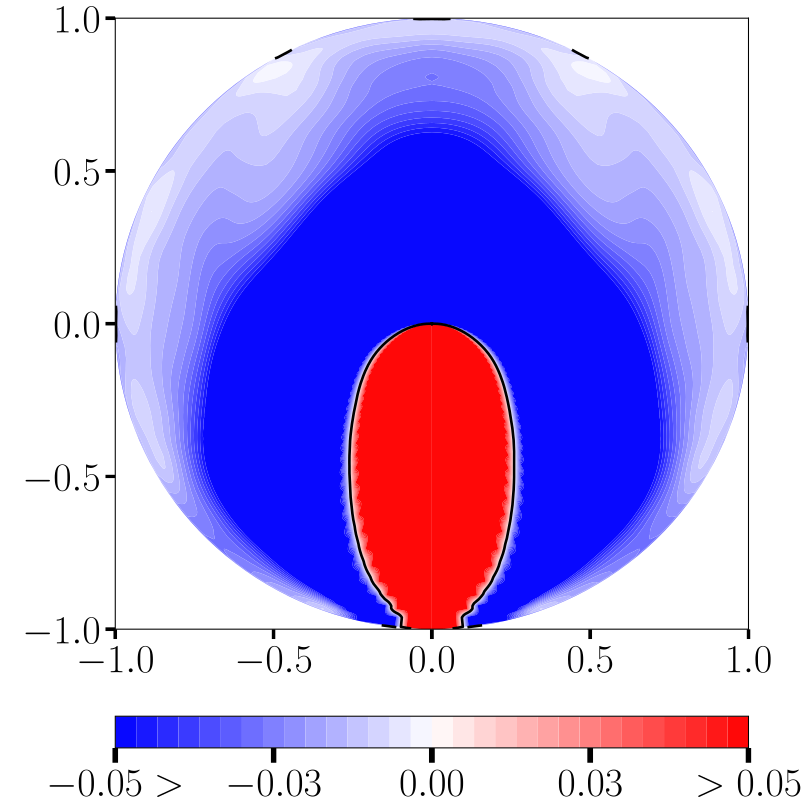}
        \caption{$\ret=360$, NWO}
        \label{fig:pUU360STD}
    \end{subfigure}
    \begin{subfigure}{0.33\linewidth}
        \includegraphics[width=0.95\linewidth]{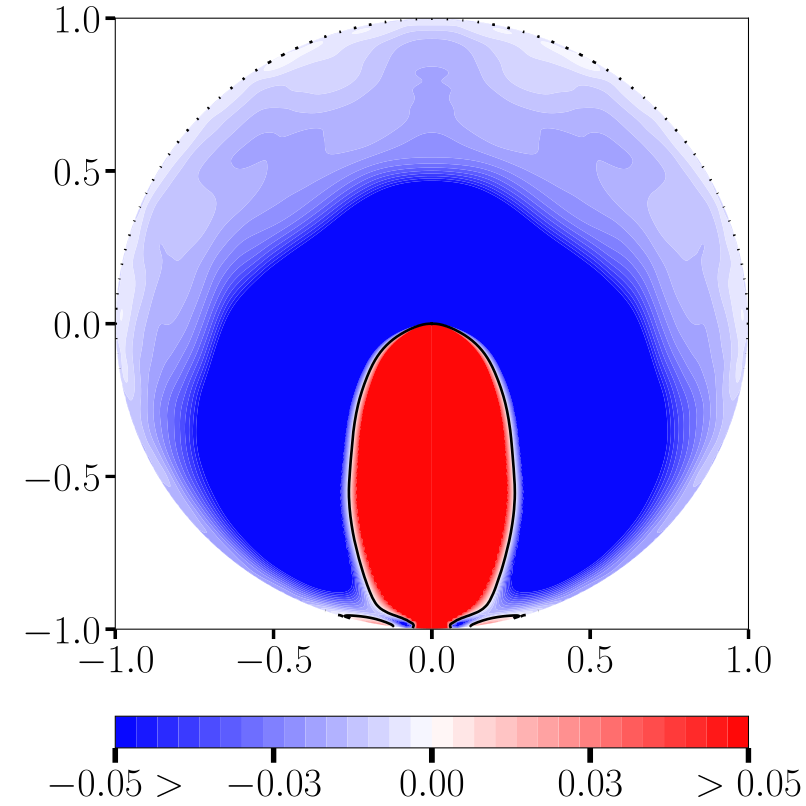}
        \caption{$\ret=720$, NWO}
        \label{fig:pUU720STD}
    \end{subfigure}%
    
    \begin{subfigure}{0.33\linewidth}
        \includegraphics[width=0.95\linewidth]{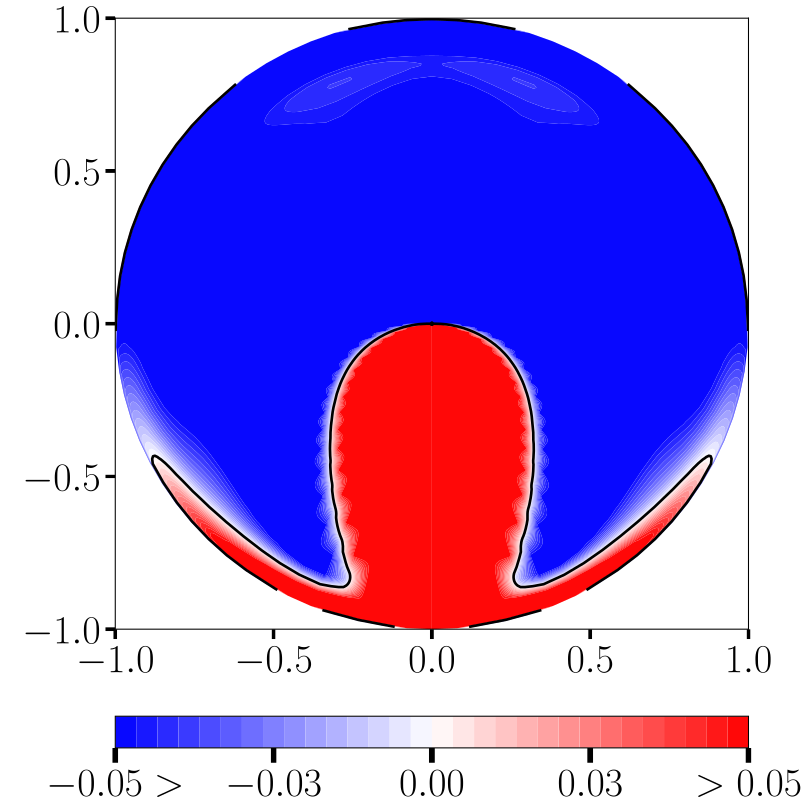}
        \caption{$\ret=170$, WWO}
        \label{fig:pUU170OSC}
    \end{subfigure}
    \begin{subfigure}{0.33\linewidth}
        \includegraphics[width=0.95\linewidth]{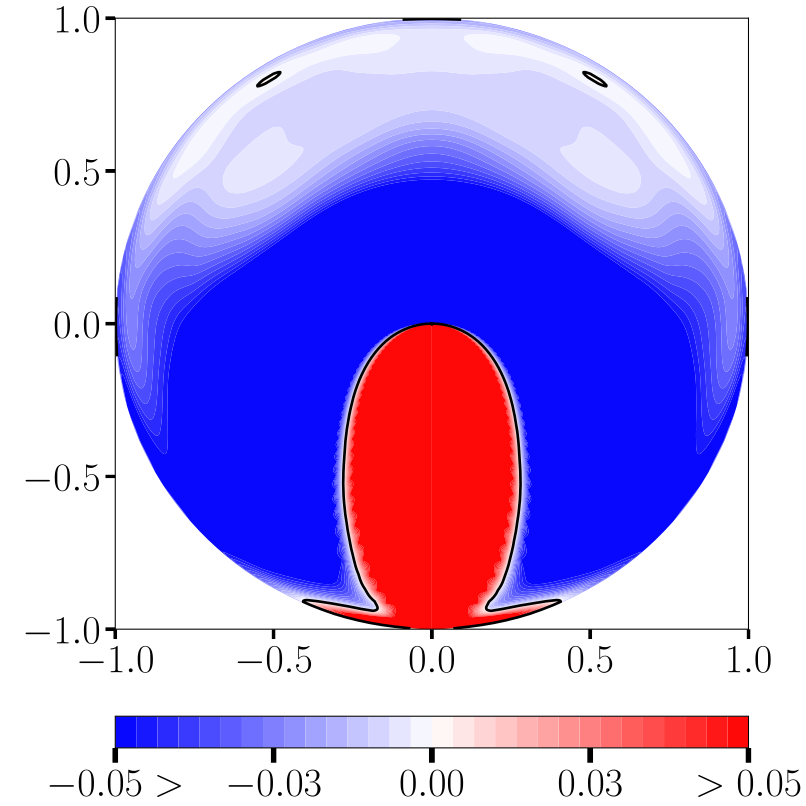}
        \caption{$\ret=360$, WWO}
        \label{fig:pUU360OSC}
    \end{subfigure}%
    \begin{subfigure}{0.33\linewidth}
        \includegraphics[width=0.95\linewidth]{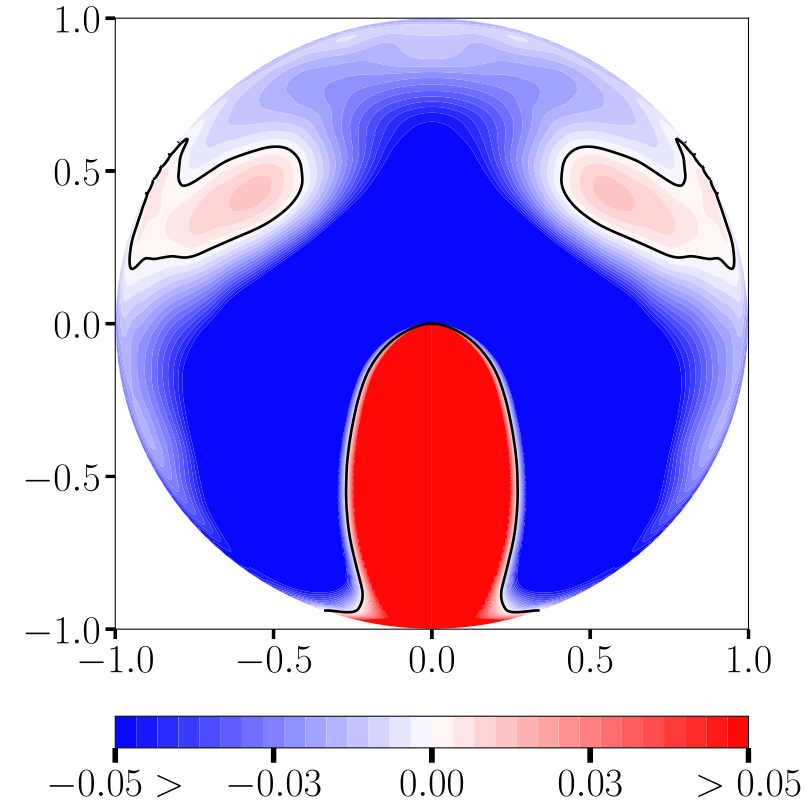}
        \caption{$\ret=720$, WWO}
        \label{fig:pUU720OSC}
    \end{subfigure}
    \caption{Azimuthal correlation coefficient of streamwise velocity at a fixed radial location, $\rho_{u_x u_x}(\Delta \,\theta,\rstar,\rstar)$, for (a,d)  $Re_\tau = 170$; (b,e)  $Re_\tau = 360$; (c,f)  $Re_\tau = 720$. Top, NWO cases; bottom, WWO cases.  
    Black contour lines indicate a level of zero correlation.}
    \label{fig:azcor}
\end{figure}
 Wall oscillations impact the azimuthal energy spectrum of streamwise velocity similarly across the Reynolds numbers when scaled with viscous units, Figure \ref{fig:DeltaUUAzm}.  The energy in the azimuthal wavelengths associated with the near-wall streak spacing, $\lambda_s^+ \approx 100-200$, is reduced in the buffer layer between ($y^+ \approx 5-30$). Since streaks and quasi-streamwise vortices are closely related, wall oscillation presumably weakens the quasi-streamwise near-wall vortices, thereby reducing their transport of streamwise momentum into the streaks. 

 At the two highest Reynolds numbers wall oscillations enhance the spectral energy density at azimuthal wavelengths greater than  $\lambda_s^*\approx \pi/4$. The maxima in Figures \ref{fig:DuuAzmRe360}, \ref{fig:DuuAzmRe720} suggest periodicity of about $\pi/2$. To visualize the structure more explicitly, Figure~\ref{fig:azcor} plots the azimuthal correlation coefficient of streamwise velocity  $\rho_{u_x u_x}(\Delta \theta,\rstar,\rstar)$ between a fixed point at $(\rstar, \Delta \theta=0)$ and all other points on the circle $\rstar=\text{constant}$, while varying $\rstar$ from $0$ to $R$ (note the difference between this set of one-point correlations and the full two-point correlation $\rho_{u_x u_x}(\Delta \theta,\rstar,r_0^{\star})$ between velocities at a fixed point $(r_0^{\star},\Delta \theta=0)$ and all other points in the pipe cross-section $(\rstar, \Delta \theta)$, as presented, e.g., in~\cite{baltzer2013structural}. Strengthening of the correlations for the two highest Reynolds numbers is a maximum at the separation angles of $\Delta\,\theta\approx 4\pi/5$ at $Re_{\tau}=360$ and $\Delta\,\theta\approx 2\pi/3$ at $Re_{\tau}=720$, consistent with the energy spectra enhancement in Figures \ref{fig:DuuAzmRe360}, \ref{fig:DuuAzmRe720} at the wavelengths corresponding to these separation angles. As expected, this mode is not observed in the NWO cases. The origin of this large-scale mode in a wall-oscillated pipe is unknown and requires further investigation.

\begin{figure}
\begin{subfigure}{0.32\linewidth}
        \includegraphics[width=0.99\linewidth]{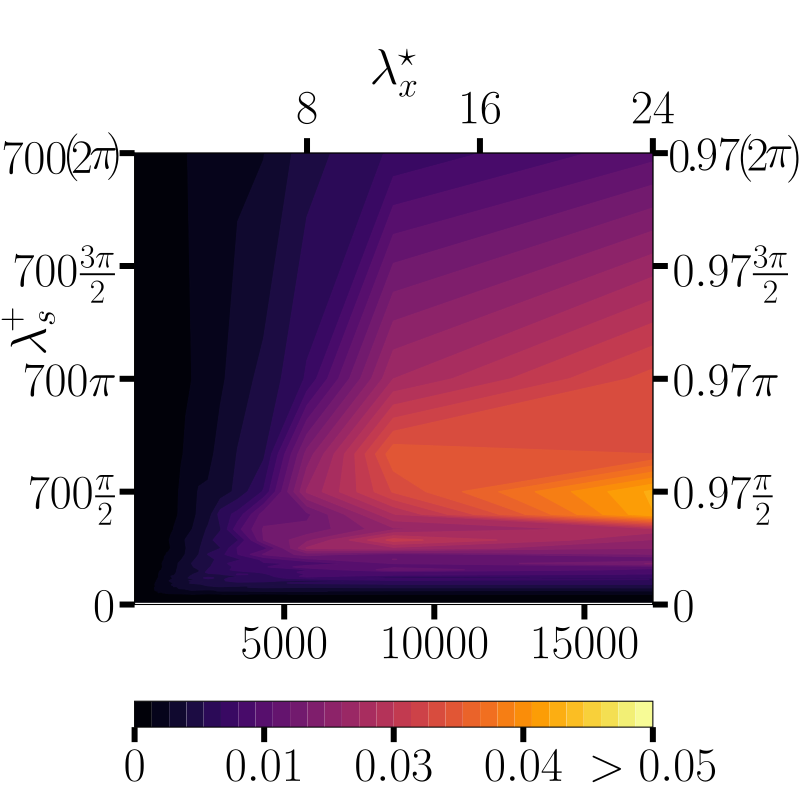}
        \caption{$y^+ = 20$, NWO}
        \label{fig:720uuAzmStwyp20NWO}
    \end{subfigure}
    \begin{subfigure}{0.32\linewidth}
        \includegraphics[width=0.99\linewidth]{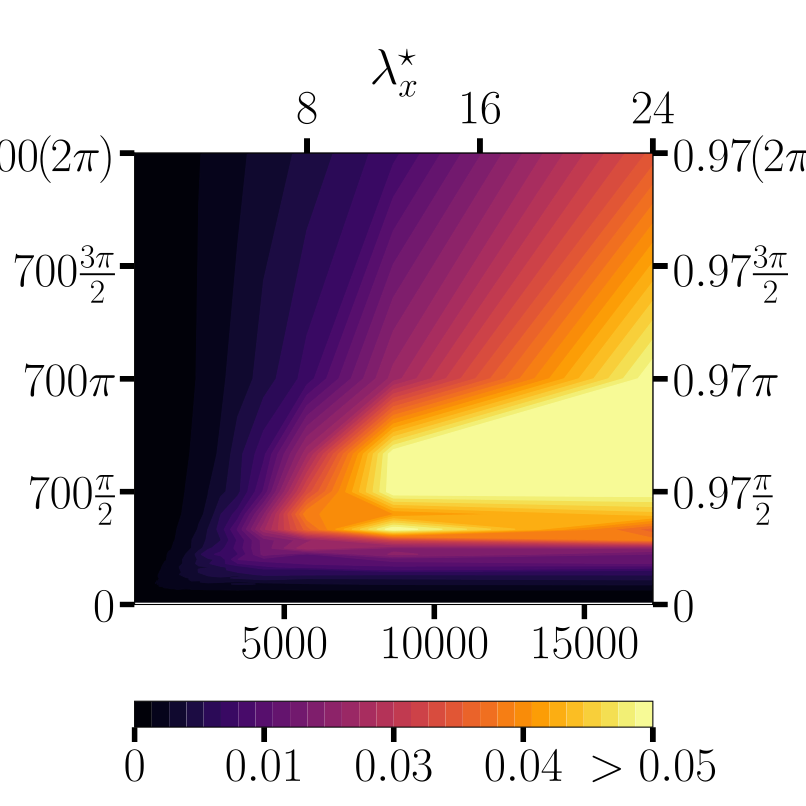}
        \caption{$y^+ = 20$, WWO}
        \label{fig:720uuAzmStwyp20WWO}
    \end{subfigure}
    \begin{subfigure}{0.32\linewidth}
        \includegraphics[width=0.99\linewidth]{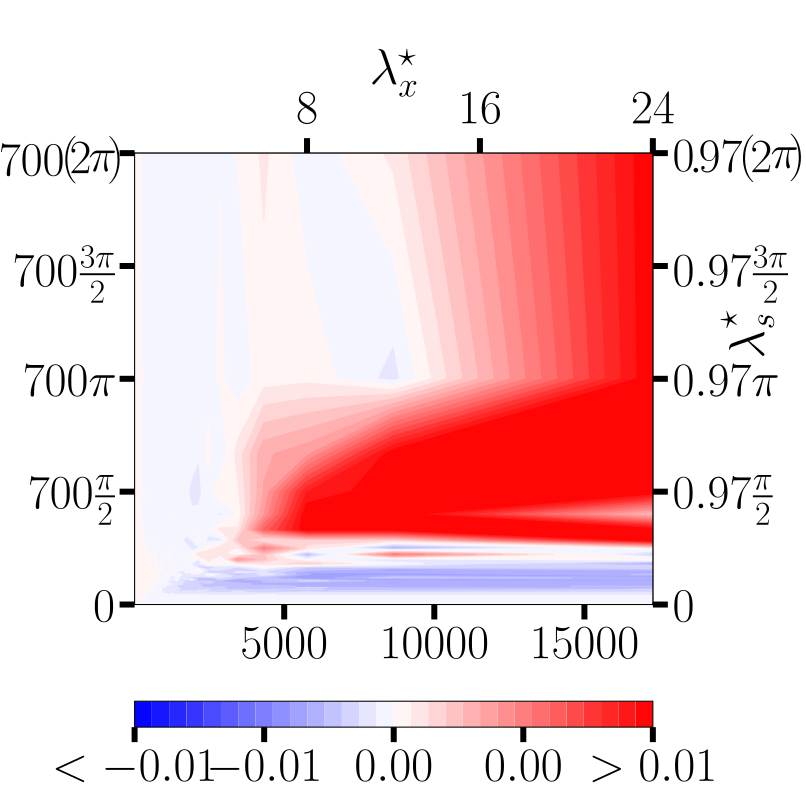}
        \caption{$y^+=20$, WWO-NWO}
        \label{fig:DuuAzmStwyp20Re720}
    \end{subfigure}
    
    \begin{subfigure}{0.32\linewidth}
        \includegraphics[width=0.99\linewidth]{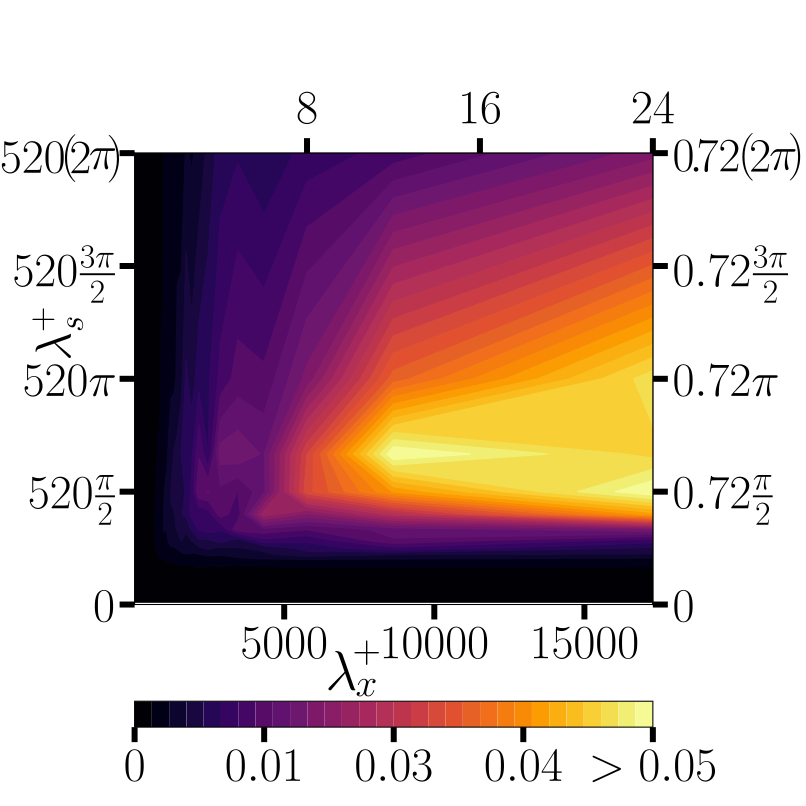}
        \caption{$y^+=200$, NWO}
        \label{fig:720AzmStwyp200NWO}
    \end{subfigure}
    \begin{subfigure}{0.32\linewidth}
        \includegraphics[width=0.99\linewidth]{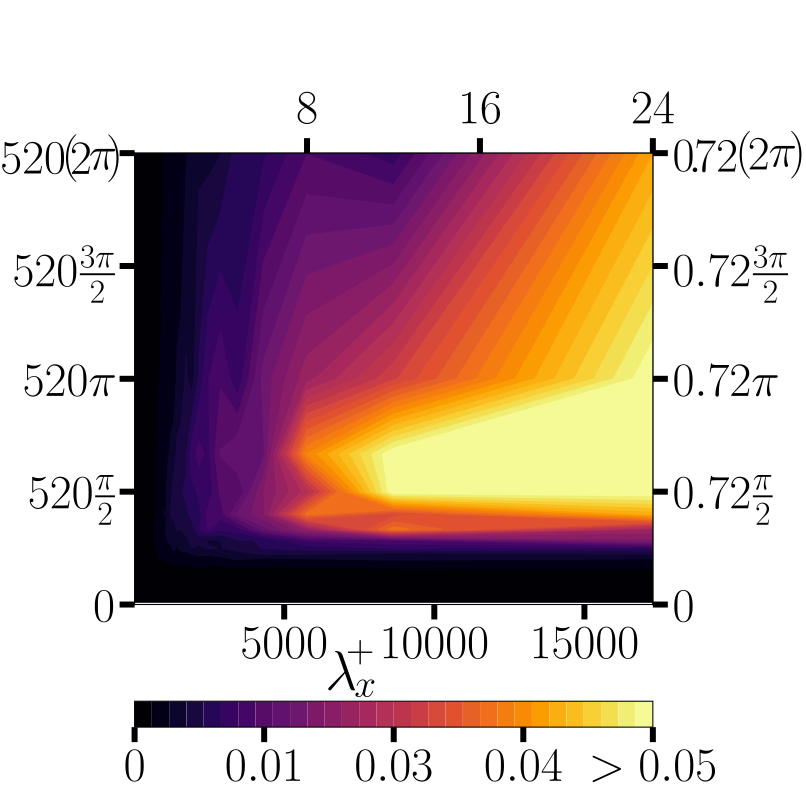}
        \caption{$y^+=200$, WWO}
        \label{fig:720AzmStwyp200WWO}
    \end{subfigure}
    \begin{subfigure}{0.32\linewidth}
        \includegraphics[width=0.99\linewidth]{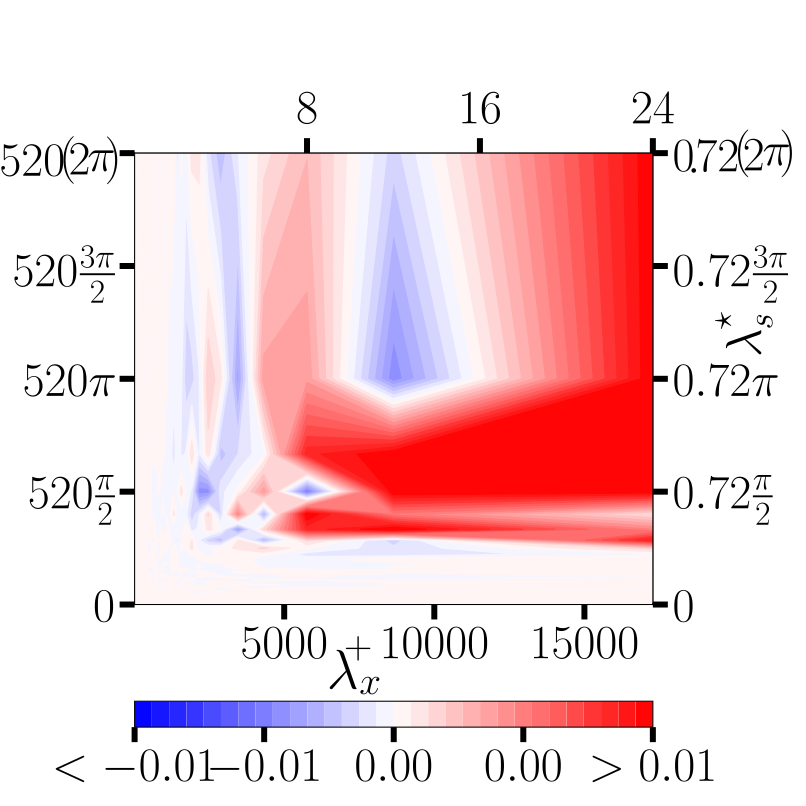}
        \caption{$y^+=200$, WWO-NWO}
        \label{fig:DuuAzmStwyp200Re720}
    \end{subfigure}
    \caption{ Two-dimensional spectra of streamwise kinetic energy, $ \Phi_{u_x u_x}(\lambda_x^+,y^+,\lambda_s^+)/u_\tau^2$, at a wall normal location of (a,b,c) $y^+=20$; (d,e,f) $y^+=200$ for $\ret = 720$. (a,d) NWO spectra; (b,e) WWO spectra; (c,f) change in spectra, $\Delta \Phi_{u_x u_x}(\lambda_x^+,y^+,\lambda_s^+)/u_\tau^2$. }
    \label{fig:720UUypAzmStw}
\end{figure}

Figure~\ref{fig:720UUypAzmStw} further elucidates on the structure of this enhanced large-scale mode in the WWO pipe by plotting the two-dimensional streamwise-azimuthal spectrum of the streamwise kinetic energy, $\Phi_{u_x u_x}(\lambda_x^+,y^+,s^+)$, as well as its change, at two wall-normal locations  of $y^+=20$ and $y^+=200$ for the highest Reynolds numbers, $\ret=720$. The absence of the large-scale mode in the buffer layer ($y^+=20$) in the NWO case and its presence in the WWO case is clearly seen. The structures energized as a result of the developed large-scale mode are very long in the streamwise direction ($\lambda_x^+\ge 10,000$) and their spanwise scale grows proportionally to the streamwise scale. While, consistent with the previous observations~\citep{guala2006large,lee2019space}, large-scale motions do preside in the NWO pipe in the outer layer ($y^+=200$), they are significantly energized by wall oscillations.

\subsection{Effect of Reynolds number on the net turbulent force}
To introduce the net turbulent force, consider the streamwise momentum equation,
\begin{equation}
    \frac{\partial u_x}{\partial t} + \frac{1}{r} \frac{\partial r u_x u_r}{\partial r} + \frac{1}{r^2} \frac{\partial u_x u_\theta}{\partial \theta} + \frac{\partial u_x u_x}{ \partial x} = - \frac{1}{\rho} \frac{\partial \,p'}{\partial x} + \nu\,\nabla^2 u_x - \frac{1}{\rho} \av{\frac{\partial\, p}{\partial \,x}}, 
    \label{eqn:stwMom}
\end{equation}
where $\nabla^2 = \left\lbrace \frac{1}{r} \frac{\partial }{\partial r} \left( r \frac{\partial}{\partial r}\right) + \frac{1}{r^2} \frac{\partial^2}{\partial \theta^2} + \frac{\partial^2}{\partial x^2}\right\rbrace$.  Averaging equation \eqref{eqn:stwMom} in the stationary time and along the homogeneous (streamwise and azimuthal) directions yields:
\begin{equation}
    0=-\frac{1}{r} \frac{d \{r \av{\upp \vpp}\}}{d r}  + \frac{\nu}{r} \frac{d}{d r} \left( r \frac{d}{d r}\right) \av{u_x} -\frac{1}{\rho} \av{\frac{\partial\, p}{\partial \,x}}
    \label{eqn:stwMomAt}.
\end{equation}

The net turbulent force (per unit mass) reduces to the first term on the right-hand side of equation (\ref{eqn:stwMomAt}):
\begin{equation}\label{eqn:netturb}
    F_{turb}(r) = - \frac{1}{r} \frac{d\{ r \av{\upp \vpp}\}}{d r}.
\end{equation}
While equation~(\ref{eqn:netturb}) defines the net turbulent force per unit mass, we will be referring to it as ``the net turbulent force'' for brevity. The net turbulent force is made non-dimensional by scaling the radial coordinate with $r = r^+ l_\tau$ and velocities with $u_i = u_i^+ \utau$, such that:
\begin{equation}\label{eqn:netturbplus}
    F_{turb}^+(r^+) = -\frac{1}{r^+} \frac{d \{r^+ \av{\upp^+ \vpp^+}\}}{d r^+},
\end{equation}
where $F_{turb}^+=F_{turb}\, l_{\tau}/u_{\tau}^2$.

The net turbulent force determines the local acceleration or deceleration of the mean flow due to turbulent shear stresses. When the flow is stationary, it balances the contributions from the mean pressure gradient and the viscous stress. The net turbulent force accelerates the mean flow below the region of maximum Reynolds shear stress (premultiplied Reynolds shear stress, $r \av{\upp \vpp}$, for a pipe) and decelerates the flow above it. 
At the location of the maximum (premultiplied) Reynolds shear stress, the net turbulent force equals zero. Utilizing Parseval's theorem (See equation (\ref{eqn:parseval})), the net turbulent force can be decomposed into a sum of contributions from the streamwise and azimuthal Fourier modes as:
\begin{equation}
    F_{turb}^+(r^+) = -\sum_{k_x} \sum_{k_\theta} \frac{1}{r^+} \frac{\partial \{r^+ \Phi_{u_x u_r} (k_x, r, k_\theta)\} }{\partial r^+}.
\end{equation}
\begin{figure}
    \centering
    \begin{subfigure}{0.33\linewidth}
        \includegraphics[width=0.99\linewidth]{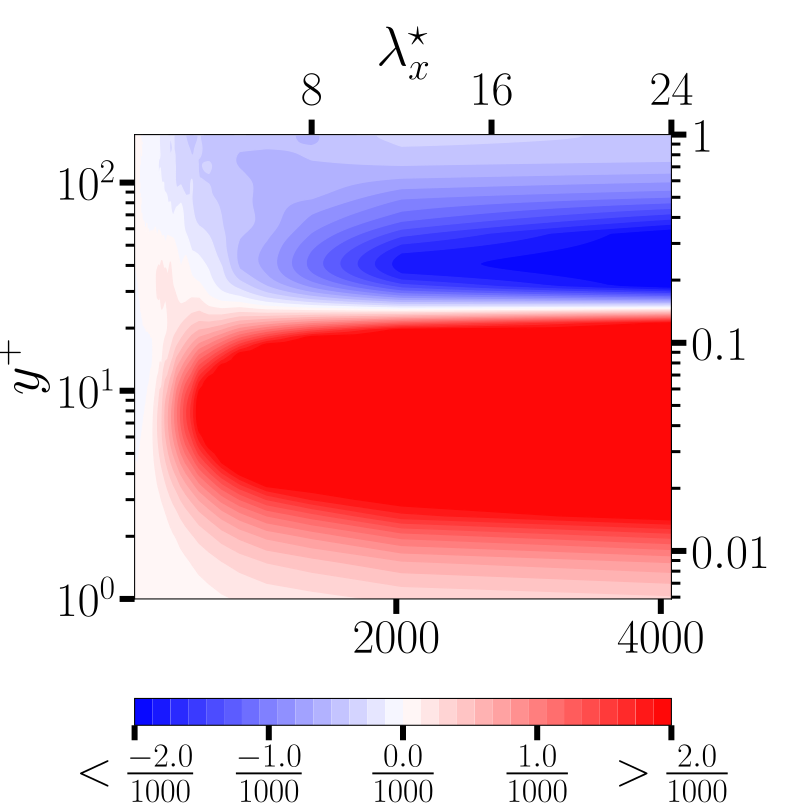}
        \caption{$\ret=170$, NWO}
        \label{fig:DuvDyStwRe170std}
    \end{subfigure}%
    \begin{subfigure}{0.33\linewidth}
        \includegraphics[width=0.99\linewidth]{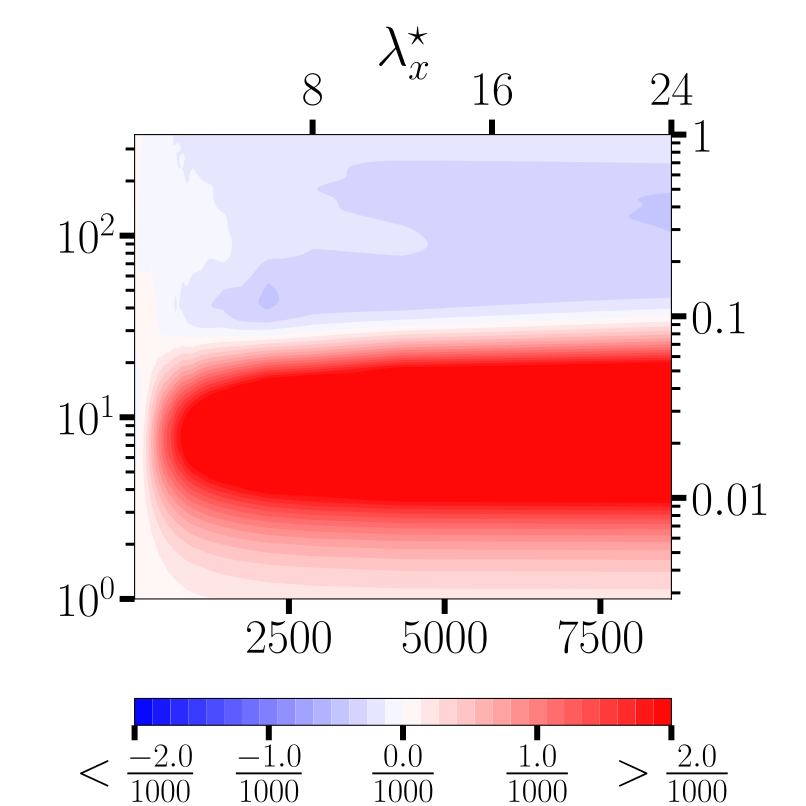}
        \caption{$\ret=360$, NWO}
        \label{fig:DuvDyStwRe360std}
    \end{subfigure}
    \begin{subfigure}{0.33\linewidth}
        \includegraphics[width=0.99\linewidth]{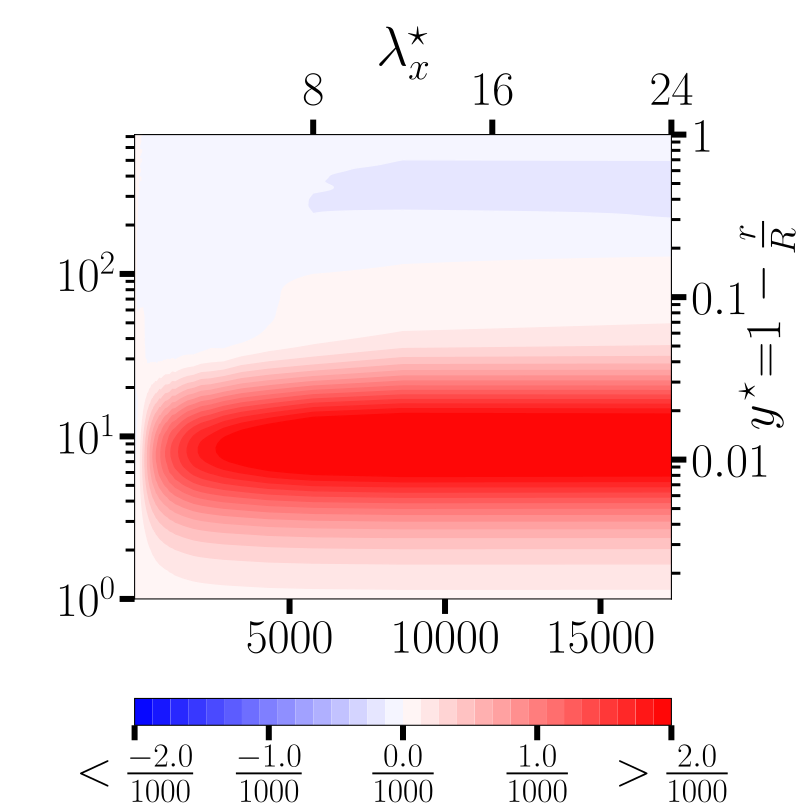}
        \caption{$\ret=720$, NWO}
        \label{fig:DuvDyStwRe720std}
    \end{subfigure}%
    
    \begin{subfigure}{0.33\linewidth}
        \includegraphics[width=0.99\linewidth]{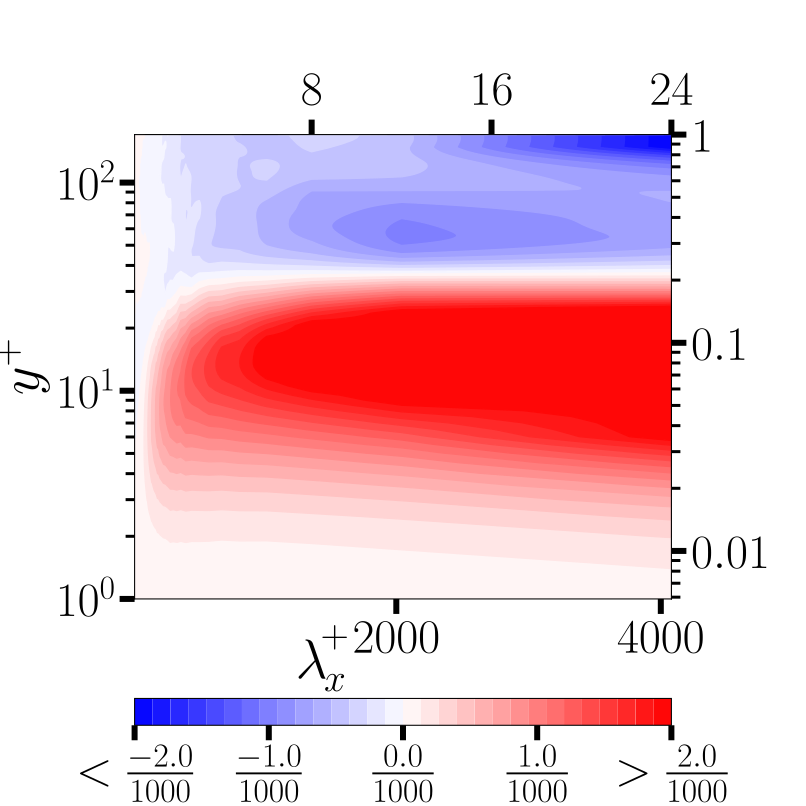}
        \caption{$\ret=170$, WWO}
        \label{fig:DuvDySTWRe360osc}
    \end{subfigure}
    \begin{subfigure}{0.33\linewidth}
        \includegraphics[width=0.99\linewidth]{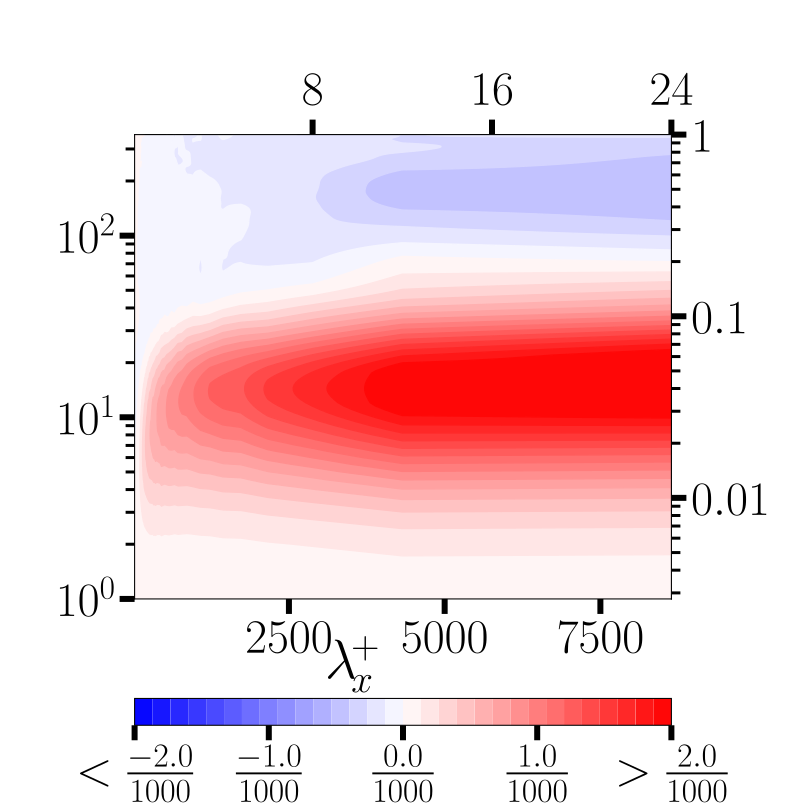}
        \caption{$\ret=360$, WWO}
        \label{fig:DuvDySTWRe720std}
    \end{subfigure}%
    \begin{subfigure}{0.33\linewidth}
        \includegraphics[width=0.99\linewidth]{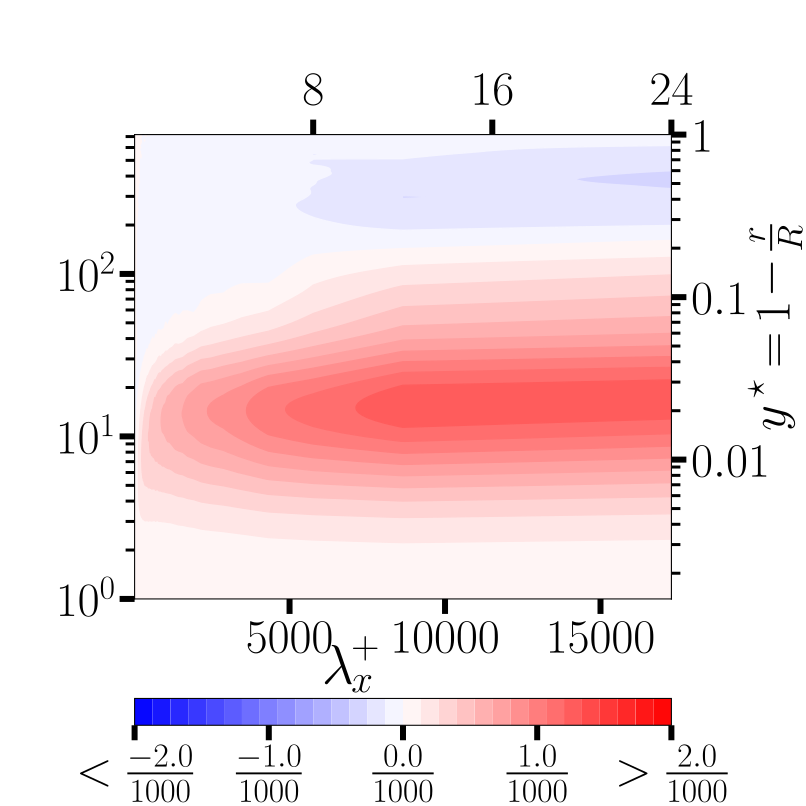}
        \caption{$\ret=720$, WWO}
        \label{fig:DuvSTWRe720osc}
    \end{subfigure}
    \caption{Streamwise spectra of the net turbulent force as a function of the wall normal coordinate and streamwise wavelength, $F_{turb}^+ (\lambda_x^+,y^+)$, for (a,d) $Re_\tau = 170$; (b,e) $Re_\tau = 360$;  (c,f) $Re_\tau = 720$. Top, NWO cases; bottom, WWO cases.
     }
    \label{fig:DeltaUVDyStw}
\end{figure}
Figure \ref{fig:DeltaUVDyStw} shows the streamwise spectra of the net turbulent force, $F_{turb}^+(\lambda_x^+,y^+)$, as a function of the wall normal coordinate and the streamwise wavelength for the NWO and WWO cases. We remark that Figure \ref{fig:DeltaUVDyStw} agrees well with the data presented in~\cite{wu2012direct} (Figure 17) for the NWO pipe flow if replotted on a log-log scale (not shown here for the sake of brevity). Consistent with previous observations~\citep{guala2006large,balakumar2007large,wu2012direct}, we find that the net force is positive across all the scales of motion $\lambda_x^+\ge 100$ in the buffer layer ($y^+<20$), amounting to an acceleration of the mean flow, and negative above it, implying retardation. While all scales of motion experience the aforementioned acceleration and deceleration, the effect is larger for large scales at all Reynolds numbers, consistent with the works of \cite{guala2006large,wu2012direct}. The impact of the wall oscillations is to reduce this effect, diminishing both the acceleration of turbulent structures near the wall and their deceleration in the outer layer. In general these changes are conducive to drag reduction, since they bring the mean velocity profile closer to its laminar shape.

\begin{figure}
    \centering
    \begin{subfigure}{0.45\linewidth}
        \includegraphics[width=0.95\linewidth]{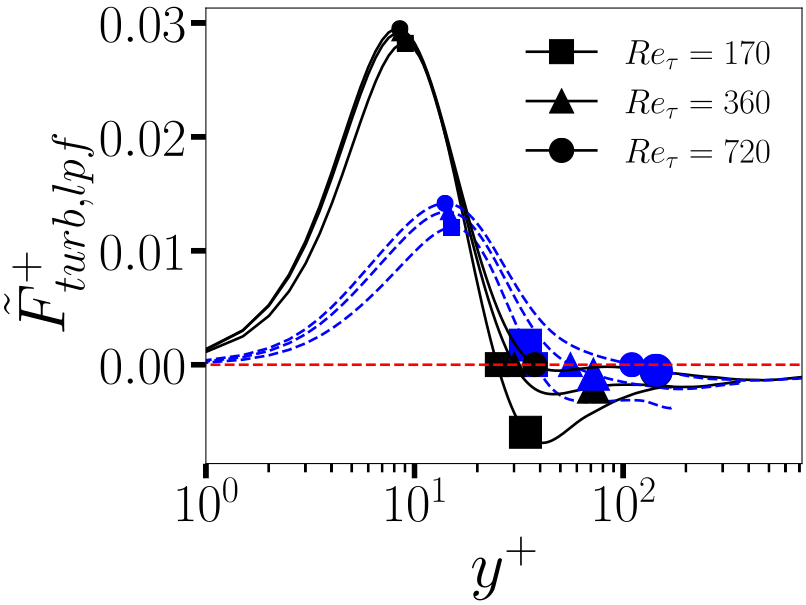}
        \caption{Low-pass filtered}
        \label{fig:fturblpf}
    \end{subfigure}
    \begin{subfigure}{0.45\linewidth}
        \includegraphics[width=0.95\linewidth]{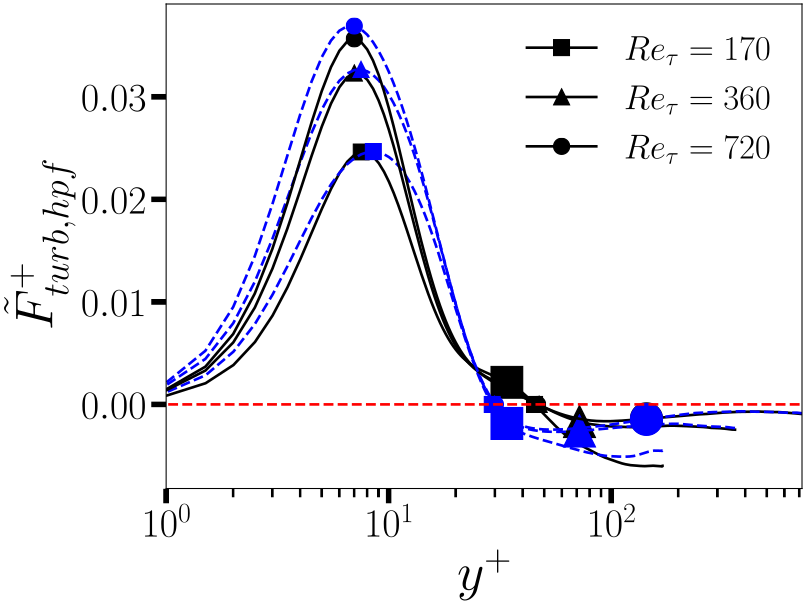}
        \caption{High-pass filtered}
        \label{fig:fturbhpf}
    \end{subfigure}
    \caption{Filtered net turbulent force profile as a function of the wall normal coordinate: (a) low-pass filtered net turbulent force, and (b) high-pass filtered net turbulent force. Black solid lines indicate NWO and blue dashed lines indicate WWO.  The marker closest to the wall (smallest) indicates the location of the maximum accelerating turbulent force, the next marker indicates the location of the zero turbulent force and the last (largest) marker indicates the top of the log layer for the given Reynolds number. 
    }
    \label{fig:fturbfiltcomp}
\end{figure}

To quantify the  modification of the net turbulent force by wall oscillations and to assess the contributions of different scales of motion, we apply 
a Gaussian low-pass filter~\citep{guala2006large,lee2019space},
\begin{equation}
\hat{g}_{lpf}(k_x) = \exp \left( -\frac{ k_x^2}{2 \sigma^2}\right),
\end{equation}
with $\sigma$ being the filter width, and  $k_x$ the streamwise wavenumber. We set the filter width such that the strength of the filter is at 50\% of its peak at a filter cutoff location, $k_{x,cutoff}=2\pi/\lambda_{x,cutoff}$. This gives the value of $\sigma^2=k_{x,cutoff}^2/(2\ln 2)$. The cutoff wavelength  $\lambda_{x,cutoff}^{+}=1000$ is chosen 
such that the scales smaller than this value are attenuated by the low-pass filter. Conversely, its high-pass filter counterpart, $\hat{g}_{hpf}(k_x) = 1 - \hat{g}_{lpf}(k_x)$, attenuates the scales with $\lambda_{x}^{+}\ge \lambda_{x,cutoff}^{+}$. We apply both filters to the net turbulent force spectra. The filtered net turbulent force is defined as
\begin{equation}
\tilde{F}_{turb,\{lpf,hpf\}}^+(r^+) = -\sum_{k_x} \sum_{k_\theta} \frac{1}{r^+} \frac{\partial \{r^+ \Phi_{u_x u_r} (k_x, r, k_\theta)\} }{\partial r^+} \hat{g}_{\{lpf,hpf\}}(k_x).
\end{equation} The cumulative low-pass and high-pass filtered contributions are documented in Figure \ref{fig:fturbfiltcomp}. Additionally, Table \ref{tab:fturbzero} records the position of the zero net turbulent force for the total (unfiltered) and filtered quantities for the NWO and WWO cases, together with the difference between the NWO and WWO locations ($\Delta$). We refer to the location of the zero net turbulent force based on the total (unfiltered) quantities as $y_{f0}^+$.

Figure \ref{fig:fturbfiltcomp} shows that the wall oscillations significantly attenuate the magnitude of the low-pass filtered net turbulent force while leaving its high-pass filtered counterpart relatively unchanged, i.e. the major effect of the net turbulent force modification is coming from relatively large scales of motion ($\lambda_x^+ > 1000$). A considerable net force reduction in large scales (low-pass filtered contribution) is found all the way from the wall and throughout the top of the log layer of the flow.  Another important effect is the shift of both the maximum and the zero net force locations upwards by wall oscillations, which is primarily seen in its low-pass filtered contribution. As can be judged from 
the Table \ref{tab:fturbzero}, the major effect on the shift indeed comes from the large scales of motion, with the small and intermediate scales (high-pass filtered) contributing less than 20\% of the total shift. Overall, the small and intermediate scales, shown in Figure \ref{fig:fturbhpf}, promote a slightly higher acceleration of the mean velocity profile in the WWO case near the wall, with the decreasing effect as the distance from the wall increases. Between the top of the buffer layer and the top of the log layer, the small and intermediate scales exhibit a stronger deceleration of the mean velocity profile in the WWO case. Overall, the results indicate that the acceleration due to large and very large scales is the most impacted by the wall oscillations.  To the contrary, large-scale net turbulent force is  enhanced above the log layer of the flow for the two highest Reynolds numbers. 
This is consistent with a reduction of the normalized centerline velocity $\bar{U}_c=U_c/U_{bulk}$, despite the growth of the bulk mean velocity $U_{bulk}$ in the WWO as compared to the NWO cases observed in Table~\ref{tble:Ub_Cf}: increased (negative) turbulent force in this region acts to decelerate the large-scale structures more significantly in the center of the pipe flow with wall oscillation.  This effect indicates a decreased effectiveness of the drag reduction mechanism  in the outer layer.  The centerline retardation  increases with the Reynolds number, pointing once again towards a reduced effectiveness of the current drag reduction mechanism at higher Reynolds numbers.

\begin{table}
    \centering
    \begin{tabular}{c c c c}
         $\ret$ & (NWO,\,WWO,\,$ \Delta$) & (NWO,\,WWO,\,$ \Delta$)   & (NWO,\,WWO, \,$ \Delta$)   \\
          &  unfiltered&  low-pass filtered  & high-pass filtered \\ \hline
         170 & (25.8,\,31.1,5.3) & (24.9,\,38.0,\,13.1) & (45.3,\,29.2,\,-16.1)  \\
         360 & (31.7,\,37.1,5.4) & (30.1,\,55.9,\,25.8) & (47.4,\,29.9,\,-17.5) \\
         720 & (39.1,\,44.8,5.7) & (37.7,\,110.1,\,72.4) & (46.7,\,30.2,\, -16.5)
    \end{tabular}
    \caption{Wall normal location of the zero net turbulent force in plus units
    for (NWO,WWO) based on unfiltered, low-pass filtered, and high-pass filtered quantities, along with the difference between the WWO and NWO locations ($\Delta$).}
    \label{tab:fturbzero}
\end{table}

\begin{figure}
    \centering
    \begin{subfigure}{0.33\linewidth}
        \includegraphics[width=0.99\linewidth]{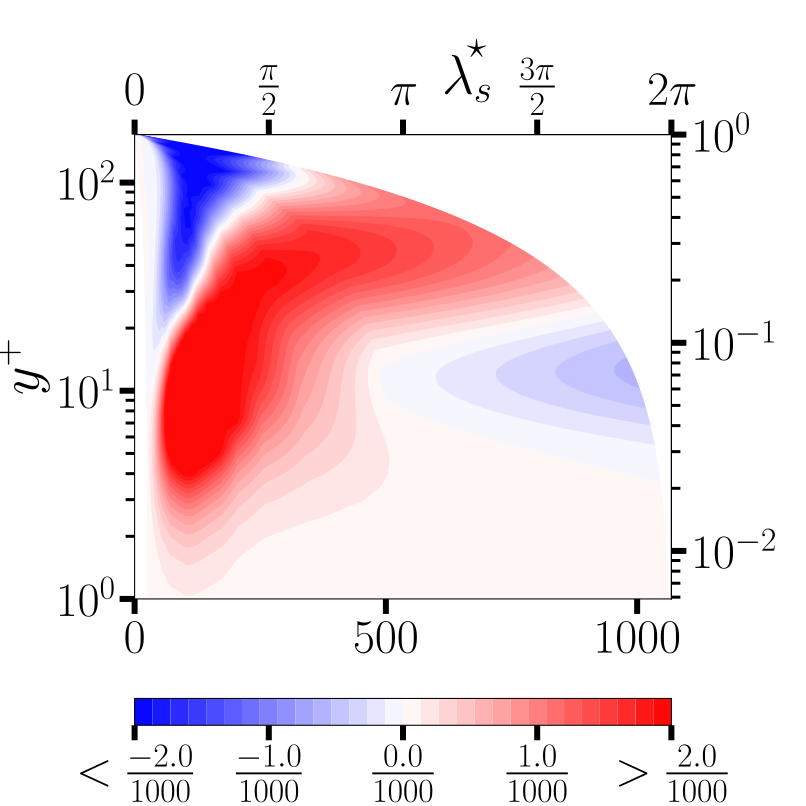}
        \caption{$\ret=170$, NWO}
        \label{fig:DuvDyAzmRe170std}
    \end{subfigure}%
    \begin{subfigure}{0.33\linewidth}
        \includegraphics[width=0.99\linewidth]{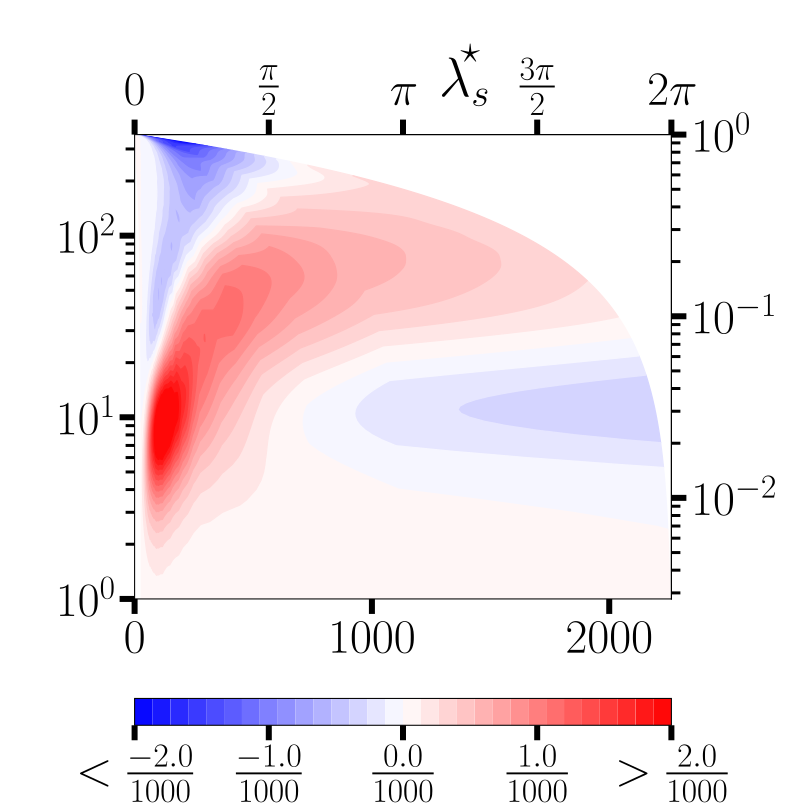}
        \caption{$\ret=360$, NWO}
        \label{fig:DuvDyAzmRe360std}
    \end{subfigure}
    \begin{subfigure}{0.33\linewidth}
        \includegraphics[width=0.99\linewidth]{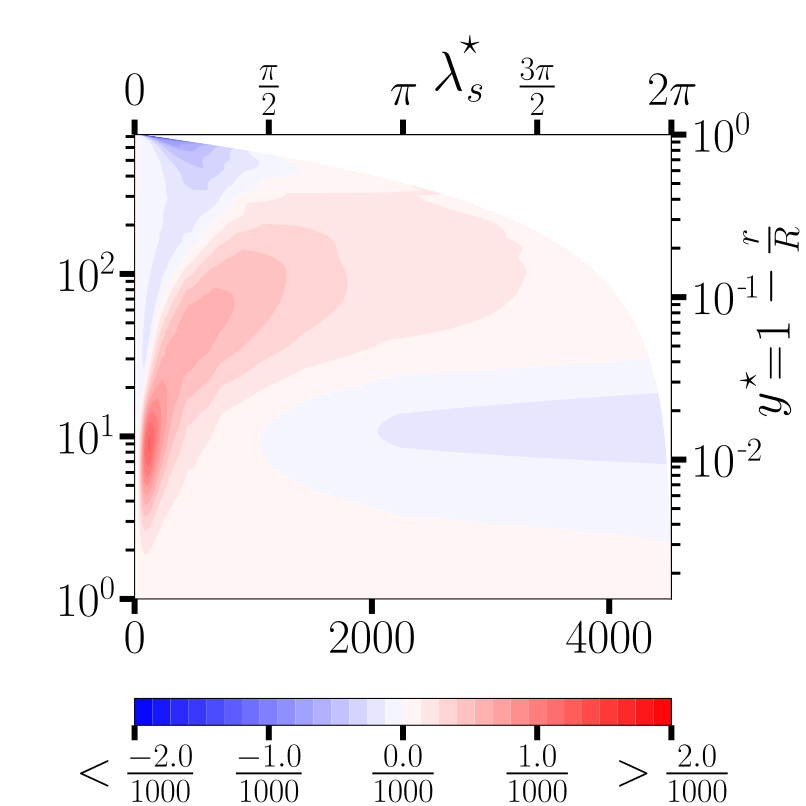}
        \caption{$\ret=720$, NWO}
        \label{fig:DuvDyAzmRe720std}
    \end{subfigure}%
    
    \begin{subfigure}{0.33\linewidth}
        \includegraphics[width=0.99\linewidth]{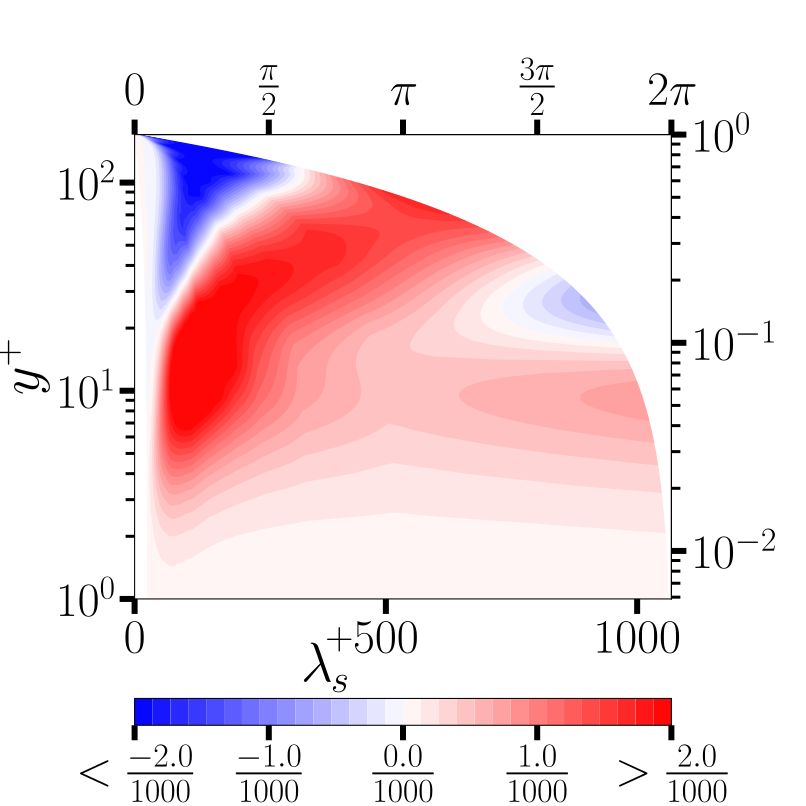}
        \caption{$\ret=170$, WWO}
        \label{fig:DuvDyAzmRe170osc}
    \end{subfigure}
    \begin{subfigure}{0.33\linewidth}
        \includegraphics[width=0.99\linewidth]{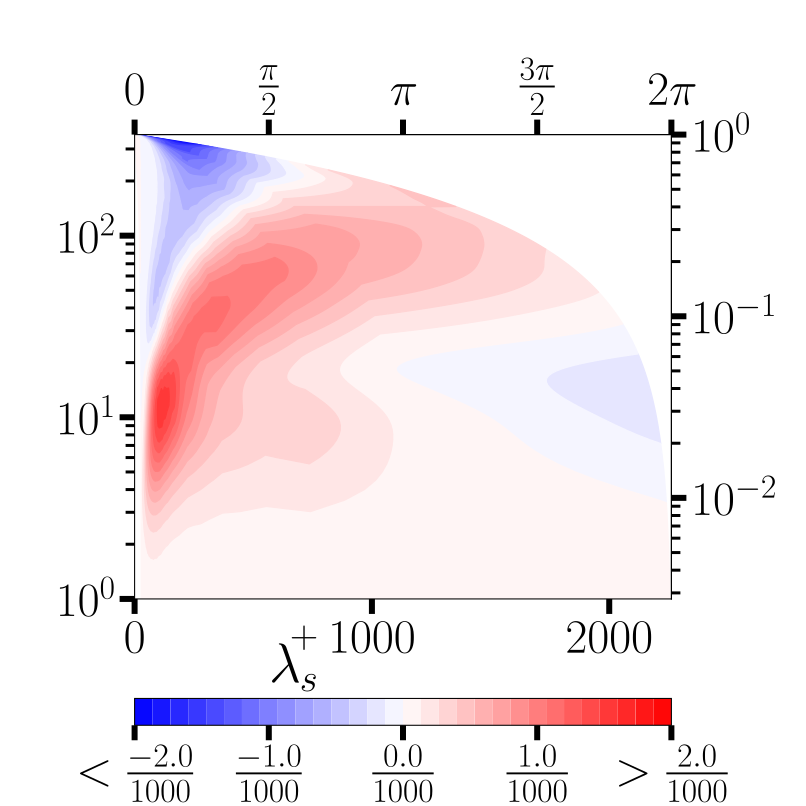}
        \caption{$\ret=360$, WWO}
        \label{fig:DuvDyAzmRe360osc}
    \end{subfigure}%
    \begin{subfigure}{0.33\linewidth}
        \includegraphics[width=0.99\linewidth]{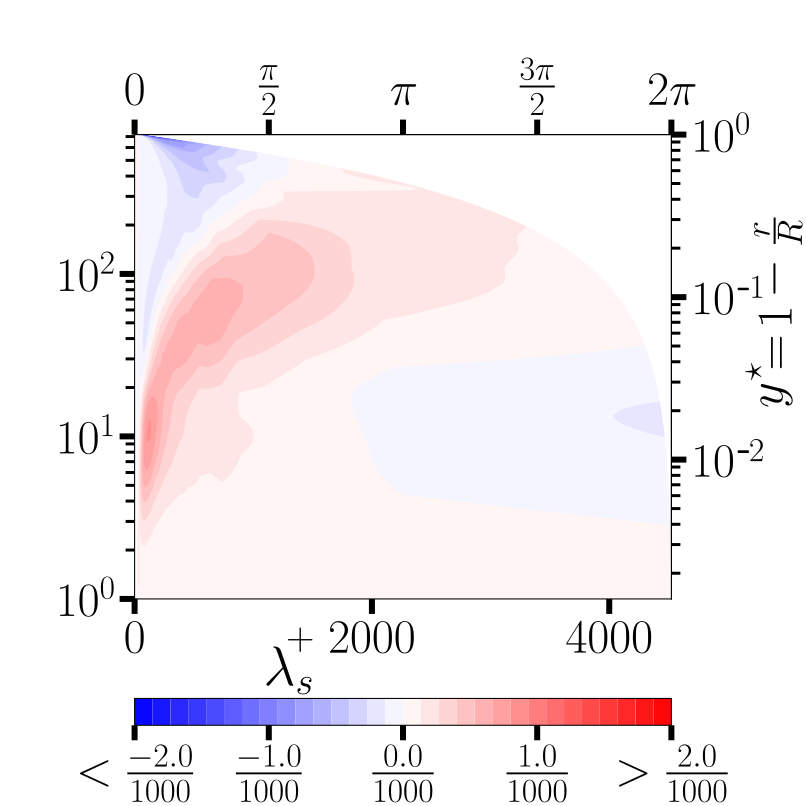}
        \caption{$\ret=720$, WWO}
        \label{fig:DuvAzmRe720osc}
    \end{subfigure}
    \caption{Azimuthal spectra of the net turbulent force as a function of the wall normal coordinate and azimuthal wavelength, $F_{turb}^+, (\lambda_s^+,y^+)$, for (a,d) $Re_\tau = 170$; (b,e) $Re_\tau = 360$;  (c,f) $Re_\tau = 720$. Top, NWO cases; bottom, WWO cases.}
    \label{fig:DeltaUVDyAzm}
\end{figure}

Figure \ref{fig:DeltaUVDyAzm} shows the spectral decomposition of the net turbulent force as a function of the azimuthal wavelength. In the NWO cases, we observe a clear peak in the net turbulent force with the azimuthal wavelength $\lambda_s^+ \approx 100$, associated with the near-wall streak spacing, at the bottom of the buffer layer ($y^+\approx 10$) in all three Reynolds numbers. Consistent with the previous observations, the turbulent motions are generally accelerated in and below the buffer layer and are decelerated above it. The effect of wall oscillations, as in the case of a streamwise spectra, is to diminish these accelerating and decelerating motions. We observe that the major reduction in the net turbulent force comes from large azimuthal scales in the buffer layer ($\lambda_s^+>1000$), whose acceleration is  retarded as a result of wall oscillations. The suppression of the net turbulent force in large streamwise and azimuthal scales due to wall oscillations is consistent with the hypothesis that wall oscillations inhibit the growth of the hairpin packets. The lack of growth in the large-scale structures forming the packets prevents steepening of the velocity gradient in the buffer layer, which leads to a lower net turbulent force in this region in the WWO cases. This inhibition of growth of the hairpin packets  may be a consequence of suppression of turbulent auto-generation mechanisms with wall oscillation. 

The budget of the net turbulent force describes the contribution of the velocity-vorticity correlations to the turbulent force \citep{klewicki1989velocity}. In cylindrical coordinates, such a decomposition applied to equation~(\ref{eqn:netturbplus})
can be shown to be:
\begin{equation}
    F_{turb}(r)^+ =\langle \vpp^+ {\omega_\theta^\dpr}^+ \rangle - \langle \wpp^+ {\omega_r^\dpr}^+ \rangle - \frac{\langle \upp^+ \vpp^+ \rangle}{r^+}.
\end{equation}
The first term on the right-hand side, $\langle \vpp^+ {\omega_\theta^\dpr}^+\rangle$, is referred to as the advective vorticity transport, the second term, $-\langle \wpp^+ {\omega_r^\dpr}^+ \rangle$, is the vortex stretching term \citep{yoon2016contribution}, and the last term, $ - \langle \upp^+ \vpp^+ \rangle/r^+$ arises due to a cylindrical geometry of the problem.  This allows for a physical interpretation of the effects causing the reduction in the net turbulent force.

\begin{figure}
    \centering
    \begin{subfigure}{0.33\linewidth}
        \includegraphics[width=0.99\linewidth]{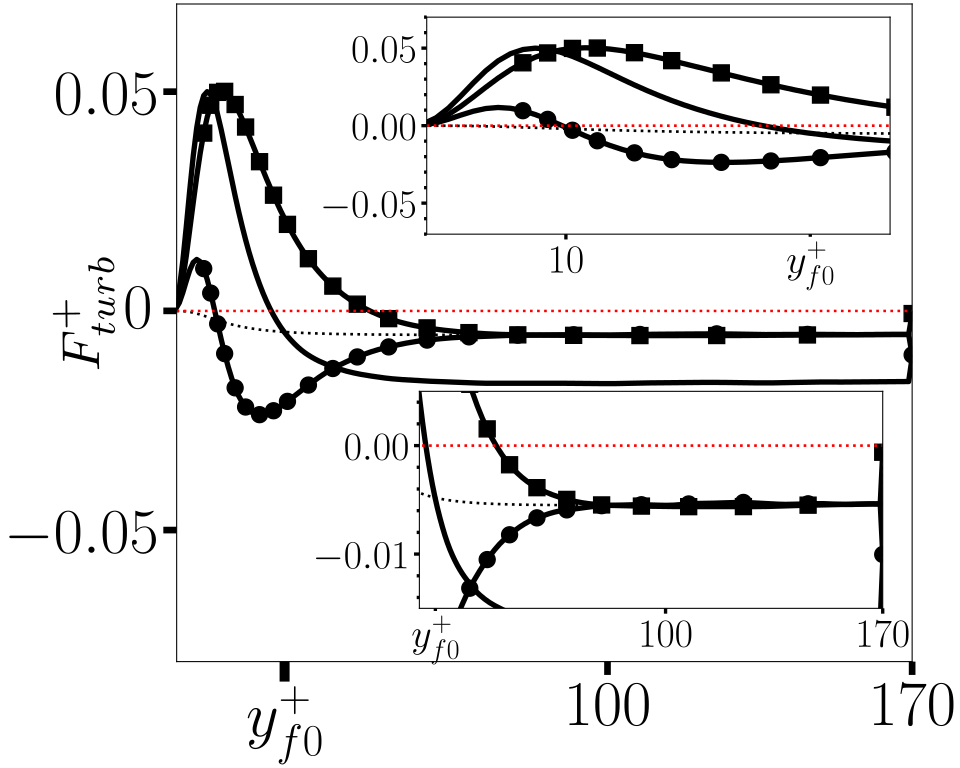}
        \caption{$\ret=170$, NWO}
        \label{fig:budgetForce170std}
    \end{subfigure}%
    \begin{subfigure}{0.33\linewidth}
        \includegraphics[width=0.99\linewidth]{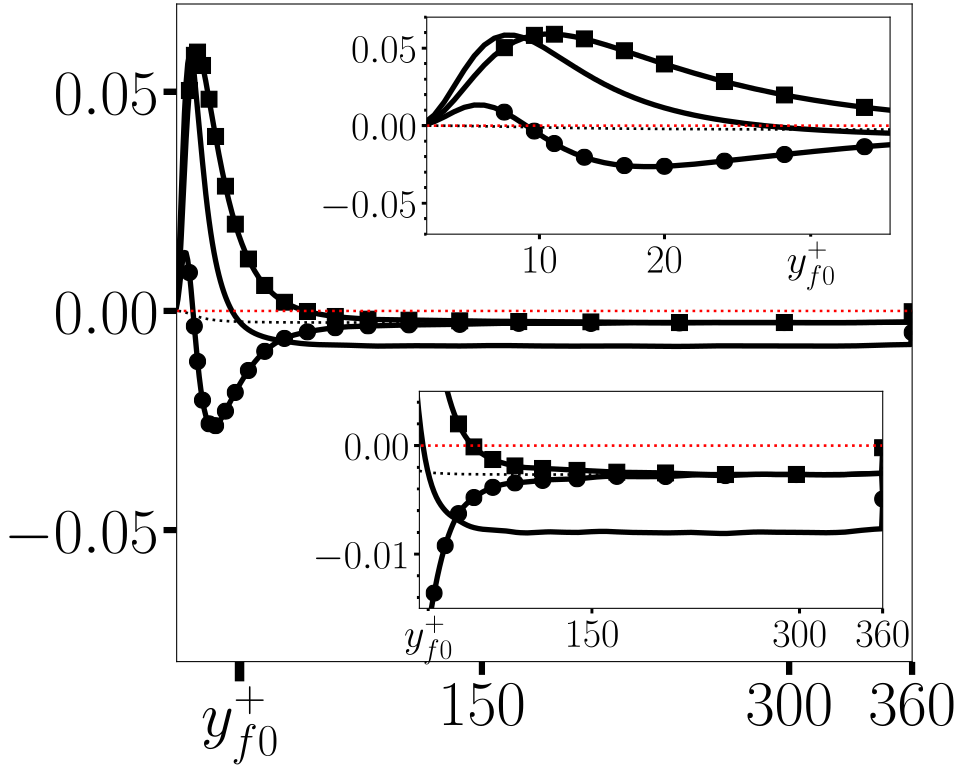}
        \caption{$\ret=360$, NWO}
        \label{fig:budgetForce360std}
    \end{subfigure}
    \begin{subfigure}{0.33\linewidth}
        \includegraphics[width=0.99\linewidth]{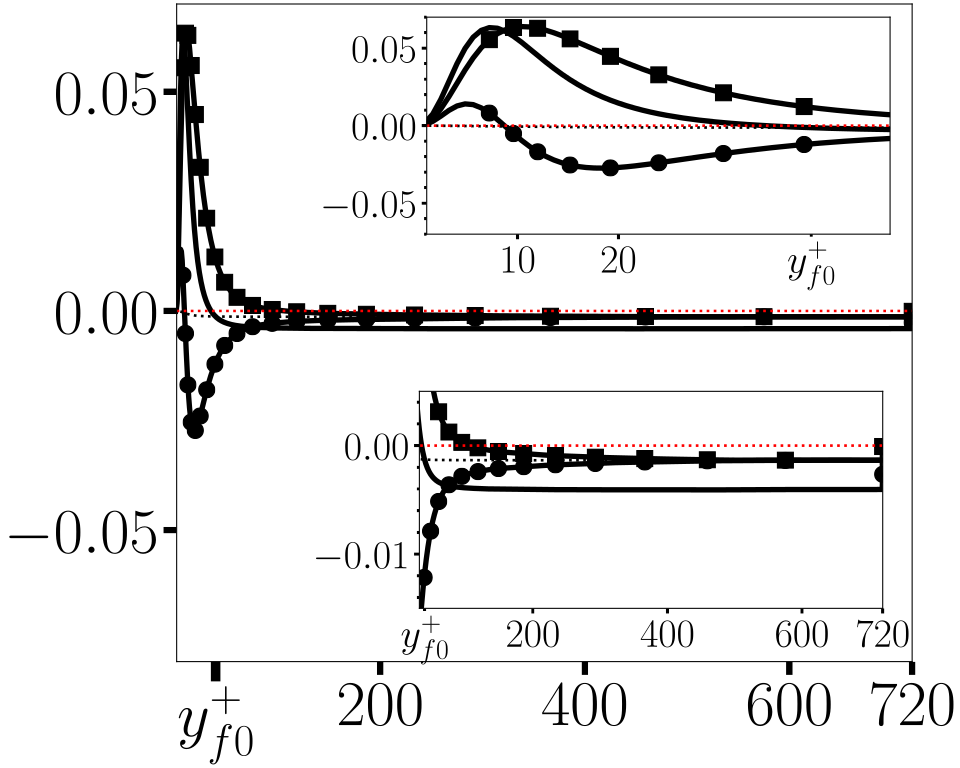}
        \caption{$\ret=720$, NWO}
        \label{fig:budgetForce720std}
    \end{subfigure}%
    
    \begin{subfigure}{0.33\linewidth}
        \includegraphics[width=0.99\linewidth]{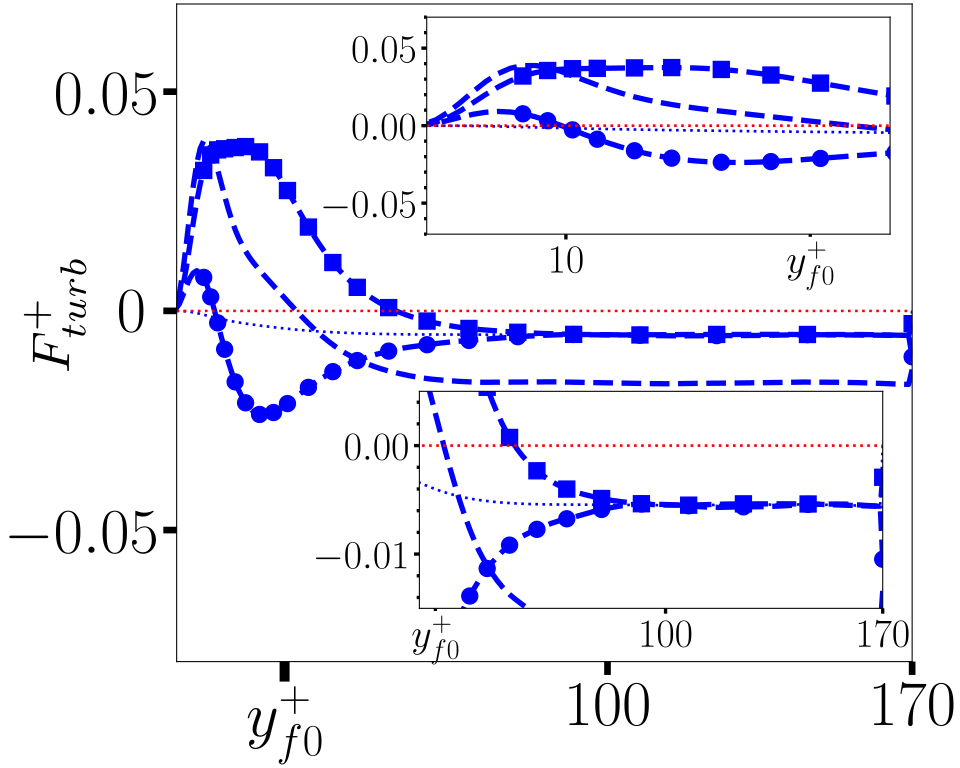}
        \caption{$\ret=170$, WWO}
        \label{fig:budgetForce170osc}
    \end{subfigure}
    \begin{subfigure}{0.33\linewidth}
        \includegraphics[width=0.99\linewidth]{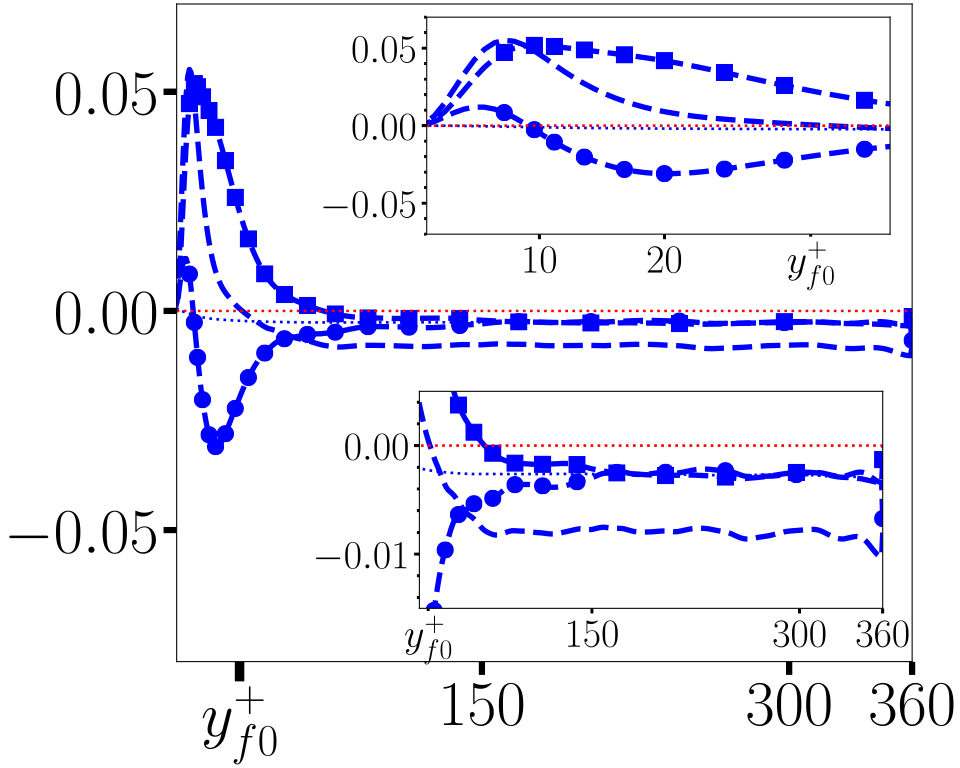}
        \caption{$\ret=360$, WWO}
        \label{fig:budgetForce360osc}
    \end{subfigure}%
    \begin{subfigure}{0.33\linewidth}
        \includegraphics[width=0.99\linewidth]{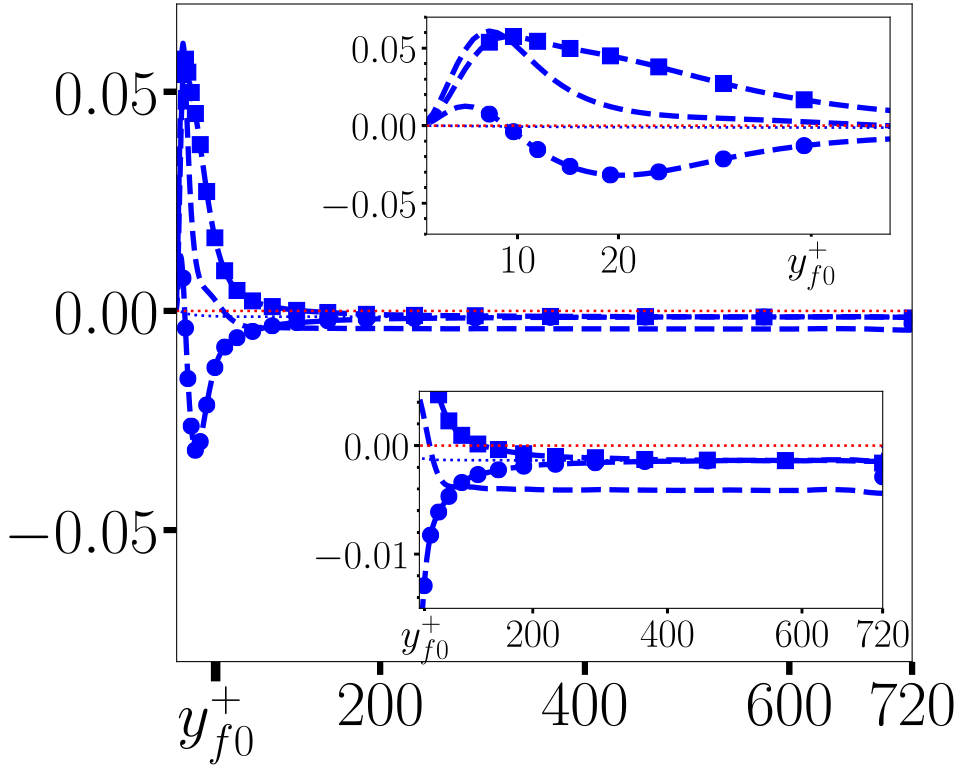}
        \caption{$\ret=720$, WWO}
        \label{fig:budgetForce720osc}
    \end{subfigure}

    \begin{subfigure}{0.33\linewidth} \includegraphics[width=0.99\linewidth]{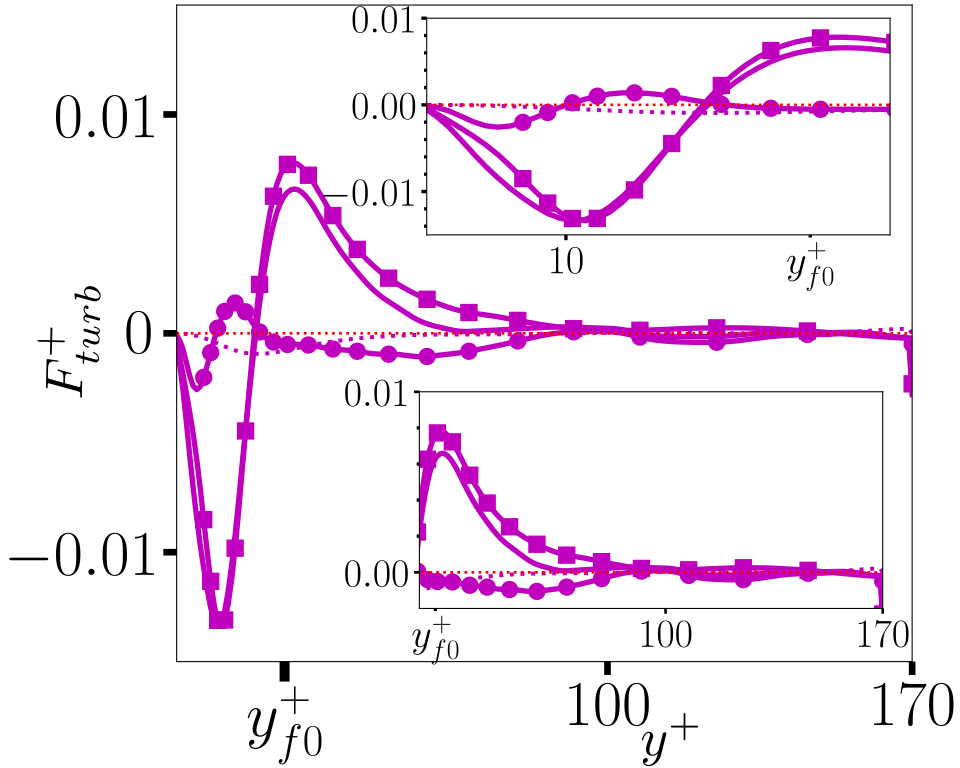}
        \caption{$\ret=170$, WWO-NWO}
        \label{fig:budgetForce170delta}
    \end{subfigure}
    \begin{subfigure}{0.33\linewidth}
        \includegraphics[width=0.99\linewidth]{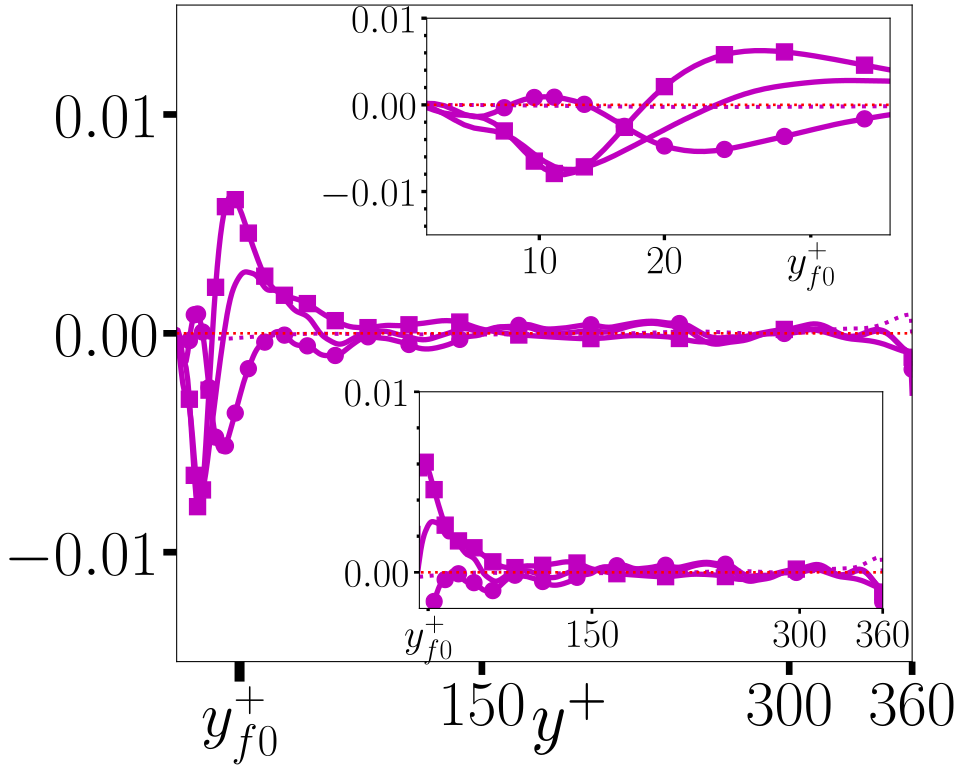}
        \caption{$\ret=360$, WWO-NWO}
        \label{fig:budgetForce360delta}
    \end{subfigure}%
    \begin{subfigure}{0.33\linewidth}
        \includegraphics[width=0.99\linewidth]{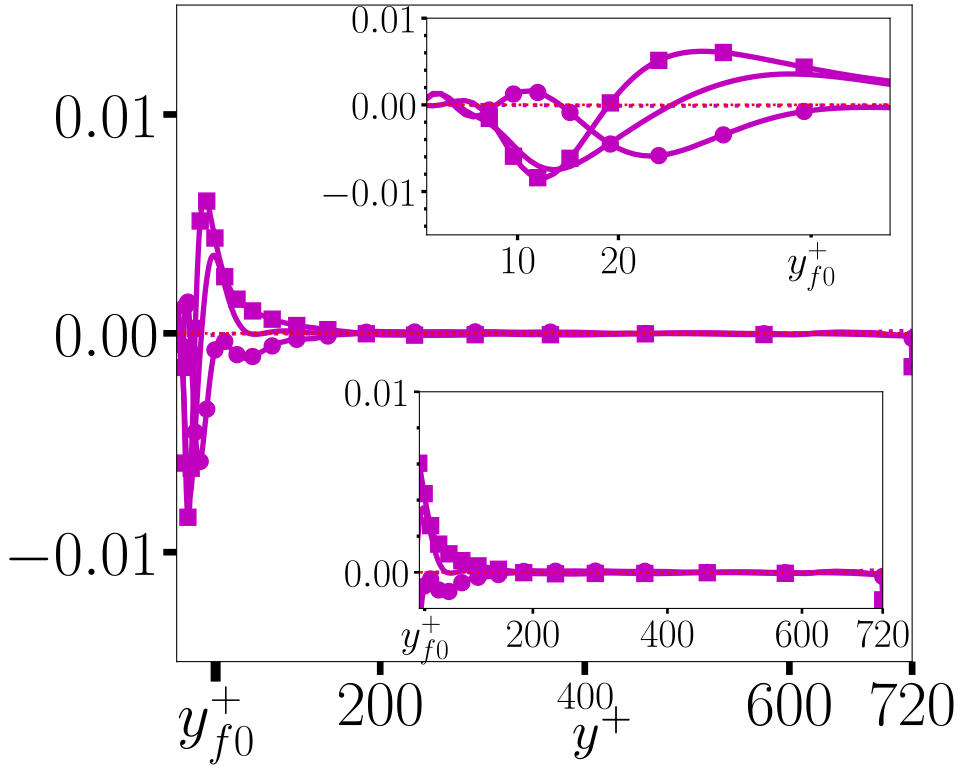}
        \caption{$\ret=720$, WWO-NWO}
        \label{fig:budgetForce720delta}
    \end{subfigure}
    \caption{Budget of the net turbulent force through the velocity-vorticity correlations.  From left to right, $Re_\tau = 170, 360,\text{ and }720$,  with the top row corresponding to $NWO$, middle row to $WWO$, and bottom row to their difference (WWO-NWO).  The lines with no marks is the total net turbulent force; lines with squares indicate $-\av{ \wpp^+ {\omega_r^\dpr}^+ }$ term; lines with circles  $\av{ \vpp^+ {\omega_\theta^\dpr}^+}$ term; and the thin dotted lines indicate  $\av{\upp^+ \vpp^+}/r^+$ term. The top inset in each subfigure plots a zoom-in of the region of acceleration, $y^+ < y_{f0}^+$, while the bottom inset shows a zoom-in of the region of deceleration, $y^+>y_{f0}^+$. The bottom axis in each subfigure also marks a location of $y_{f0}^+$ for each Reynolds number.  $y_{f0}^+$ is taken from the NWO case. Pink dotted horizontal line indicates the zero value. }
    \label{fig:budgetForce}
\end{figure}

Figure \ref{fig:budgetForce} shows the decomposition of the net turbulent force into its corresponding velocity-vorticity components.  
It can be seen that the the vortex stretching term, $-\av{\wpp^+ {\omega_r^\dpr}^+}$~\citep{chen2014experimental} is the primary contributor to the flow acceleration (positive $F_{turb}^+$) in both NWO and WWO flows. This term however is markedly reduced in the WWO as compared to the NWO flows in the region of $y^+<20$ promoting  reduction of the near-wall velocity gradient, while it is increased in the buffer, logarithmic and wake region with wall oscillation. The amount of reduction in the vortex stretching between WWO and NWO cases diminishes with the Reynolds number. The location at which the change switches from negative to positive (thus accelerating the flow) appears to saturate at $y^+ \approx 20$ with $Re$, which is driven closer and closer to the wall as a fraction of $y_p^+$.  
The second budget term, $\av{\vpp^+ {\omega_\theta^\dpr}^+}$, the advective vorticity transport, has a  smaller effect on the acceleration of the near-wall streamwise velocity, and in fact mostly acts to decelerate the flow, as compared to the vortex stretching term. Its contribution to a difference between the NWO and WWO cases is also less significant as compared to the vortex stretching term.
It is interesting to note the overall effect of the pipe flow geometry (the third budget term, $-\av{\upp^+ \vpp^+}/r^+$) to diminish $F_{turb}^+$ and thus to decelerate the mean flow at all wall-normal locations. Its value is however significantly smaller than that of the other two terms, and, conceivably, its contribution with the Reynolds number also reduces.
\subsection{Effect of Reynolds number on turbulent contribution to bulk mean velocity}
The Fukagata-Iwamoto-Kasagi (FIK) identity relates the wall shear stress to a componentwise contribution of different dynamical effects in a turbulent flow~\citep{fukagata2002contribution}. 
To assess the effect on drag reduction in the current setup, where the wall shear stress is fixed but the volumetric flow rate is allowed to vary, it is more appropriate to express the FIK identity in terms of the bulk mean velocity~\citep{marusic2007laminar, yakeno2014modification}. Such an expression for the bulk mean velocity  in a pipe flow (expressed in wall units) can be derived as 
\begin{equation}
    \ubulk^+ = \frac{\ret}{4} -\ret \int_0^1 \av{\upr^+ \vpr^+} \rstar^2 d\rstar,
    \label{eqn:fik0}
\end{equation}
where, recall, $r^{\star}=r/R$ is a radial coordinate scaled with the outer units. For the wall-oscillated pipe, the expression~(\ref{eqn:fik0}) is largely unchanged, albeit it is shown in the Appendix~\ref{sec:derivation_fik} that the fluctuating component of a triply-decomposed Reynolds stress, $\av{\upp^+ \vpp^+}$, can be used instead of $\av{\upr^+ \vpr^+}$, yielding
\begin{equation}
    \ubulk^+ = \frac{\ret}{4} -\ret \int_0^1 \av{\upp^+ \vpp^+} \rstar^2 d\rstar. 
    \label{eqn:fik}
\end{equation}
Appendix~\ref{sec:derivation_fik} additionally shows that the presence of non-zero azimuthal wall velocity boundary conditions does not change the derivation. The first term in equation (\ref{eqn:fik}) corresponds to the laminar contribution to the bulk mean velocity (i.e. a contribution from a corresponding parabolic flow profile which were to develop under the same mean pressure gradient in a laminar flow), and the second term corresponds to the turbulent contribution. Since laminar contribution scaled with $Re_{\tau}$ is identical between the NWO and WWO cases (the spanwise Stokes' layer due to the transverse wall motion is decoupled from the streamwise boundary layer in a laminar solution~\citep{panton1984incompressible,coxe2022stokes}), we turn our attention to the turbulent contribution, which is the only term responsible for the difference in the bulk mean velocity between the two flows. We can represent a turbulent contribution as a limiting value of the cumulative distribution function evaluated at the pipe centerline $r^\star=0$ as:
\begin{equation}\label{eqn:UbulkT}
    \ubulk^{t\,+} = {\ubulk^{t,cum}}^+ (r^{\star}=0)= -\ret \int_0^1 \av{\upp^+ \vpp^+} \rstar^2 d\rstar,
\end{equation}
with the cumulative  distribution function (termed as a ``cumulative turbulent contribution'') defined as shown in Appendix \ref{sec:derivation_fik}:
\begin{equation}\label{eq:cumulative}
    {\ubulk^{t,cum}}^+ (r^{\star}) = -\ret \int_{r^{\star}}^1 \av{\upp^+ \vpp^+} \rstar^2 d\rstar. 
\end{equation}
With this definition, the value of the cumulative turbulent contribution at the pipe wall is zero, ${\ubulk^{t,cum}}^+ (r^{\star}=1)=0$, consistent with the physical meaning of this term. This integral can be written equivalently as a function of the normalized wall normal coordinate $y^{\star} = 1-\rstar$~\citep{fukagata2002contribution} as:
\begin{equation}\label{eqn:UbulkTCum}
    {\ubulk^{t,cum}}^+ (y^{\star}) = -\ret \int_{1-y^{\star}}^1 \av{\upp^+ \vpp^+} {\rstar}^2 d\rstar.
\end{equation}

Applying Parseval's theorem (equation \eqref{eqn:parseval}) to the equations (\ref{eqn:UbulkT}) and (\ref{eqn:UbulkTCum}), we can express the turbulent total and cumulative contributions to the bulk mean velocity through the sums of their corresponding spectral contributions:
\begin{equation}\label{eqn:UbulkTSpectra}
    \ubulk^{t\,+} = -\ret \int_{0}^1 \sum_{k_x} \sum_{k_\theta}  \Phi_{\upp^+ \vpp^+}(k_x,\rstar,k_\theta)  {\rstar}^2 d\rstar,
\end{equation}
\begin{equation}
    {\ubulk^{t,cum}}^+ (y^{\star}) = -\ret \int_{1-y^{\star}}^1 \sum_{k_x} \sum_{k_\theta}  \Phi_{\upp^+ \vpp^+}(k_x,\rstar,k_\theta)  {\rstar}^2 d\rstar.
\end{equation}

Figure \ref{fig:FIKcontribution} shows the cumulative turbulent contribution to the bulk mean velocity, as well as the spectra of the total turbulent contribution for the NWO and WWO cases, together with their change. 
From Figure \ref{fig:FIKcumReynolds}, one can observe that the major increase in the bulk mean velocity in a controlled flow as compared to the uncontrolled flow comes from the buffer and the log layer of the flow, with the peak around the top of the log layer, and the cumulative contribution decreasing in the outer layer. Figure \ref{fig:FIKstwtotal} shows an increased contribution of large streamwise scales to the mean flow retardation by turbulence, with the effect of wall oscillation to suppress this retardation in the intermediate scales and increase it in larger scales. Interestingly, an azimuthal spectra presented in Figure \ref{fig:FIKazmtotal} shows a clear peak in $-\ubulk^{t\,+}$ at azimuthal scales around $\lambda_s^+\approx 1000$ which increase with $\ret$.  These scales, representative of the hairpin packet organization~\citep{adrian2000vortex,adrian2007hairpin}, contribute the most to the turbulent drag. Figure \ref{fig:FIKstwReynolds} shows a remarkable collapse of the $\Delta \,\ubulk^{t\,+}$ streamwise spectra across all three Reynolds numbers, indicating that these are the same scales of motion (in wall units) that are responsible for drag reduction, irrespective of the Reynolds number. Specifically, length scales in the range of $500\leq \lambda_x^+ \leq 5000$ reduce drag with wall oscillation. An important result is that larger scales of motion in WWO flows act to increase drag. This explains the decreased effectiveness of the wall oscillation mechanism (at least with the chosen oscillation parameters) at higher Reynolds numbers: there are more large-scale motions that develop at higher $Re$, and they are the ones which negatively effect drag reduction. The azimuthal spectra of $\Delta \,\ubulk^{t\,+}$ shown in Figure  \ref{fig:FIKazmReynolds} shows a reasonable amount of collapse but not to the same extent as found in the streamwise spectra. The lowest Reynolds number, $Re_{\tau}=170$, seems to be very effective at reducing drag in azimuthal scales corresponding to the individual hairpins ($\lambda_s^+\approx100-200$), perhaps because the Reynolds number is too low to effectively form larger structures composed from the agglomeration of hairpins. The two higher Reynolds numbers, $Re_{\tau}=360$ and  $Re_{\tau}=720$, affect a larger range of azimuthal wavenumbers ($100\le\lambda_s^+\le 700$).
Larger azimuthal scales, $\lambda_s^+>1000$, negatively contribute to drag reduction. 

To characterize both the length scales and the wall normal location of turbulent motions contributing to drag reduction, Figure \ref{fig:DeltaFikSpec} plots the streamwise and azimuthal spectra of the cumulative turbulent contribution as a function of wall normal coordinate. It can be clearly seen that the streamwise scales of motion with the wavelengths between $500\leq \lambda_x^+ \leq 5000$ contribute to drag reduction (positive {$\Delta\,{\ubulk^{t,cum}}^+ (\lambda_x^+,y^+)$) all throughout the vertical extent of the pipe for all three Reynolds numbers. It can also be noted that larger wavelengths, while still acting to cumulatively reduce drag in the log layer, overtake and lead to a drag increase in the outer layer. This effect is absent at the lowest Reynolds number, $Re_{\tau}=170$, and is the strongest at the highest Reynolds number, $Re_{\tau}=720$.

\begin{figure}
    \centering
    \begin{subfigure}{0.32\linewidth}
        \includegraphics[width=0.98\linewidth]{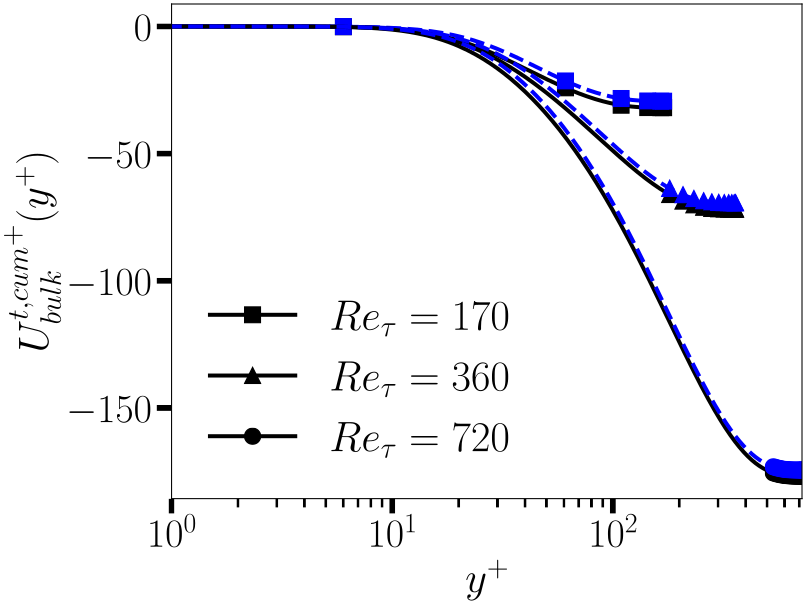}
        \caption{\centering\footnotesize{Cumulative turbulent contribution}}
        \label{fig:FIKcumtotal}
    \end{subfigure}
    \begin{subfigure}{0.32\linewidth}
        \includegraphics[width=0.98\linewidth]{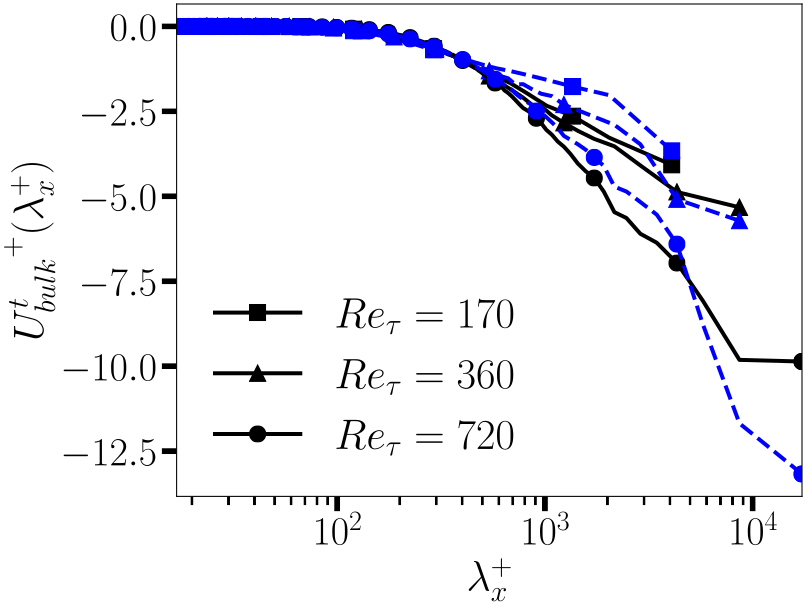}
        \caption{\centering\footnotesize{Streamwise spectra of total turbulent contribution}}
        \label{fig:FIKstwtotal}
    \end{subfigure}
    \begin{subfigure}{0.32\linewidth}
        \includegraphics[width=0.98\linewidth]{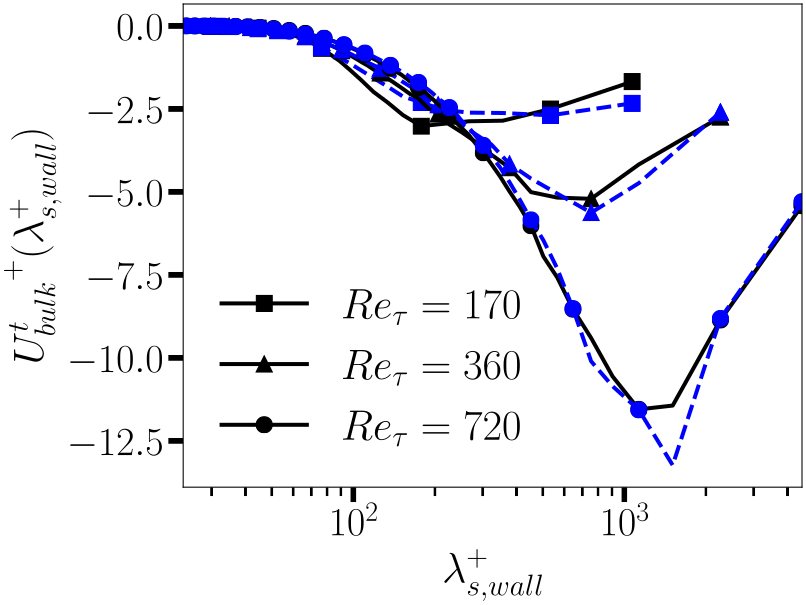}
        \caption{\centering\footnotesize{Azimuthal spectra of total turbulent contribution}}
        \label{fig:FIKazmtotal}
    \end{subfigure}
    \begin{subfigure}{0.32\linewidth}
        \includegraphics[width=0.98\linewidth]{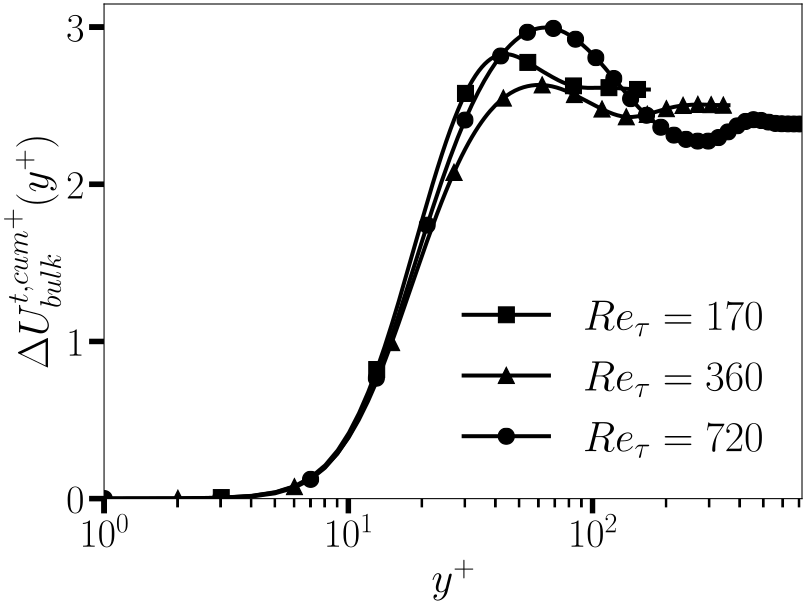}
        \caption{\centering\footnotesize{Change in cumulative turbulent contribution}}
        \label{fig:FIKcumReynolds}
    \end{subfigure}
    \begin{subfigure}{0.32\linewidth}
        \includegraphics[width=0.98\linewidth]{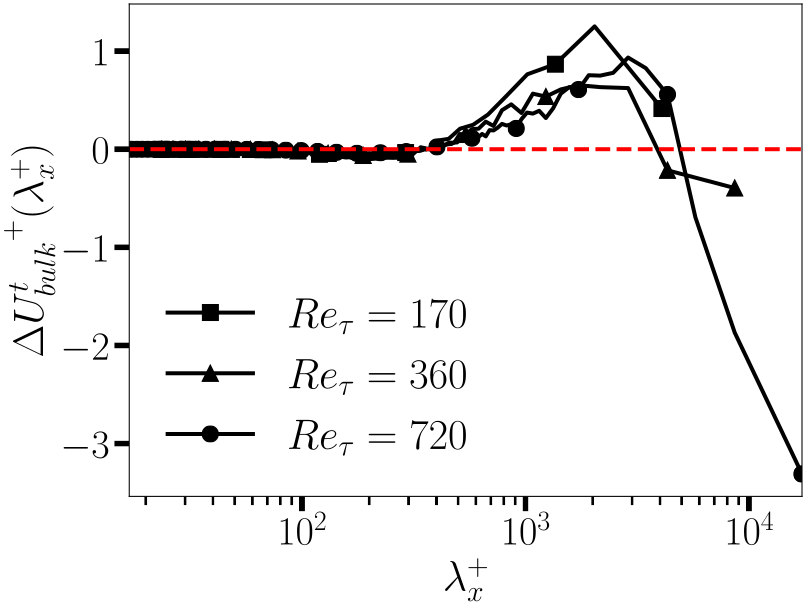}
        \caption{\centering\footnotesize{Change in streamwise spectra }}
        \label{fig:FIKstwReynolds}
    \end{subfigure}
    \begin{subfigure}{0.32\linewidth}
        \includegraphics[width=0.98\linewidth]{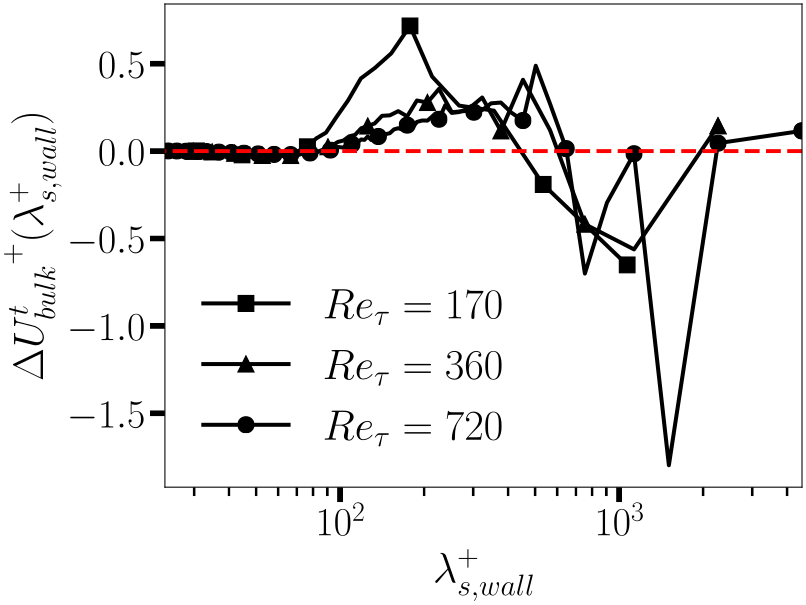}
        \caption{\centering\footnotesize{Change in azimuthal spectra}}
        \label{fig:FIKazmReynolds}
    \end{subfigure}
    \caption{Turbulent contribution to the bulk mean velocity. (a,b,c) Contributions for NWO (black solid lines) and WWO (blue dashed lines) cases; (d,e,f) change between NWO and WWO cases. (a,d) Cumulative turbulent contribution as a function of wall normal coordinate; (b,e)  streamwise spectra of the total turbulent contribution; (c,f) azimuthal spectra of the total turbulent contribution. The notation  $\lambda_{s,wall}^+$ refers to $\lambda_{s}^+$ evaluated at the wall. The red dashed line in (e,f) denotes the zero level.}
    \label{fig:FIKcontribution}
\end{figure}

\begin{figure}
    \centering
    \begin{subfigure}{0.33\linewidth}
        \includegraphics[width=0.99\linewidth]{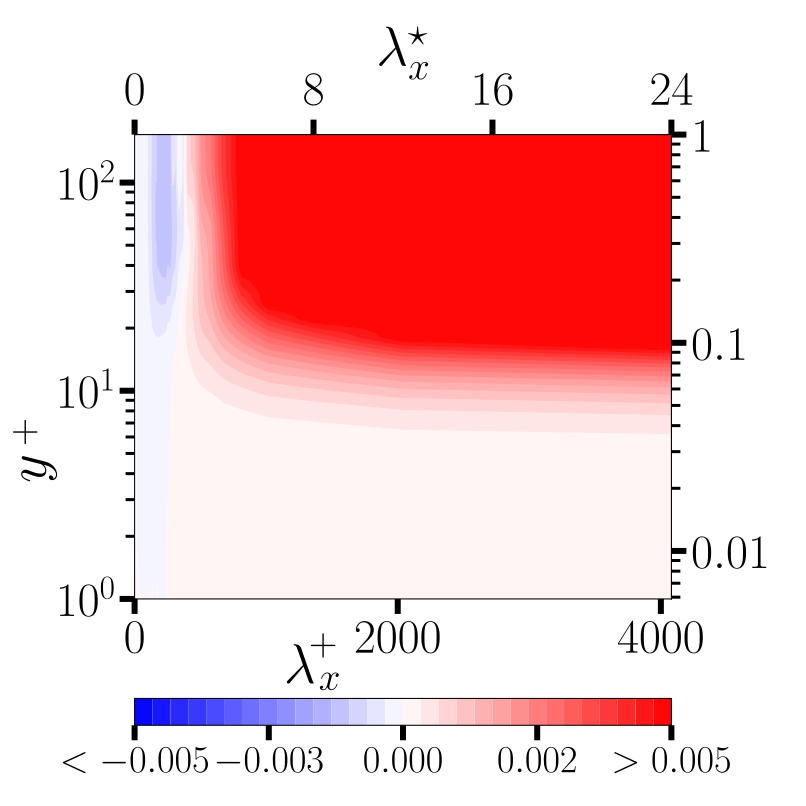}
        \caption{$\ret=170$, Streamwise}
        \label{fig:FikStwRe170cum}
    \end{subfigure}%
    \begin{subfigure}{0.33\linewidth}
        \includegraphics[width=0.99\linewidth]{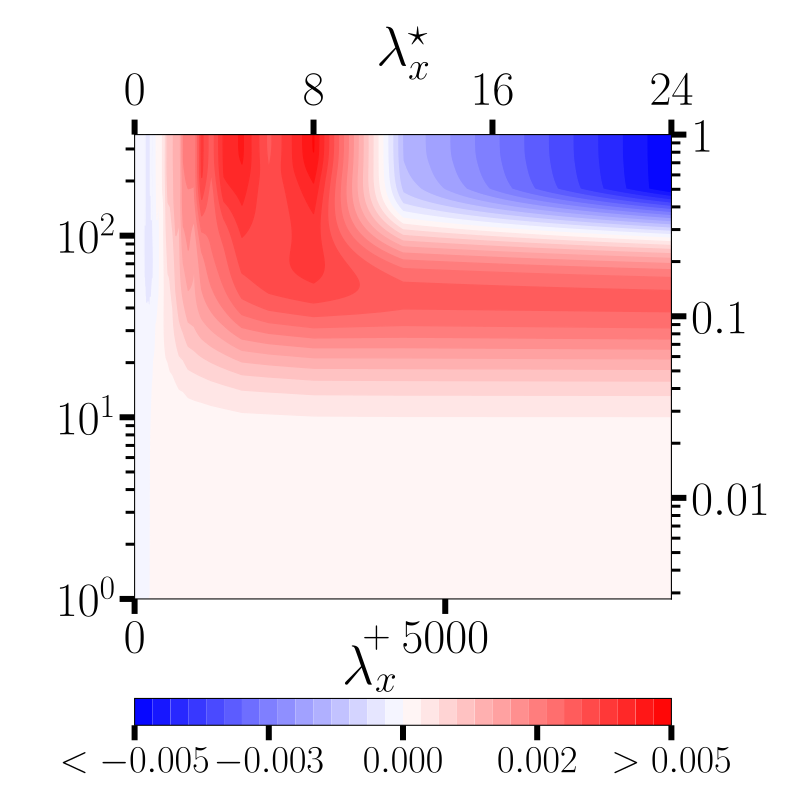}
        \caption{$\ret=360$, Streamwise}
        \label{fig:FikStwRe360cum}
    \end{subfigure}
    \begin{subfigure}{0.33\linewidth}
        \includegraphics[width=0.99\linewidth]{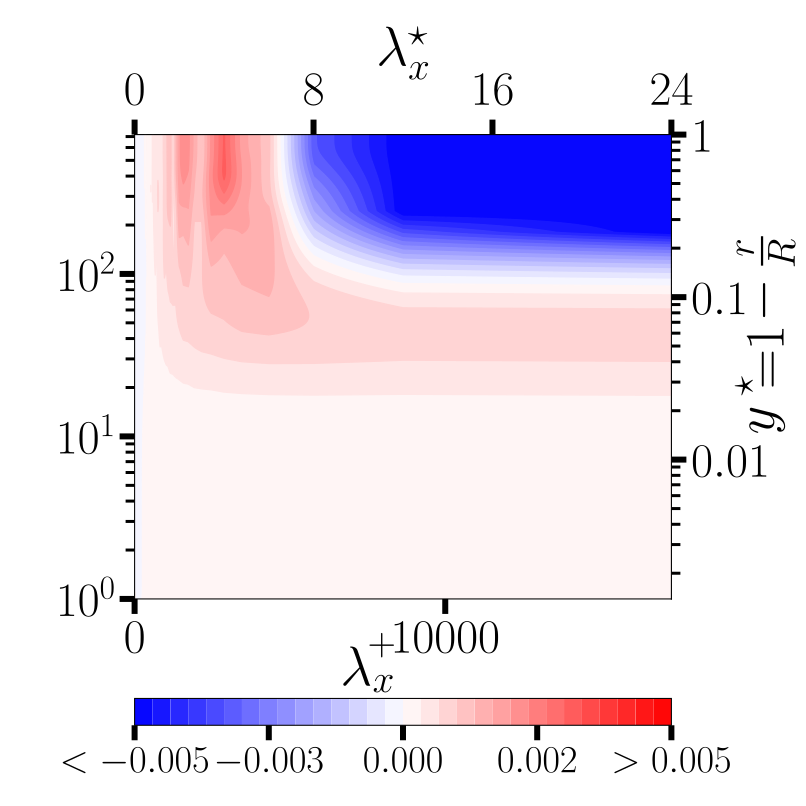}
        \caption{$\ret=720$, Streamwise}
        \label{fig:FikStwRe720cum}
    \end{subfigure}%
    
    \begin{subfigure}{0.33\linewidth}
        \includegraphics[width=0.99\linewidth]{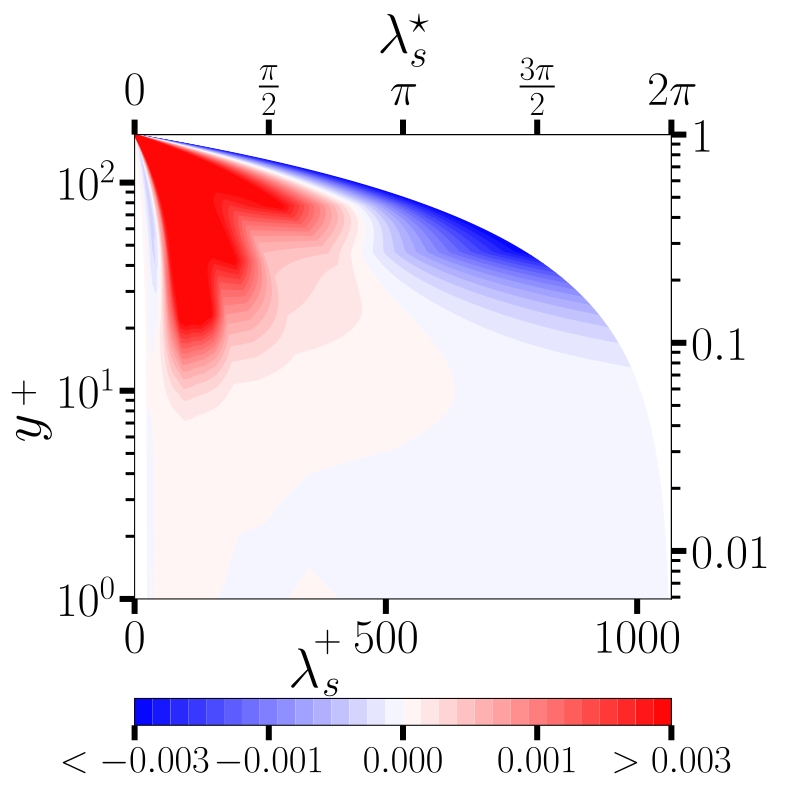}
        \caption{$\ret=170$, Azimuthal}
        \label{fig:FikAzmRe170cum}
    \end{subfigure}
    \begin{subfigure}{0.33\linewidth}
        \includegraphics[width=0.99\linewidth]{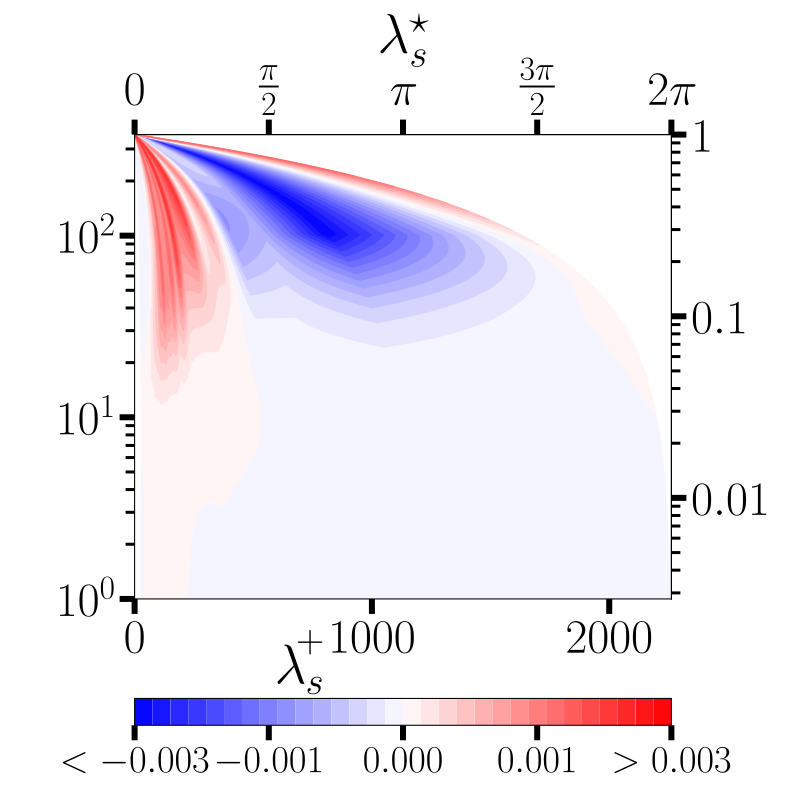}
        \caption{$\ret=360$, Azimuthal}
        \label{fig:FikAzmRe360cum}
    \end{subfigure}%
    \begin{subfigure}{0.33\linewidth}
        \includegraphics[width=0.99\linewidth]{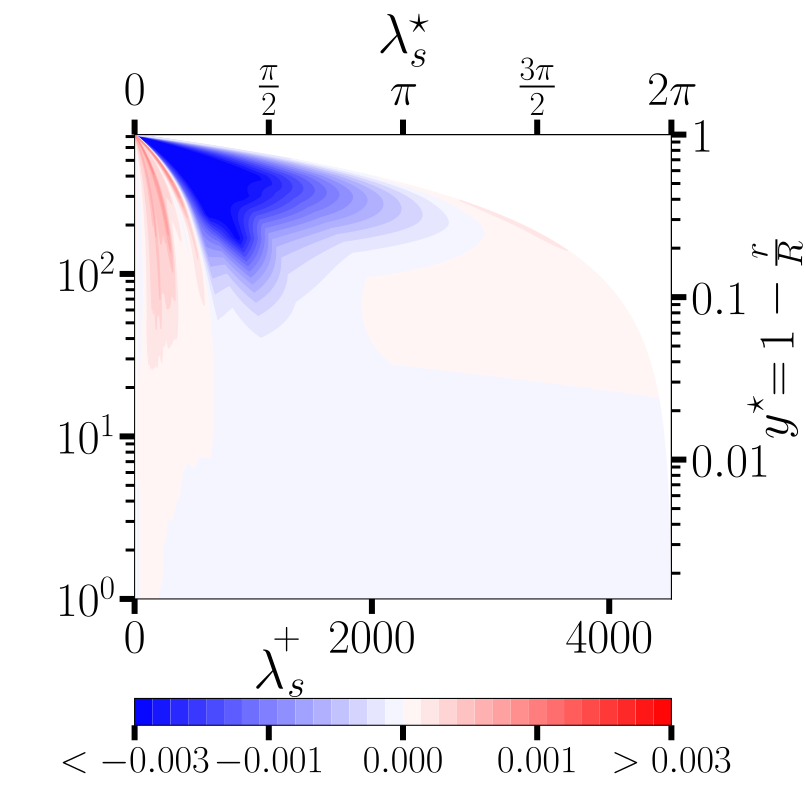}
        \caption{$\ret=720$, Azimuthal}
        \label{fig:FikAzmRe720cum}
    \end{subfigure}
    \caption{Change in the cumulative turbulent contribution spectra (normalized by the laminar contribution) as a function of wall normal location and the wavelength: (a,b,c) streamwise spectra, $\Delta\,4 {\ubulk^{t,cum}}^+ (\lambda_x^+,y^+)/\ret$; (d,e,f) azimuthal spectra, $\Delta\, 4 {\ubulk^{t,cum}}^+ (\lambda_s^+,y^+)/\ret$.  From top to bottom: (a,d) $\ret = 170$, (b,e) $\ret = 360$, and (c,f) $\ret = 720$.
    }
    \label{fig:DeltaFikSpec}
\end{figure}
For the azimuthal spectra, it is observed  
 that the azimuthal length scales between $50 \leq \lambda_s^+ \leq500$ act to increase the flow rate (reduce drag) while larger structures reduce the flow rate (increase drag), essentially independent of a wall normal location. Interestingly, drag-reducing azimuthal scales organize themselves into a fractal-like pattern (red ``fingers'') visible in Figures~\ref{fig:FikAzmRe170cum}, \ref{fig:FikAzmRe360cum}, \ref{fig:FikAzmRe720cum}. A fractal-like pattern is consistent with the attached eddy hypothesis of the near-wall turbulence~\citep{townsend1951structure,hwang2015statistical} and the trends in hairpin packet organization~\citep{adrian2007hairpin}; the fact that the drag-reducing motions adhere to this pattern suggests a link between the drag reduction mechanisms and a weakening of the hairpin packets.  Similarly to the streamwise spectra, 
 we observe that
 increasing Reynolds number introduces larger azimuthal scales of motion (relative to the viscous scale) which hinder the effectiveness of the selected oscillation parameters to reduce turbulent drag. 

 To compare with the analysis performed by~\cite{hurst2014effect,yao2019reynolds} in a turbulent channel flow with wall oscillation, we decompose the turbulent contribution to the bulk mean velocity into the corresponding ``inner'' (accelerating layer) and ``outer'' (decelerating layer) components, such that  
 \begin{eqnarray}
    U_{bulk, "inner"}^{t\,+} = -\ret \int_{1-y_{f0}^{\star}}^{1} \av{\upp^+ \vpp^+}  {\rstar}^2 d\rstar \label{eqn:UbulkTSpectra1}\\
    U_{bulk, "outer"}^{t\,+} = -\ret \int_{0}^{1-y_{f0}^{\star}} \av{\upp^+ \vpp^+}{\rstar}^2 d\rstar \label{eqn:UbulkTSpectra2},
\end{eqnarray}
where $y_{f0}^{\star}=y_{f0}/R$ is the location of the zero net turbulent force. We remark that the zero net turbulent force location coincides with the peak Reynolds shear stress location in a channel flow~\citep{chen2018quantifying,yao2019reynolds}, however it is slightly different in a pipe flow due to a curvature effect~\citep{wu2012direct}. In a pipe flow, the net turbulent force attains zero when $r \av{\upp \vpp}$ reaches its maximum, and not $\av{\upp \vpp}$, see Eq.~(\ref{eqn:netturb}).
The classification of the ``inner'' and ``outer'' layers based on the location of the zero net turbulent force is different from the classical demarcation of the inner and outer layers as being directly affected and unaffected by viscosity, respectively~\citep{sreenivasan1997persistence,adrian2000vortex,jimenez2018coherent} (see also Table~\ref{tab:locations}). Figure~\ref{fig:FIKlayers} presents the corresponding component contributions to the bulk mean velocity (normalized by the laminar component) for the NWO and WWO pipe flows. The ``inner'' and ``outer'' turbulent contributions are computed from Eqs.~(\ref{eqn:UbulkTSpectra1})--(\ref{eqn:UbulkTSpectra2}), taking the NWO value for $y_{f0}^{\star}$ as a reference for both the NWO and WWO cases at each Reynolds number. Consistent with the previous observations, we see that the ``outer'' layer contributes more significantly to the mean flow retardation from the laminar flow, and this contribution increases with the Reynolds number. The effect of wall oscillations is to reduce the contributions to $U_{bulk}$ from both the ``inner'' and the ``outer'' layers. Figure~\ref{fig:FIKdeltaubulk} presents the difference of the component contributions between the NWO and WWO cases normalized by $U_{bulk, NWO}$. We observe that the reduction of effectiveness of wall oscillations in increasing the bulk mean velocity of controlled flow as compared to the uncontrolled flow with $\ret$ mostly comes from the ``outer'' (decelerating) layer, consistent with Figure~\ref{fig:FIKcumReynolds}. The analysis presented here is different from~\cite{hurst2014effect,yao2019reynolds} in that we keep the wall shear stress (or pressure gradient) constant, while they keep the bulk mean velocity (or bulk flow rate) constant. To adhere to their analysis, Figures~\ref{fig:FIKlayers_cf},~\ref{fig:FIKlayers_dr} present the corresponding component contributions to the skin friction coefficient, $C_f=2\av{\tau_w}/\rho \ubulk^2$ (equation~\ref{eqn:fik3} in Appendix~\ref{sec:derivation_fik}), together with the change between the NWO and WWO cases. Concerning the contribution to skin friction, these results are very close to the data presented in~\cite{hurst2014effect,yao2019reynolds} for channel flows, albeit we also see a change in a laminar contribution to skin friction between the NWO and WWO cases due to a change in the bulk flow rate with and without wall oscillation in our setup. Interestingly, while the loss of effectiveness mostly comes from the ``outer'' (decelerating) layer when the change in bulk mean velocity is concerned, it mostly comes from the ``inner'' (accelerating) layer when the change in skin friction coefficient is considered, the latter conclusion drawn by~\cite{yao2019reynolds} in their channel flow study. In both cases, loss of effectiveness is attributed to the large scales, which are either not effectively suppressed (as found in channel flows) or even energized (as found here in regard to pipe flows), both in the inner and outer layers of the flow.

\begin{figure}
    \centering
    \begin{subfigure}{0.24\linewidth}
        \includegraphics[width=0.98\linewidth]{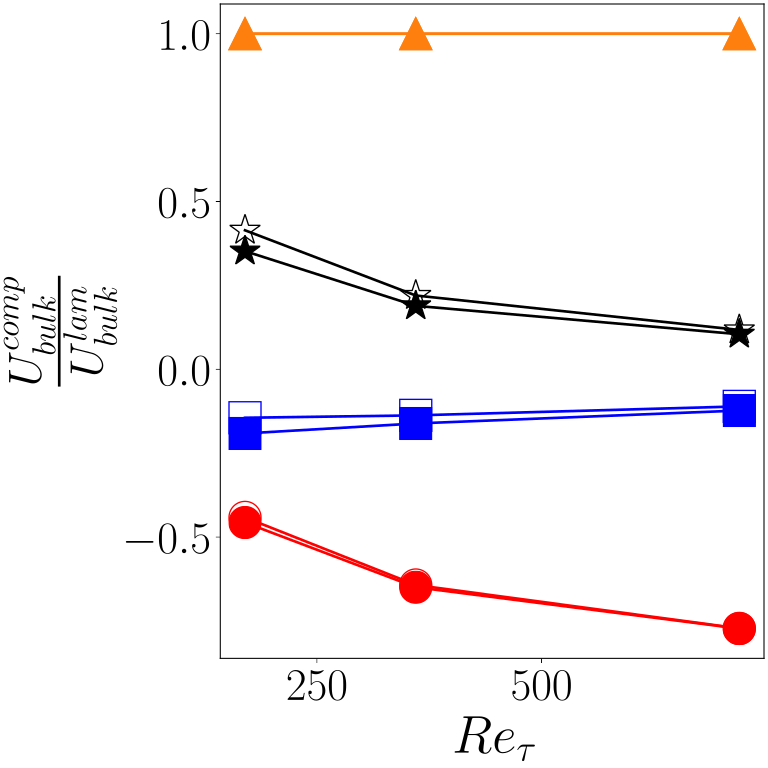}
        \caption{\centering\footnotesize{ $\ubulk^{comp}/\ubulk^{lam}$}}
        \label{fig:FIKlayers}
    \end{subfigure}
    \begin{subfigure}{0.24\linewidth}
        \includegraphics[width=0.98\linewidth]{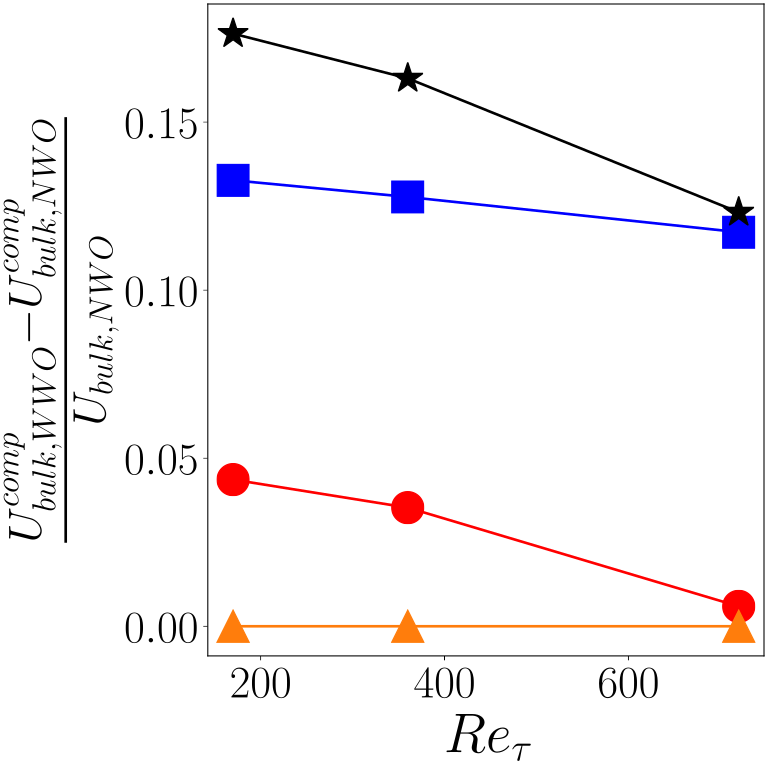}
        \caption{\centering\footnotesize{Change in $\ubulk^{comp}$} }
        \label{fig:FIKdeltaubulk}
    \end{subfigure}
    \begin{subfigure}{0.24\linewidth}
        \includegraphics[width=0.98\linewidth]{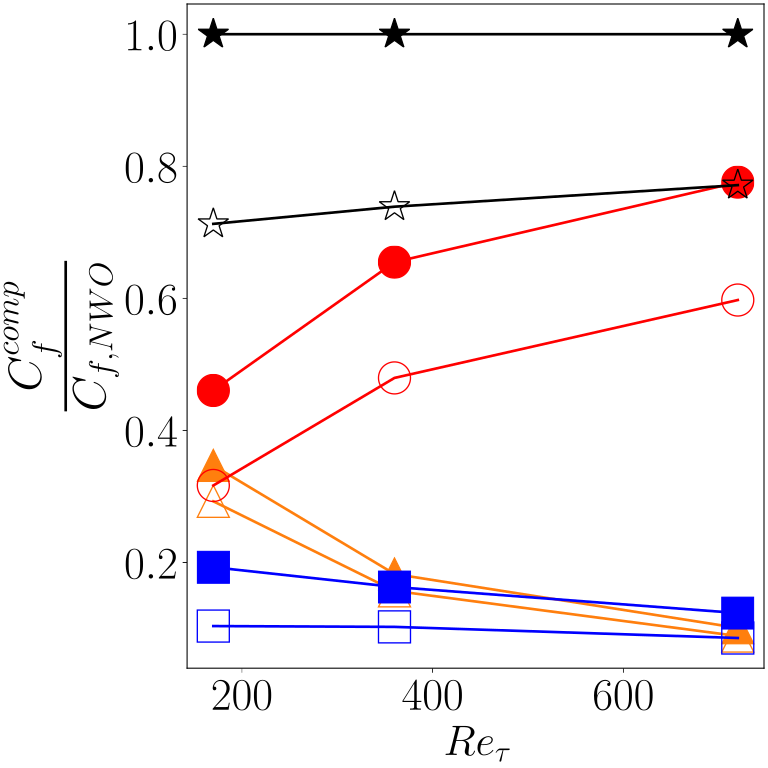}
        \caption{\centering\footnotesize{ $C_{f}^{comp}/C_{f,NWO}$} }
        \label{fig:FIKlayers_cf}
    \end{subfigure}
    \begin{subfigure}{0.24\linewidth}
        \includegraphics[width=0.98\linewidth]{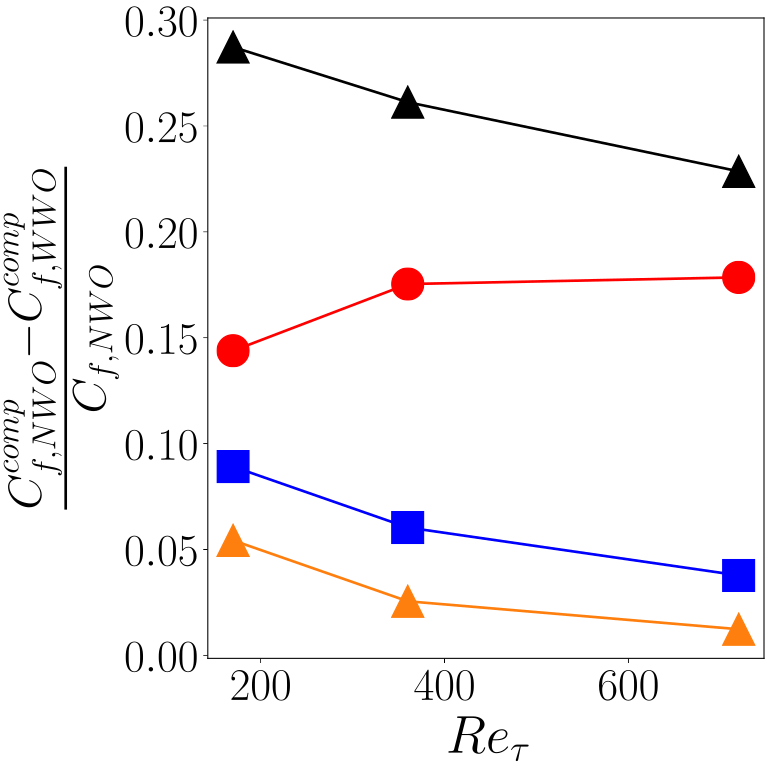}
        \caption{\centering\footnotesize{Change in $C_f^{comp}$}}
        \label{fig:FIKlayers_dr}
    \end{subfigure}
    \caption{Component contributions to (a,b) bulk mean velocity; (c,d) skin friction coefficient. (a,c) The normalized component contributions (filled symbols, NWO; open symbols, WWO); (b,d) change between the NWO and WWO cases.  Line colors indicate: red, ``outer'' layer turbulent contribution; blue, ``inner'' layer turbulent contribution; orange, laminar contribution; black, total contribution.}
    \label{fig:FIKlayer}
\end{figure}


 
\section{Conclusions}\label{sec:conclusions}

The current study documents the results of direct numerical simulation of a turbulent pipe flow with and without transverse wall oscillation for three Reynolds numbers, $Re_{\tau}=170, 360$ and $720$. It is found that wall oscillation results in an increase of a flow rate by almost 20\% and, consequently, achieves a drag reduction of approximately 30\% at the lowest Reynolds number; however, this effect decreases as the Reynolds number increases. One-dimensional and two-dimensional spectra of streamwise kinetic energy, net turbulent force and the turbulent contribution to the bulk mean velocity are analyzed to explain this effect.

It is found that the primary effect of wall oscillation is to reduce the energy and the net turbulent force in the intermediate- to large- streamwise and azimuthal scales of motion in the buffer layer of the flow. To the contrary, energy is increased in the large-scale structures in the log layer and the wake region.  
At the lowest Reynolds number, $Re_\tau = 170$, the inner layer extends through $\approx 65\%$ of the domain while it comprises $\approx 15\%$ of the domain at $\ret = 720$. Since the overall attenuation of energetic structures is limited to the inner layer of the flow, this explains the reduced effectiveness of the wall oscillation as a drag reduction mechanism as the Reynolds number increases. This effect is well illustrated by Figure~\ref{fig:fturbfiltcomp}, where low-pass filtered and high-pass filtered net turbulent force is plotted as a function of wall-normal coordinate. This figure shows that (a) most of the attenuation comes from the effect of the scales of $\lambda_x^+>1000$ (above the filter cut-off length),
and (b) the reduction of the magnitude of the net turbulent force is confined to the buffer layer and the log layer of the flow. The reduction of the magnitude of the net turbulent force by wall oscillation results in a reduced flow acceleration in the near-wall layer and an increased acceleration in the log layer,  making the velocity profile less blunt and more ``laminar-like''. This is offset by an increase of the net turbulent force magnitude (promoting flow deceleration) above the log layer, leading to a lower ratio of the centerline velocity to the bulk velocity in WWO cases, which reduces effectiveness of drag reduction at higher Reynolds number flows.  From the velocity-vorticity budget decomposition of the net turbulent force, it is observed that this effect mostly comes from a suppression of the vortex stretching within the Stokes' layer of the flow and its augmentation above the Stokes' layer and below $y^+\approx 100$. A reduced vortex stretching in the Stokes' layer inhibits a lift-up of the hairpins and formation of their necks, thus suppressing the hairpin auto-generation. 

From superimposing the analysis of the streamwise and azimuthal spectra, together with the wall normal location of the effected length scales, one can deduce the shape of the structures most affected by drag reduction. It can be seen that a significant energy reduction occurs at streamwise scales at and slightly above $\lambda_x^+\approx 1000$  and azimuthal scales of at and slightly above $\lambda_s^+\approx 100$, which corresponds to the scales of motions typically associated with the hairpin packets~\citep{adrian2000vortex,adrian2007hairpin}.  
Since streaks and quasi-streamwise vortices are closely related, wall oscillation presumably weakens the quasi-streamwise near-wall vortices, thereby reducing their transport of streamwise momentum into the streaks. This is consistent with the observations of~\cite{yao2019reynolds} who reported a suppression of Reynolds shear stresses at $\lambda_s^+<400$ in a turbulent channel flow with wall oscillation. This points towards a suppression  of 
hairpin packets by wall oscillation being one of the main mechanisms of drag reduction. It is hypothesized that the auto-generation mechanism  of turbulence~\citep{zhou1999mechanisms, kim2008dynamics, kempaiah20203} is suppressed by the wall oscillation, thus attenuating the formation and growth of the hairpin packets. 
Interestingly, the shorter streamwise scales of motion, $\lambda_x^+<500$, are amplified, which suggests that the wall oscillation mechanism does not suppress the energy in the individual hairpins but rather hinders their regeneration abilities. Large streamwise scales, $\lambda_x^+>5000$, and large azimuthal scales, $\lambda_s^+>1000$, are also   found to be amplified by wall oscillation, both in the buffer layer and above.  Such amplification of the large-scale azimuthal energy in the buffer layer may be associated with the large-scale mode observed in the current study for the wall-oscillated pipe flow cases with $\ret=360$ and 720, potentially created by the sloshing motions spurred by the wall oscillation. In the outer layer, the amplified structures correspond to the very-large-scale motions~\citep{guala2006large, balakumar2007large} of the high Reynolds number flows. Interestingly, while \cite{yao2019reynolds} reported a reduced effectiveness of  WWO control in suppressing large-scale azimuthal motions ($\lambda_s^+> 1000$) in channel flows, they did not observe an \textit{amplification} of such scales, as the current study does, which might point towards a particular influence of the large-scale mode, potentially distinct to pipe flows.

A convincing evidence of the effect of different scales of motion on drag reduction comes from the   spectral analysis of the Fukagata-Iwamoto-Kasagi (FIK) identity~\citep{fukagata2002contribution}; specifically, of the turbulent contribution to the bulk mean velocity.  To this end, Figure~\ref{fig:FIKcontribution} demonstrates a remarkable collapse of the difference in its spectra between the NWO and WWO pipes across all three Reynolds numbers, showing that the drag reduction is limited to the streamwise wavelengths of $\lambda_x^+ < 5000$ independent of the Reynolds number. The wavelengths with $\lambda_x^+> 5000$, exerting increasingly larger dominance  in higher Reynolds number flows, act to increase drag. This brings us to a conclusion that the wall oscillation mechanism with the parameters investigated in the current paper, which are optimized for controlling the \textit{near-wall  turbulent cycle}~\citep{jung1992suppression,choi1998drag,  quadrio2004critical}, may not be effective for drag reduction in high Reynolds number flows. Possibly, new drag reduction mechanisms that specifically target large- and very-large scales of motions need to be developed. This can perhaps be achieved by reducing the frequency of the wall oscillation as suggested by \cite{marusic2021energy}. It is also possible that completely new drag reduction mechanisms need to be devised to target high Reynolds number flows.   

\section*{Acknowledgements} 
This research was supported by the NSF CAREER award \# CBET-1944568 and by the Ira A. Fulton Professorship endowment. The computational time has been provided by NSF ACCESS supercomputing resources.

\section*{Declaration of Interests} The authors report no conflict of interest.

\appendix
\section{Validation}\label{sec:validation}

This section presents a validation of the spectral-element code Nek5000 in application to DNS of turbulent pipe flows within the current computational setup. Additional validation is available in previous studies in~\cite{duggleby2007dynamical, el2013direct, merrill2016spectrally}.

\subsection{Turbulent pipe flow with no wall oscillation (NWO)}\label{sec:validation_nwo}

In this section, validation of the current DNS results for a turbulent pipe flow with no wall oscillation (NWO case) is presented. Figure \ref{fig:stressValid} illustrates a comparison of statistical quantities (mean streamwise velocity and fluctuating Reynolds stresses) with the previously available computational~\citep{el2013direct} and experimental~\citep{eggels1994fully, chin2015turbulent} data. Good agreement of statistics with the previously published data is observed. Figure \ref{fig:azmspec720} compares a calculated pre-multiplied energy spectra of the streamwise, radial, and azimuthal velocity fluctuations for $\ret=720$ case with the DNS data of \cite{wu2012direct}. Again, a favorable agreement is demonstrated. 
\begin{figure}
    \centering
    \begin{subfigure}[t]{.5\linewidth}
    \includegraphics[height=1.75in,center]{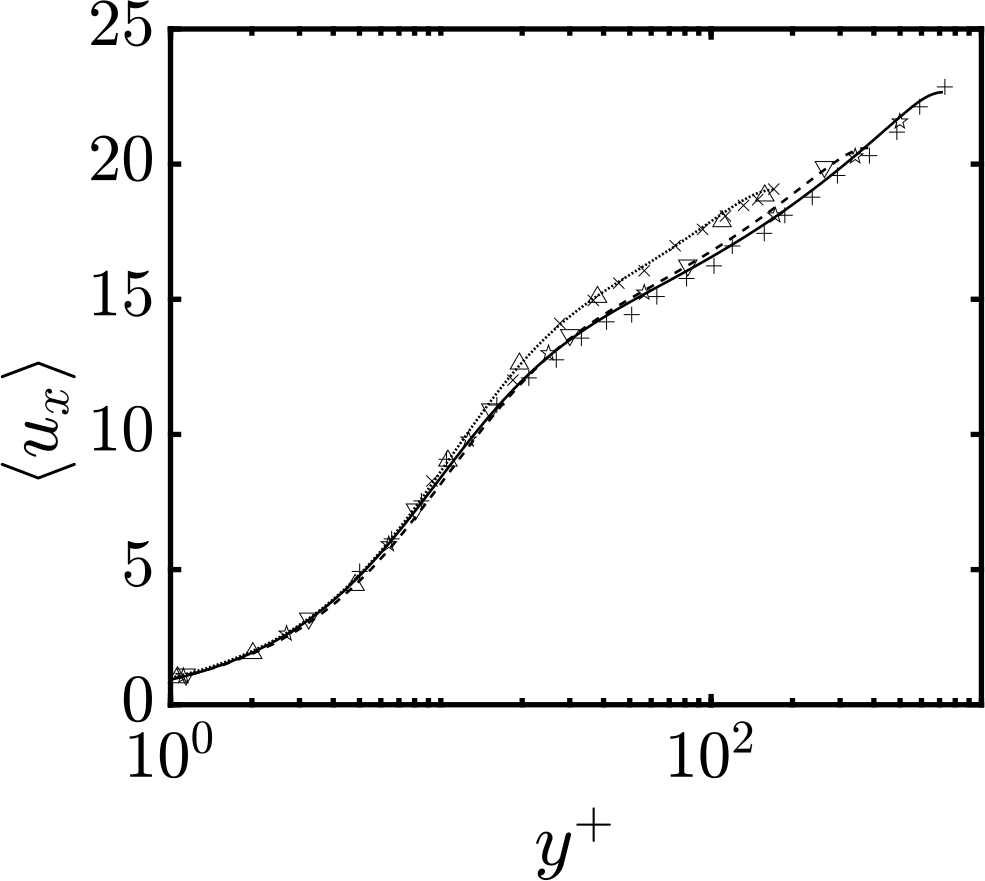}
    \caption{Mean streamwise velocity}
    \label{fig:ubarv}
    \end{subfigure}%
    \begin{subfigure}[t]{.5\linewidth}
    \includegraphics[height=1.75in,center]{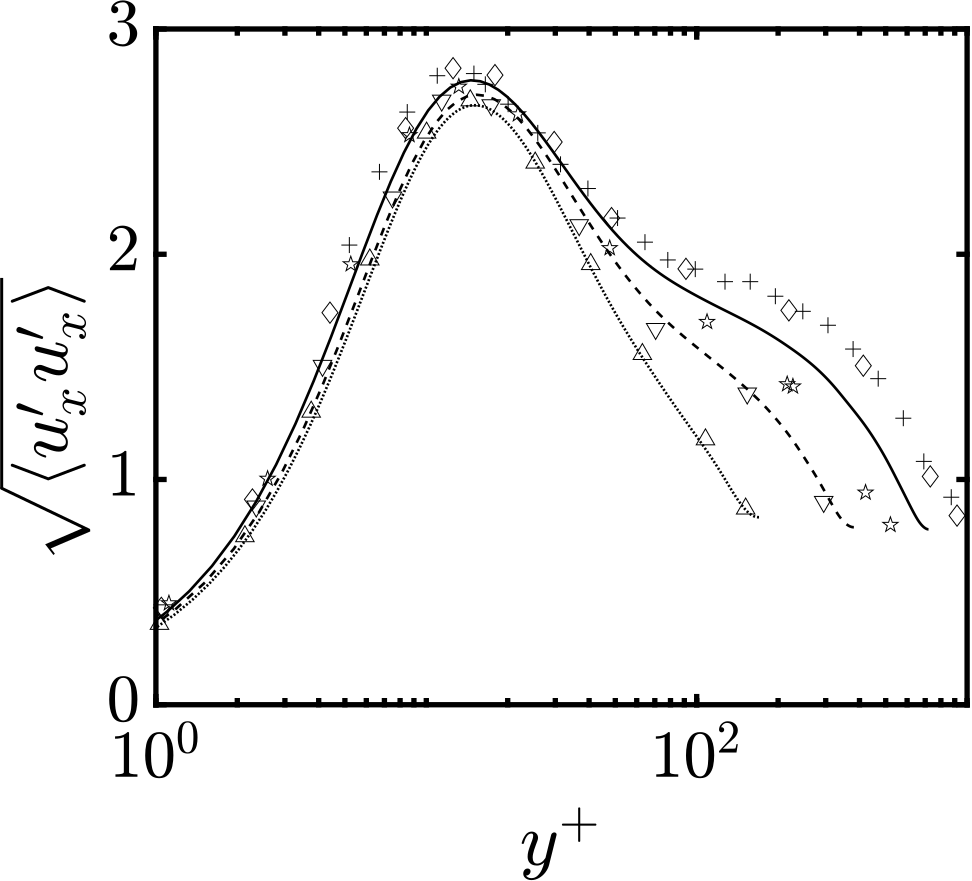}
    \caption{Streamwise velocity fluctuations}
    \label{fig:uumsv}
    \end{subfigure}
    ~
    \begin{subfigure}[t]{.5\linewidth}
    \includegraphics[height=1.75in,center]{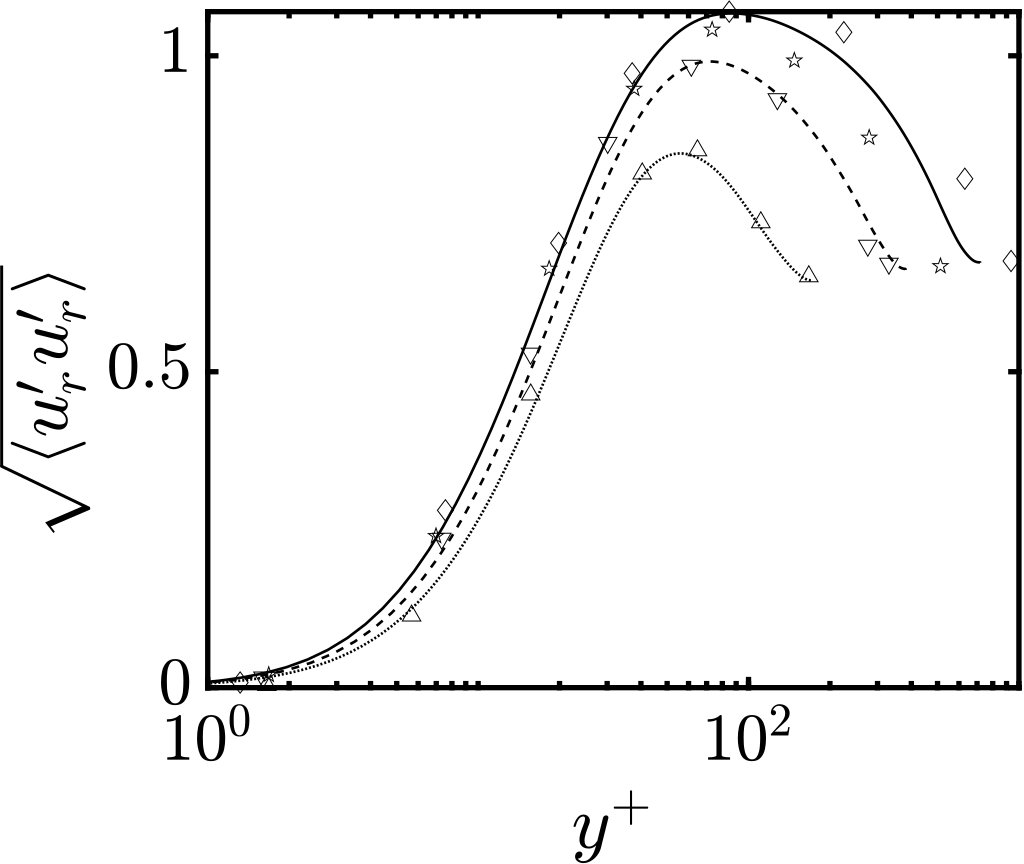}
    \caption{Radial velocity fluctuations}
    \label{fig:vvmsv}
    \end{subfigure}%
    \begin{subfigure}[t]{.5\linewidth}
    \includegraphics[height=1.75in,center]{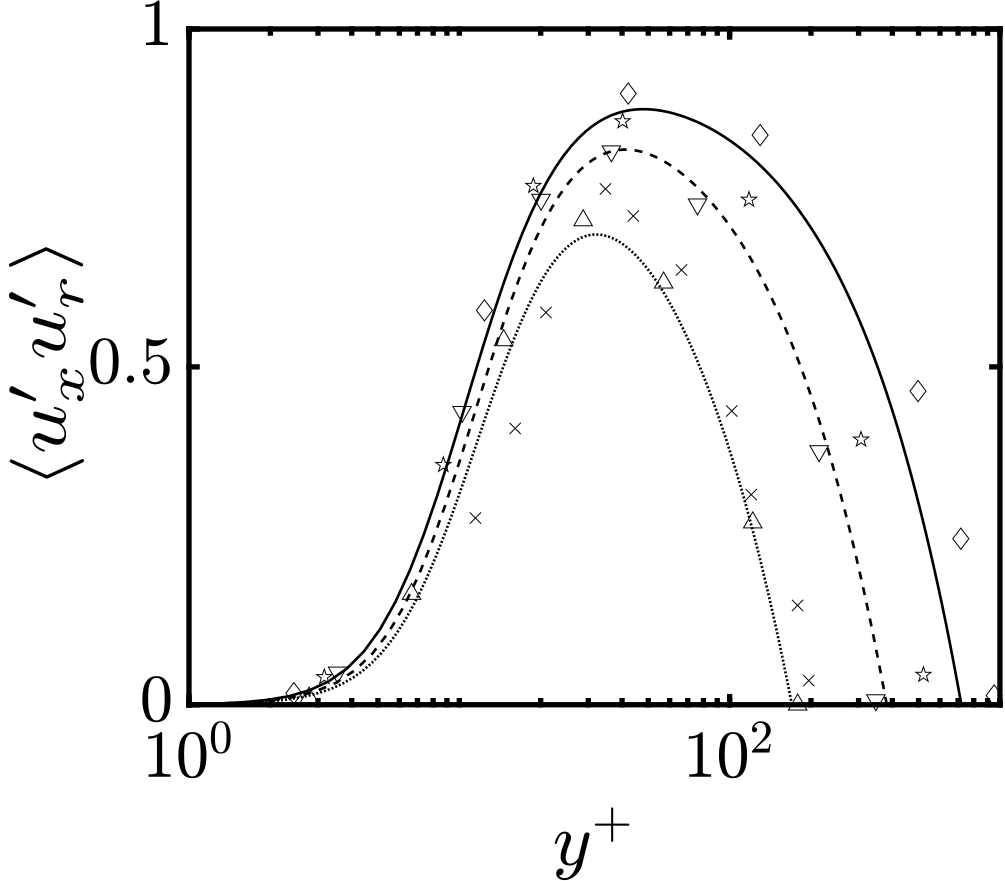}
    \caption{Reynolds shear stress.}
    \label{fig:uvmsv}
    \end{subfigure}
    \caption[]{Validation of statistical quantities for the DNS of turbulent pipe flow with Nek5000 (no wall oscillation): (a) mean streamwise velocity, (b) streamwise velocity fluctuations, (c) radial velocity fluctuations, and (d) Reynolds shear stress. Lines, current DNS: $\ret = 170$,  \tikz[baseline=-0.5ex]\draw[thick, dotted] (0,0) -- (.3,0);; $\ret = 360$, \tikz[baseline=-0.5ex]\draw[thick, dash dot] (0,0) -- (.3,0);; $\ret = 720$,  \tikz[baseline=-0.5ex]\draw[thick] (0,0) -- (.3,0);. Symbols,  \cite{el2013direct} (DNS): $\ret = 180, \vartriangle$; $\ret=360, \triangledown$; $\ret=550, \FiveStarOpen$; $\ret=1000 , \lozenge$ \text{(LES)}; \cite{eggels1994fully} (PIV): $\ret = 200, \times$;  \cite{chin2015turbulent} (hot wire): $\ret = 1000, + $. }
    \label{fig:stressValid}
\end{figure}

\begin{figure}
    \centering
    \begin{subfigure}[t]{.33\linewidth}
        \includegraphics[height=1.4in,center]{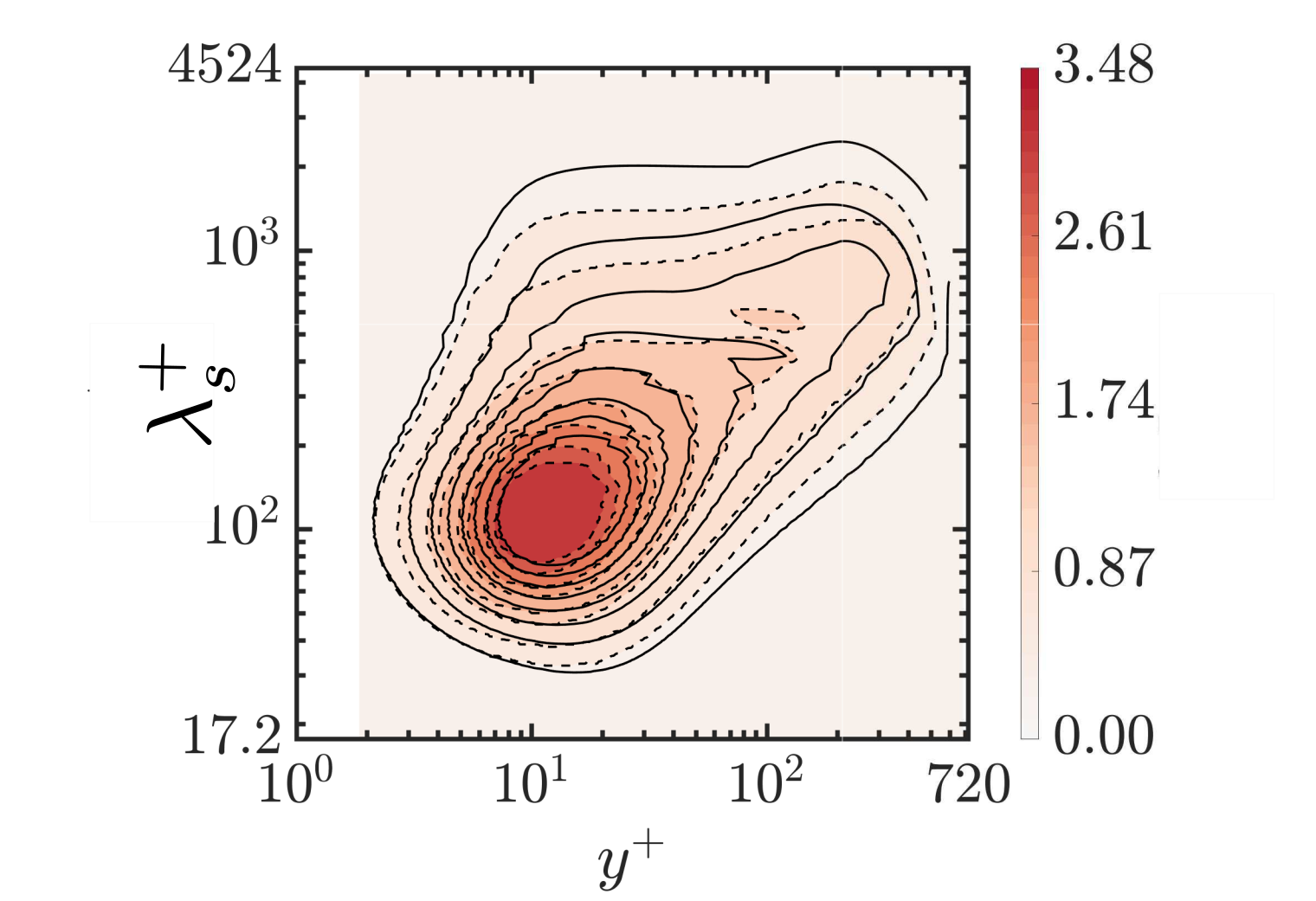}
        \caption{\centering  Streamwise velocity, $k_{\theta}\,\Phi_{u_x u_x}/u_{\tau}^2$ ($j = x$)}
        \label{fig:kSuuBaltz}
    \end{subfigure}%
    \begin{subfigure}[t]{.33\linewidth}
        \includegraphics[height=1.4in,center]{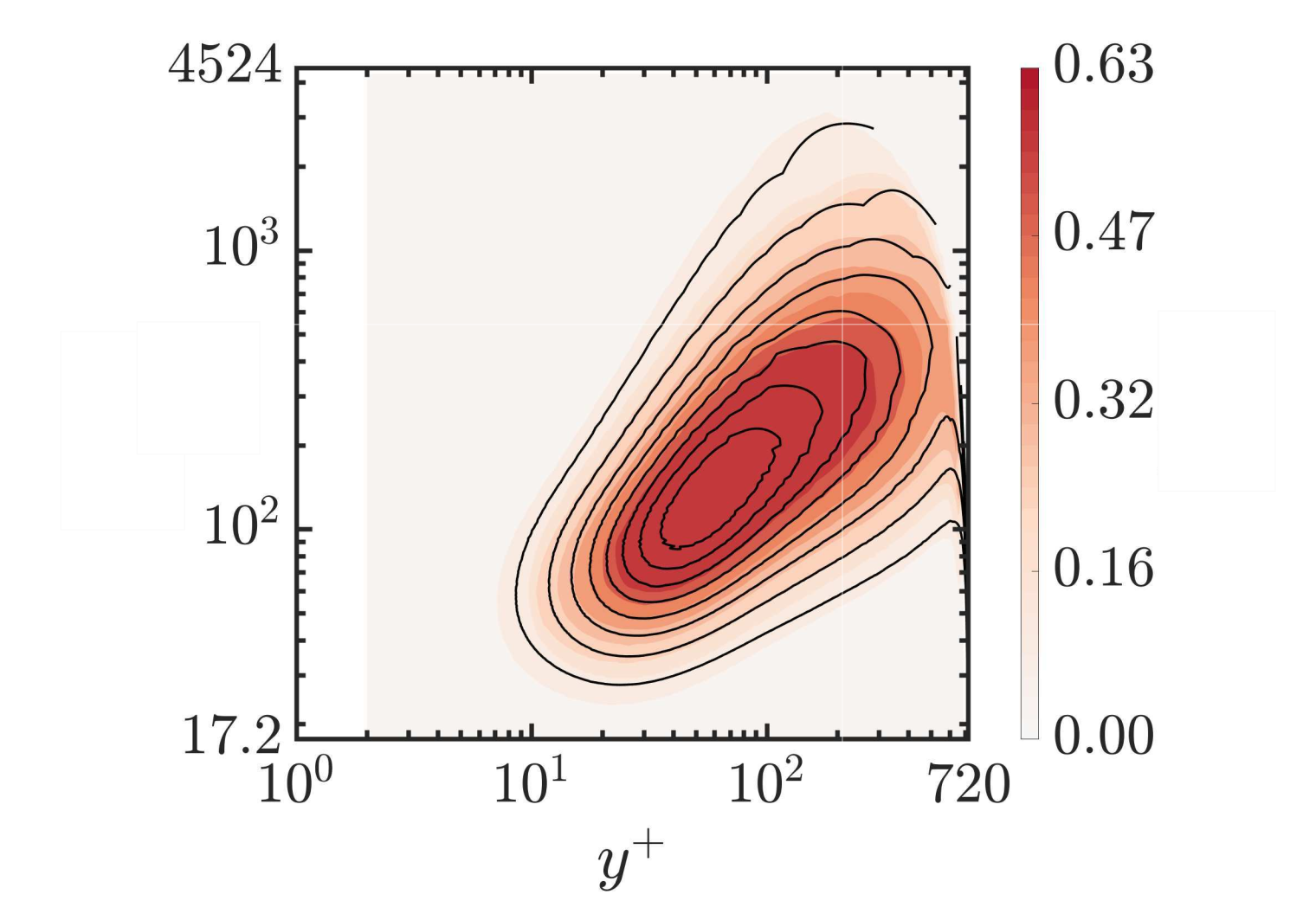}
        \caption{\centering  Radial velocity, $k_{\theta}\,\Phi_{u_r u_r}/u_{\tau}^2$ ($j = r$)}
        \label{fig:kSvvBaltz}
    \end{subfigure}
    \begin{subfigure}[t]{.33\linewidth}
        \includegraphics[height=1.4in,center]{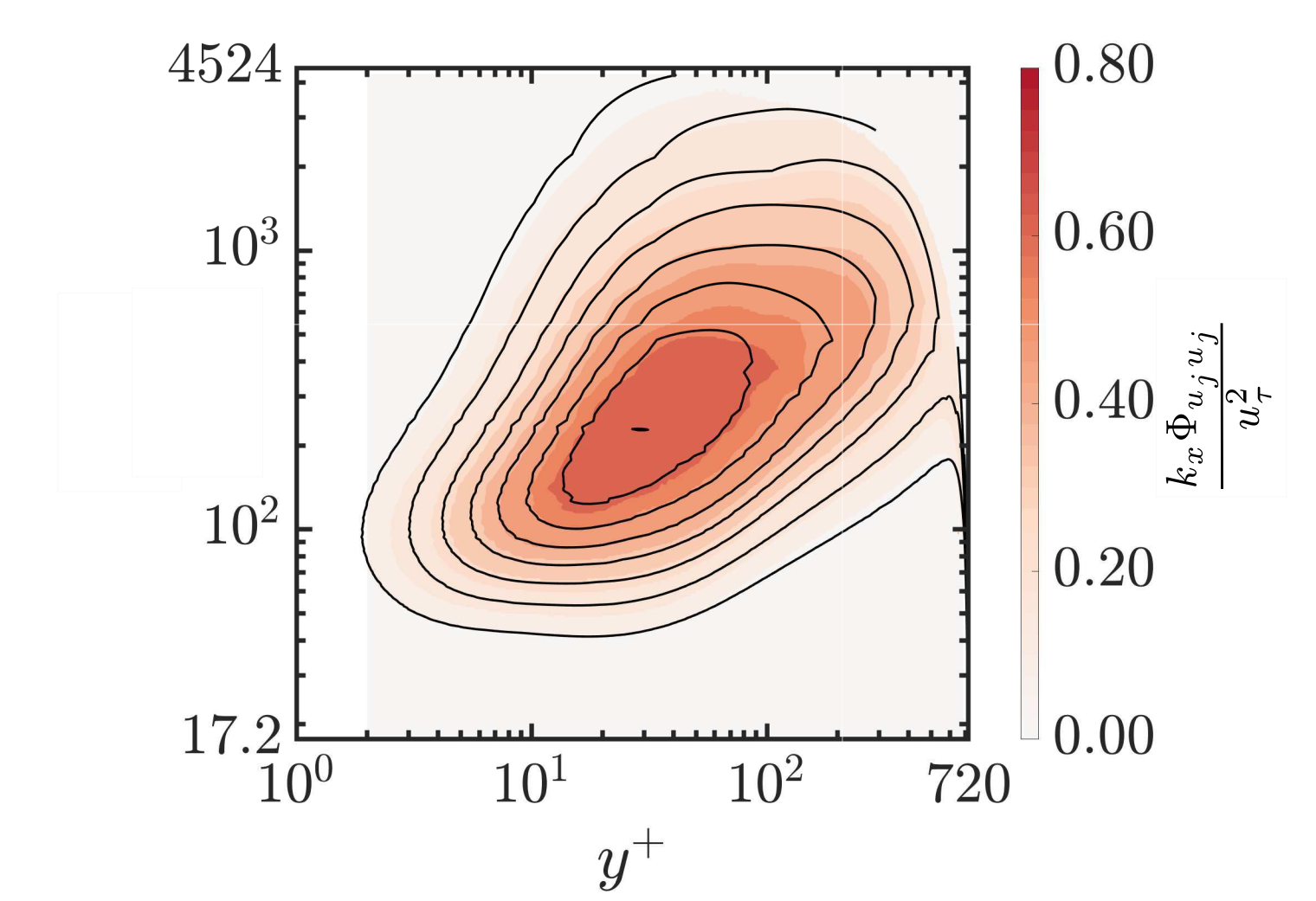}
        \caption{\centering  Azimuthal velocity, $k_{\theta}\,\Phi_{u_\theta u_\theta}/u_{\tau}^2$ ($j = \theta$)}
        \label{fig:kSwwBaltz}
    \end{subfigure}
    \caption{Premultiplied azimuthal spectrum for the $\ret = 720$ case as compared with the work of \cite{wu2012direct} for (a) streamwise velocity, $k_{\theta}\,\Phi_{u_x u_x}/u_{\tau}^2$; (b) radial velocity, $k_{\theta}\,\Phi_{u_{r}u_{r}}/u_{\tau}^2$; and (c) azimuthal velocity, $k_{\theta} \,\Phi_{u_{\theta}u_{\theta}}/u_{\tau}^2$. Color snapshots, current DNS; black contour lines, data of \cite{wu2012direct}. Black contour lines are spaced by 0.1 starting from the minimum value for the radial and azimuthal velocity spectra, and by 0.4 for the streamwise velocity spectra.}
    \label{fig:azmspec720}
\end{figure}

\subsection{Turbulent pipe flow with wall oscillation (WWO)}\label{sec:validation_wwo}
This section presents a validation of the current DNS simulations for the case of a turbulent pipe flow with wall oscillation  (WWO). Figure~\ref{fig:oscpipeval} documents a comparison of the single-point velocity statistics with the previously available data. In particular, we compare the present DNS results with the DNS of a turbulent pipe flow with wall oscillation at $\ret=150$ \citep{duggleby2007effect}, and DNS of a turbulent channel flow with wall oscillation at $\ret=1000$ \citep{agostini2014spanwise}.  The latter dataset is chosen for comparison, since no data for a turbulent pipe flow with wall oscillations is available past $\ret=180$~\citep{ricco2021review}. Figure~\ref{fig:oscpipeval} shows that the computed statistics in the WWO cases is within the range of the previously published data.

\begin{figure}
    \centering
    \begin{subfigure}{0.49\linewidth}
        \includegraphics[width=0.98\linewidth]{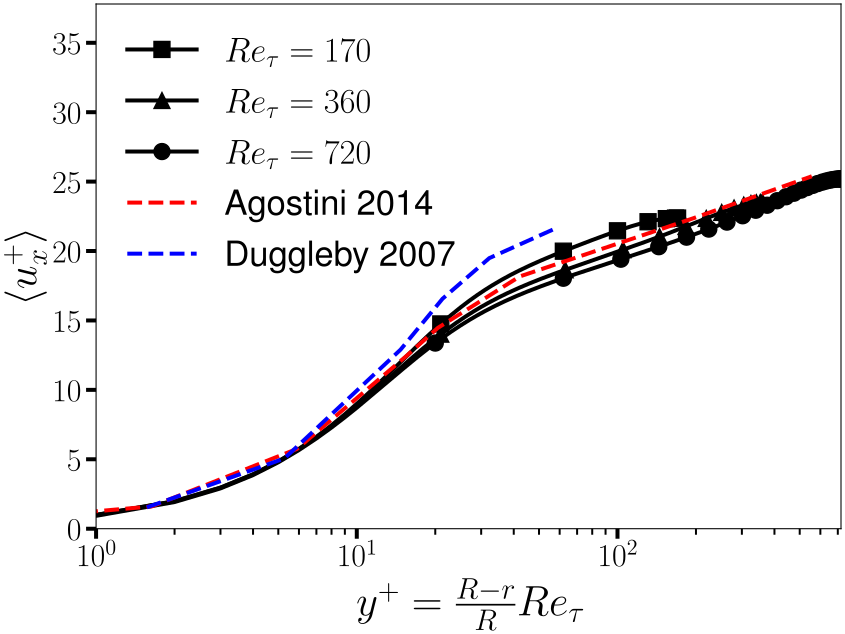}
        \caption{Mean streamwise velocity}
        \label{fig:ubarwwo}
    \end{subfigure}%
    \begin{subfigure}{0.49\linewidth}
        \includegraphics[width=0.98\linewidth]{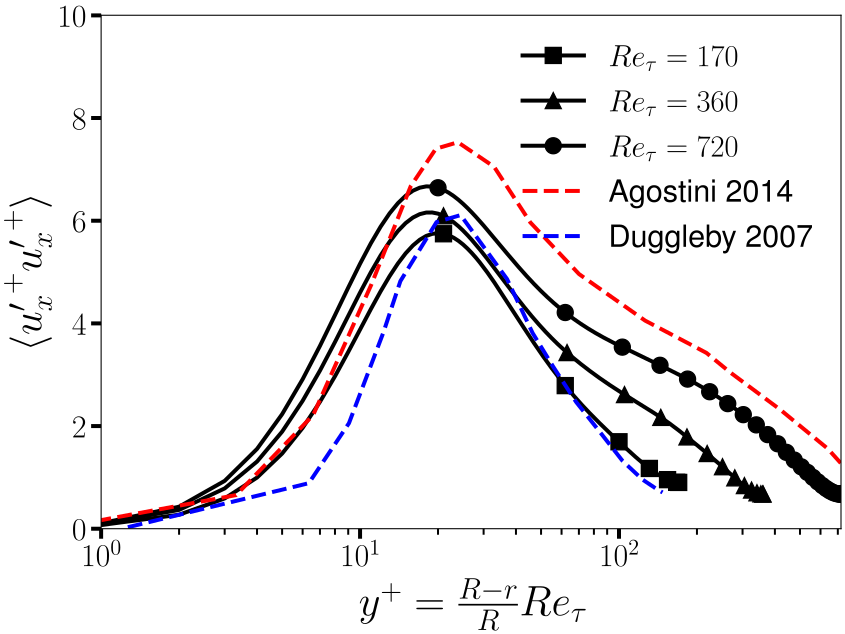}
        \caption{Streamwise velocity fluctuations}
        \label{fig:uubarwwo}
    \end{subfigure}
    \begin{subfigure}{0.49\linewidth}
        \includegraphics[width=0.98\linewidth]{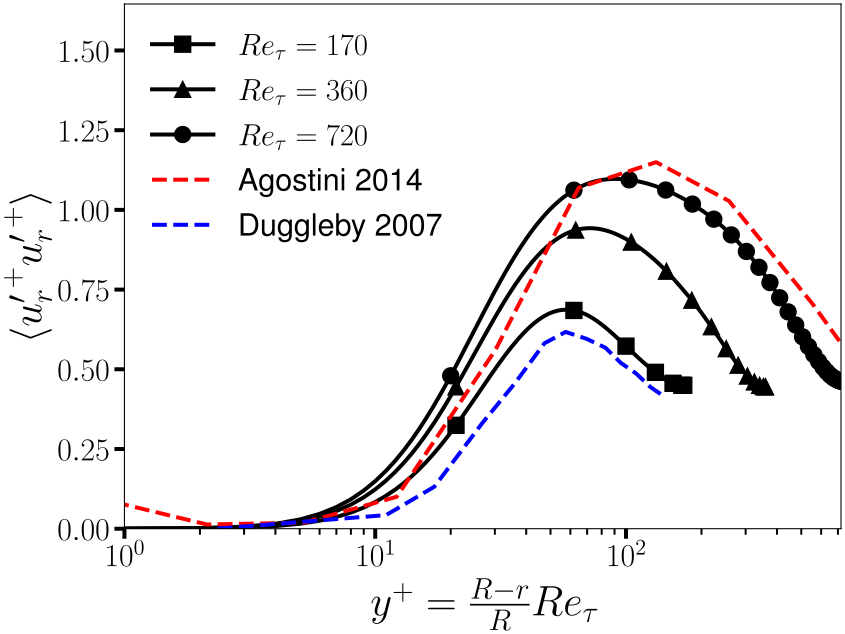}
        \caption{Radial velocity fluctuations.}
        \label{fig:vvbarwwo}
    \end{subfigure}%
    \begin{subfigure}{0.49\linewidth}
        \includegraphics[width=0.98\linewidth]{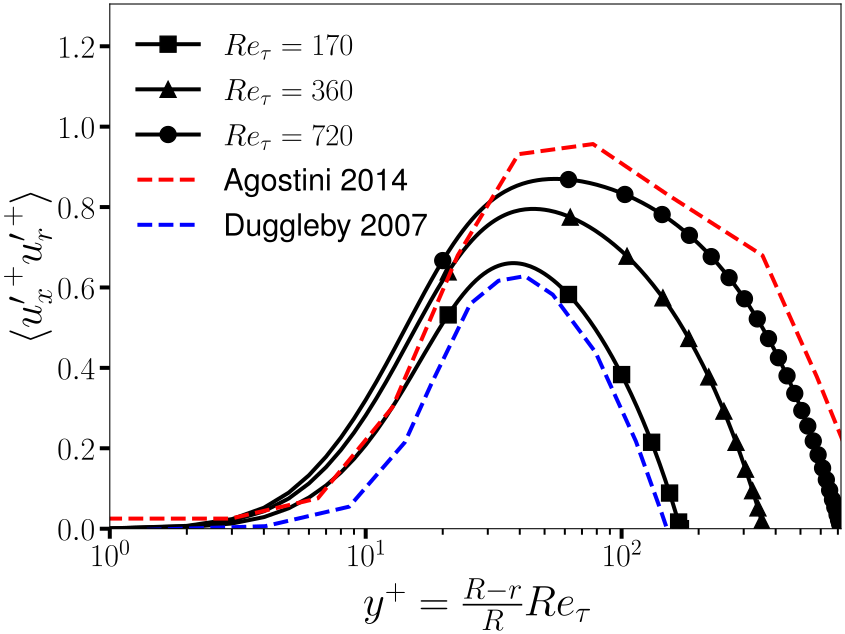}
        \caption{Reynolds shear stress}
        \label{fig:uvbarwwo}
    \end{subfigure}
    \caption{Validation of statistical quantities for the DNS of turbulent pipe flow with Nek5000 (with wall oscillation): (a) mean streamwise velocity, (b) streamwise velocity fluctuations, (c) radial velocity fluctuations, and (d) Reynolds shear stress. Black lines with symbols, current DNS at $\ret=170$, 360 and 720 (See the legend); blue dashed line, DNS of pipe flow with wall oscillation at $Re_{\tau}=150$ \citep{duggleby2007effect};  red dashed line, DNS of channel flow with wall oscillation at $Re_{\tau}=1000$ \citep{agostini2014spanwise}.}
     \label{fig:oscpipeval}
\end{figure}

\section{FIK identity for the bulk mean velocity in a turbulent pipe flow with oscillating walls}\label{sec:derivation_fik}

This appendix derives an analogue of the Fukagata-Iwamoto-Kasagi (FIK) identity~\citep{fukagata2002contribution} for the bulk mean velocity in a turbulent pipe flow with and without the oscillating walls. 
We start with the ensemble-averaged streamwise momentum equation (\ref{eqn:stwMomAt}), 
multiply it by $r$ and integrate across the vertical coordinate  as $\int_r^R (\cdot) \,r\,dr$ to yield:
\begin{equation}
     \nu\, r  \frac{d \av{u_x}}{d r}= r \av{\upr \vpr}  + \frac{r^2}{2\rho}\av{\frac{\partial\, p}{\partial \,x}}.
    \label{eqn:stwStress}
\end{equation}
Application of the boundary conditions at $r=R$, together with the equation (\ref{eqn:dpdx}) and the definition of the mean wall shear stress $\av{\tau_w}=\left(-\rho\,\nu\,d\av{u_x}/d r\right)\big|r=R$ was used to arrive at (\ref{eqn:stwStress}). 

We proceed in the same way, multiplying equation (\ref{eqn:stwStress}) by $r$ and integrating it as $\int_r^R (\cdot) \,r\,dr$ again. We obtain the following relation:
\begin{equation}\label{eqn:fikdim}
    2\nu \int_r^R \av{u_x} r dr +  \nu \av{u_x} r^2= -\frac{R^4}{8\rho} \av{\frac{\partial\, p}{\partial \,x}}\left( 1 - \left(\frac{r}{R}\right)^4 \right)  - \int_r^R \av{\upr \vpr} r^2 dr.
\end{equation}
Recasting equation (\ref{eqn:fikdim}) into the non-dimensional coordinates $\rstar=r/R$, $u_i^+=u_i/\utau$, and utilizing equation (\ref{eqn:dpdx}) once again yields:
\begin{equation}\label{eqn:fiknon}
    2\int_\rstar^1 \av{u_x^+} \rstar d\rstar+ \av{u_x^+} \rstar^2 = \frac{\ret}{4} (1-\rstar^4)  - \ret \int_\rstar^1 \av{\upr^+ \vpr^+} \rstar^2 d\rstar.
\end{equation}

The first term on the right-hand side of equation~(\ref{eqn:fiknon}) represents the cumulative laminar contribution to the bulk mean velocity, while the last term corresponds to the cumulative turbulent contribution, already presented in equation (\ref{eq:cumulative}). Evaluating equation~(\ref{eqn:fiknon}) at $\rstar=0$ and using the definition of bulk mean velocity (\ref{eqn:Ubulk}) cast into a non-dimensional form as $ U_{bulk}^+=2\int_0^1 \av{u_x^+} \rstar d\rstar$ gives the FIK identity:
\begin{equation}\label{eqn:fik2}
    U_{bulk}^+ = \frac{Re_\tau}{4} - \ret \int_0^1 \av{\upr^+ \vpr^+} \rstar^2 d\rstar,
\end{equation}
In a wall oscillated flow with a temporally-periodic mean, as per triple decomposition (\ref{eqn:triple})~\citep{hussain1970mechanics}, the fluctuating Reynolds stress $\av{\upr^+ \vpr^+} =\av{u_x^{\phi+} u_r^{\phi+}}+\av{\upp^+ \vpp^+}$ contains a phase-dependent component,  $\av{u_x^{\phi+} u_r^{\phi+}}$,   and an uncorrelated turbulent fluctuating component,  $\av{\upp^+ \vpp^+}$. We show in Figure \ref{fig:uphivphi} that the contribution from a phase-dependent component, $\av{u_x^{\phi+} u_r^{\phi+}}$, to the total Reynolds stress is negligible, so that it can be well approximated by the fluctuating component only, $\av{\upr^+ \vpr^+} \approx\av{\upp^+ \vpp^+}$, yielding
\begin{equation}\label{eq:newfik}
    U_{bulk}^+ = \frac{Re_\tau}{4} - \ret \int_0^1 \av{\upp^+ \vpp^+} \rstar^2 d\rstar
\end{equation}
for the oscillated pipe flow. We note that in the absence of wall oscillation, where a phase-dependent component $\av{u_x^{\phi+} u_r^{\phi+}}=0$, both Reynolds stresses are identically equal:  $\av{\upr^+ \vpr^+} =\av{\upp^+ \vpp^+} $.
Equation~(\ref{eq:newfik}) is the same as equation (\ref{eqn:fik}) shown previously. Note that, since only streamwise mean momentum equation is used in the derivation of (\ref{eq:newfik}), while non-zero boundary conditions for the oscillating pipe wall are set on the azimuthal velocity, this does not change the derivation.

\begin{figure}
    \centering
    \begin{subfigure}{0.32\linewidth}
        \includegraphics[width=.95\linewidth]{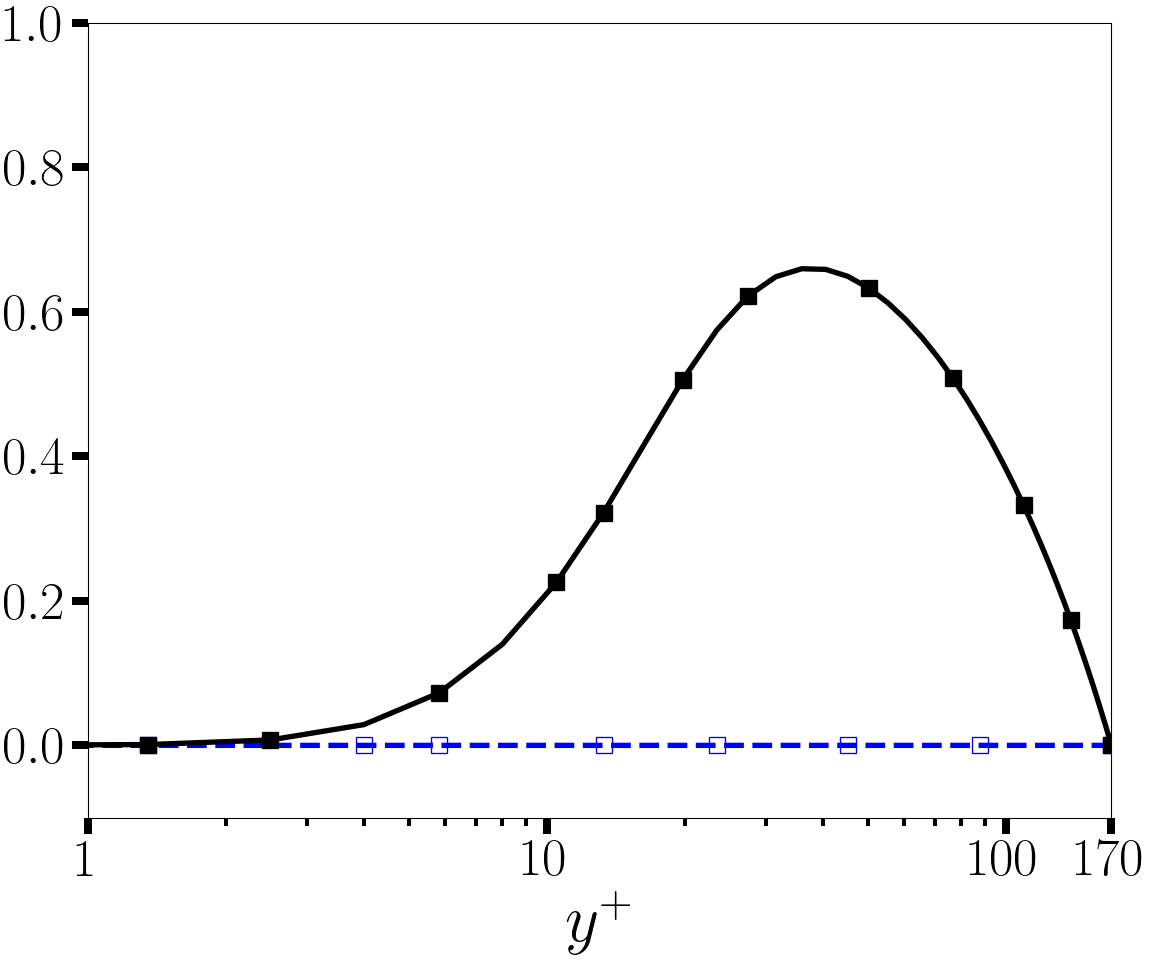}
        \caption{}
    \end{subfigure}
    \begin{subfigure}{0.32\linewidth}
        \includegraphics[width=.95\linewidth]{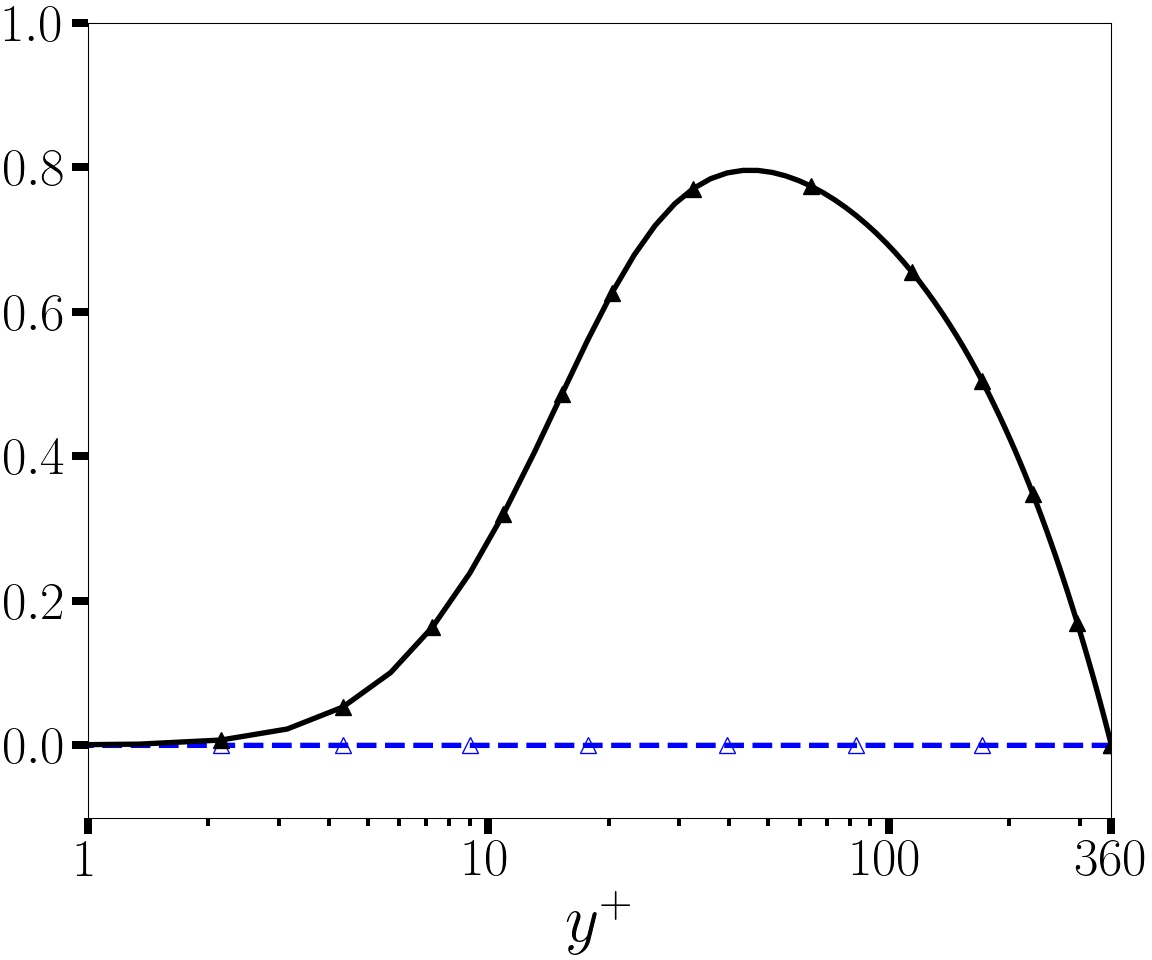}
        \caption{}
    \end{subfigure}
    \begin{subfigure}{0.32\linewidth}
        \includegraphics[width=.95\linewidth]{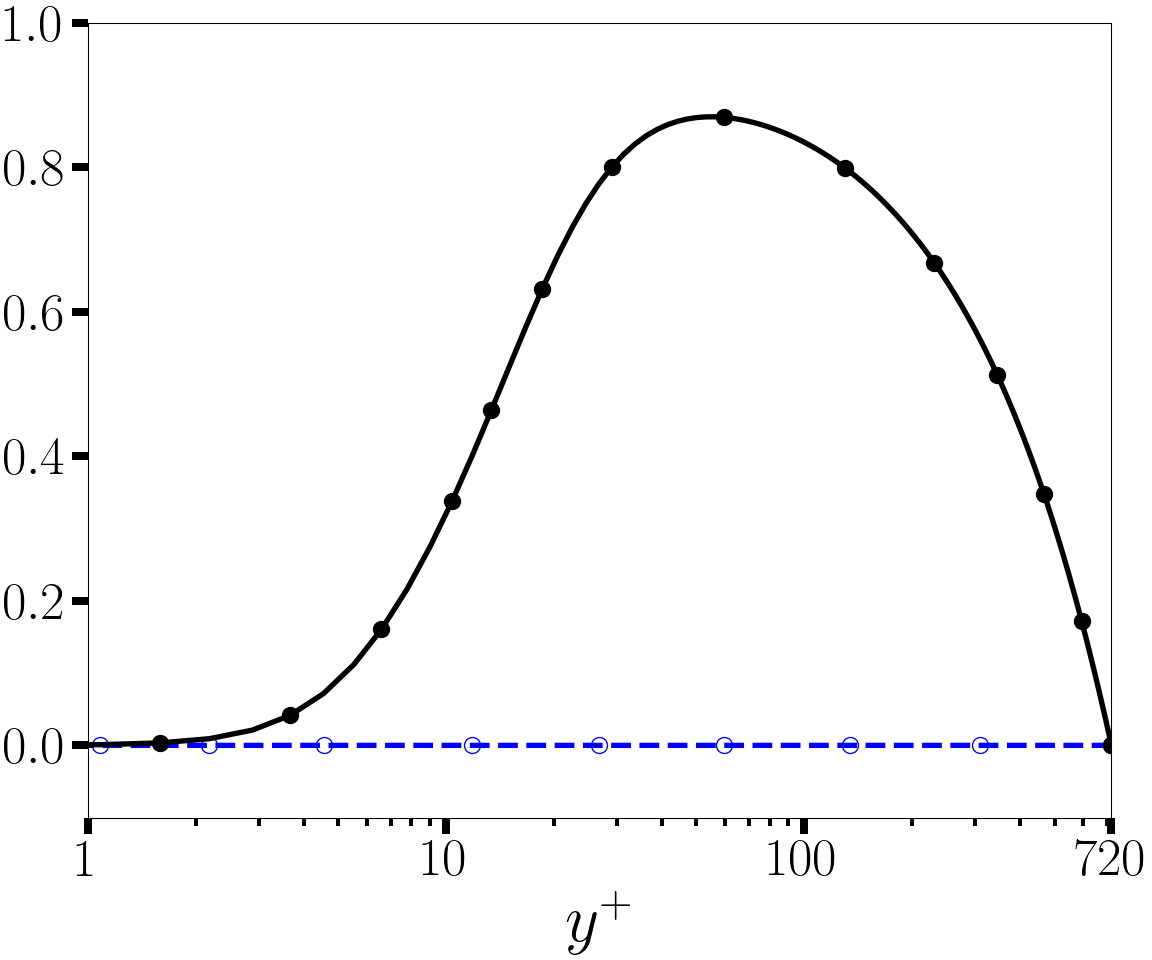}
        \caption{}
    \end{subfigure}
    \caption{Comparison of an uncorrelated turbulent fluctuating component of the Reynolds shear stress (filled markers, $\av{\upp^{+} \vpp^{+}}$) to  a phase-dependent component (empty markers, $\av{u_x^{\phi+} u_r^{\phi+}}$).  All 3 Reynolds number are represented by (a) $Re_\tau = 170$ with squares, (b) $Re_\tau = 360$ with triangles, and (c) $Re_\tau = 720$ with circles.  Black solid lines represent the . }
    \label{fig:uphivphi}
\end{figure}

The presented derivation can be easily extended to the skin friction coefficient, $C_f=2\av{\tau_w}/\rho \,\ubulk^2$, by evaluating equation~(\ref{eqn:fikdim}) at $r=0$, utilizing equation (\ref{eqn:dpdx}) to relate mean pressure gradient to the wall shear stress, and the definition of $Re_{bulk} =2 U_{bulk} R/\nu$, to yield
\begin{equation}\label{eqn:fik3}
    C_f = \frac{16}{\reb} +  8\int_0^1 \av{u_x^{\star} u_r^{\star}} \rstar^2 d\rstar,
\end{equation}
where we used the definitions $u_x^{\star}=\upp/\ubulk$, $u_r^{\star}=\vpp/\ubulk$ for velocities scaled in the outer units.

\bibliographystyle{jfm}
\bibliography{Bibliography,biblocal}

\end{document}